 \documentclass[twocolumn]{aastex6}
\usepackage{graphicx}
\usepackage{amsmath}

\begin{document}

\shortauthors{Ribas et al. 2017}
\shorttitle{Disks in Taurus, Ophiuchus, and Chamaeleon I}


\title{Far-infrared to millimeter data of protoplanetary disks: dust growth in the\\Taurus, Ophiuchus, and
  Chamaeleon~I star-forming regions$^*$.} \thanks{*\emph{Herschel} is an ESA space
  observatory with science instruments provided by European-led Principal
  Investigator consortia and with important participation from NASA.}


\author{\'Alvaro Ribas\altaffilmark{1},
  Catherine C. Espaillat\altaffilmark{1},
  Enrique Mac\'ias\altaffilmark{1}, 
  Herv\'e Bouy\altaffilmark{2},
  Sean Andrews\altaffilmark{3},
  Nuria Calvet\altaffilmark{4},
  David A. Naylor\altaffilmark{5},
  Pablo Riviere-Marichalar\altaffilmark{6},
  Matthijs H. D. van der Wiel\altaffilmark{5, 7}, and
  David Wilner\altaffilmark{3}
}



\altaffiltext{1}{Department of Astronomy, Boston University, Boston, MA 02215, USA; aribas@bu.edu}
\altaffiltext{2}{Laboratoire d'Astrophysique de Bordeaux, Univ. Bordeaux, CNRS, F-33615 Pessac, France}
\altaffiltext{3}{Harvard-Smithsonian Center for Astrophysics, Cambridge, MA 91023 USA}
\altaffiltext{4}{Astronomy Department, University of Michigan, Ann Arbor, MI 48109, USA}
\altaffiltext{5}{Institute for Space Imaging Science, Department of Physics \& Astronomy, University of Lethbridge, Canada}
\altaffiltext{6}{Instituto de Ciencia de Materiales de Madrid (CSIC). Calle Sor Juana In\'es de la Cruz 3, E-28049 Cantoblanco, Madrid, Spain}
\altaffiltext{7}{ASTRON, the Netherlands Institute for Radio Astronomy, PO Box 2, 7990AA Dwingeloo, The Netherlands}

\begin{abstract}

  Far-infrared and (sub)millimeter fluxes can be used to study dust in protoplanetary disks, the
  building blocks of planets. Here, we combine observations from the \emph{Herschel Space
    Observatory} with ancillary data of 284 protoplanetary disks in the Taurus, Chamaeleon~I, and
  Ophiuchus star-forming regions, covering from the optical to mm/cm wavelengths. We analyze their
  spectral indices as a function of wavelength and determine their (sub)millimeter slopes when
  possible. Most disks display observational evidence of grain growth, in agreement with previous
  studies. No correlation is found between other tracers of disk evolution and the millimeter
  spectral indices. A simple disk model is used to fit these sources, and we derive posterior
  distributions for the optical depth at 1.3\,mm and 10\,au, the disk temperature at this same
  radius, and the dust opacity spectral index $\beta$. We find the fluxes at 70\,$\mu$m to correlate
  strongly with disk temperatures at 10\,au, as derived from these simple models. We find tentative
  evidence for spectral indices in Chamaeleon~I being steeper than those of disks in
  Taurus/Ophiuchus, although more millimeter observations are needed to confirm this trend and
  identify its possible origin. Additionally, we determine the median spectral energy distribution
  of each region and find them to be similar across the entire wavelength range studied, possibly
  due to the large scatter in disk properties and morphologies.

\end{abstract}


\section{Introduction} \label{sec:intro}

Planetary systems form out of disks of gas and dust around young stars. However, the large number of physical
processes taking place within them \citep[e.g., accretion, photoevaporation, interaction with companions, dust
growth and settling, radial migration,][]{Takeuchi2002, Dalessio2006, Ireland2008, Alexander2014} require that
we consider several factors for their study. For this purpose, multi-wavelength observations of protoplanetary
disks can be used to better understand their properties.

The (sub)mm wavelength range is of particular interest for various reasons: at sufficiently long
wavelengths, disks become optically thin, and an estimate of their dust mass can be directly
obtained (via some assumptions) by simply measuring their flux
\citep[e.g.][]{Beckwith1990}. Although the bulk of the disk mass in the system is in gaseous phase,
fiducial (or measured, when available) gas-to-dust ratios provide an indirect estimate of the total
mass in the disk. This is a crucial parameter for planet formation theories because it determines
the available reservoir for this process. Using this method, surveys of star-forming regions with
(sub)mm facilities such as SMA and ALMA have determined that protoplanetary disks have typical
masses of 0.1-0.5\,\% of that of their host star \citep[e.g.][]{Andrews2005, Andrews2013,
  Pascucci2016}. On the other hand, dust growth represents the initial stage of planet formation;
the observed spectral index at these wavelengths can be linked to the dust opacity in the disk,
informative of its properties and grain sizes \citep[e.g.][]{Miyake1993, Draine2006}. In fact, the
comparison of the millimeter spectral index of the interstellar medium (ISM) with that of
protoplanetary disks has already revealed significant dust growth in these disks, implying the
presence of mm/cm-sized grains in many of them \citep[e.g.][]{Dalessio2001, Lommen2010,
  Ricci2010_Taurus, Ricci2010_Ophiuchus, Ubach2012}.  The combination of the mm spectral index with
additional information at other wavelengths, such as the spectral index at near/mid infrared (IR)
wavelengths or silicate features may also point to links between the evolution of the inner and
outer regions of the disks. As an example, \citet{Lommen2010} identified a tentative correlation
between the strength of the 10\,$\mu$m silicate feature and the 1-3\,mm spectral index for a sample
of T Tauri and Herbig~Ae/Be stars, suggesting a connection between the evolution of the inner and
outer regions of disks, although a later study by \citet{Ricci2010_Ophiuchus} found no signs of such
a correlation for disks in the Taurus and Ophiuchus star-forming regions.  Despite the obvious
interest of this wavelength regime, disks have relatively weak emission at millimeter wavelengths
and many of them currently lack this type of data (or, at least, sufficient observations to provide
robust estimates of their spectral indices).

At somewhat shorter wavelengths, the \emph{Herschel Space Observatory}
\citep[\emph{Herschel},][]{Herschel} observed large areas of the sky in the far-IR and sub-mm,
including several star-forming regions \citep[e.g. the Gould Belt
Survey,][]{Andre2010}. \emph{Herschel} probed the range between 50 and 150 \,$\mu$m, which is
sensitive to dust settling \citep[e.g.][]{Dalessio2006}, but also provided fluxes at longer
wavelengths (up to $\sim\,700$\,$\mu$m) probing deeper into the disk mid-plane.  Various studies
have already analyzed different aspects of \emph{Herschel} data in star-forming regions, both from
the photometric \citep[e.g.][]{Winston2012, Howard2013_Taurus, Olofsson2013, Ribas2013, Spezzi2013,
  Bustamante2015, Rebollido2015} and spectroscopic \citep[][]{Cieza2013_digit, Dent2013,
  Riviere-Marichalar2016} points of view.  On the other hand, a large comparative analysis of
\emph{Herschel} data of protoplanetary disks in different star-forming regions is still missing.

In this work, we compile multi-wavelength data of protoplanetary disks, including
homogeneous \emph{Herschel} photometry and spectroscopy, in three nearby star-forming
region: Taurus \citep[1-2\,Myr and $\sim$140\,pc,][]{Torres2007,Andrews2013}, Ophiuchus
\citep[0.3-5\,Myr and $\sim$140\,pc,][]{Wilking2008, OrtizLeon2017}, and Chamaeleon~I
\citep[2-6\,Myr and $\sim$160\,pc,][]{Whittet1997, Luhman2007}. The proximity of these
regions and the amount of available ancillary data guarantee good coverage of the spectral
energy distributions (SEDs) of several of their disks. Sec.~\ref{sec:observations}
describes our sample, data compilation, and processing. In Sec.~\ref{sec:obs_results} we
analyze different aspects of (sub)mm spectral indices and investigate observational
evidence of dust growth in these SEDs. In Sec.~\ref{sec:models}, we provide further
analysis by fitting the compiled data with a simple disk model. Sec.~\ref{sec:median_seds}
discusses and compares the median SEDs of these regions. Finally, our conclusions are
presented in Sec.~\ref{sec:conclusions}.

\section{Sample and data compilation} \label{sec:observations}

Our goal was to compile a representative sample of the disk population in the Taurus, Ophiuchus, and
Chamaeleon~I molecular clouds, while also ensuring a good coverage of their SEDs from the optical to
the far-IR, as well as the millimeter range when possible. We considered the 161 Taurus objects studied by
\citet{Furlan2011}, 134 objects in Ophiuchus in \citet[][]{McClure2010}, and the 84 objects in
Chamaeleon~I analyzed in \citet{Manoj2011}. These studies presented and analyzed \emph{Spitzer}/IRS
spectra of these disks, and performed a detailed study of the properties of their inner
regions. They also provided homogeneous compilations of the stellar properties of these
objects. Based on this and our intention to model these sources in more detail in a future study, we
selected these three sub-samples as our initial sample. To avoid disks with significant contribution
from their envelopes, we discarded envelope-dominated SEDs (as identified in these studies), which
were present both in the Ophiuchus and Chamaeleon~I samples. Our final sample comprises 315 objects:
161 in Taurus, 83 in Ophiuchus, and 71 in Chamaeleon~I.

\subsection{Herschel data}

Due to the different methods used to process \emph{Herschel} data in various studies and the inherent
difficulties of obtaining photometric and spectroscopic measurements in the presence of conspicuous background
emission (the cold dust in molecular clouds emits strongly at far-IR wavelengths), a coherent comparison of
these data is complex and has not yet been explored. To guarantee a homogeneous data set, observations of the
three regions were processed in the same manner.

\subsubsection{Herschel Photometry}

We processed a number of scan and cross scan maps available in the \emph{Herschel} Science
Archive to achieve a satisfactory coverage of the three regions considered in this
study. All of them were obtained by the the \emph{Herschel} Gould Belt Survey (P.I.:
Philippe Andr\'e), except for one set of observations in Ophiuchus (P.I.: Peter
Abraham). The corresponding OBSIDs, instruments, wavelengths, and pointing coordinates are
summarized in Table~\ref{tab:obsids_large} in Appendix~A.  After this process, a total
number of 18 objects in our sample lie outside the coverage of the large maps used in this
study. For these, we queried the \emph{Herschel} Science Archive to retrieve additional
(smaller) observations that contained these objects. We found PACS detections for 11 of
these sources. The corresponding OBSIDs and information for these data are also listed in
Appendix~A.

Maps at the three PACS wavelengths (70, 100, and 160\,$\mu$m) were processed using the
\emph{JScanam} algorithm \citep{Jscanam} within HIPE \citep[Herschel Interactive Processing
Environment,][]{Ott2010} version 14, combining scan and cross scan maps. In the particular case of
OBSIDs 1342202254 (scan), 1342202090 (cross scan 1) and 1342190616 (cross scan 2), these three maps
cover the same region of the sky, but \emph{JScanam} can only process scan\,+\,cross scan pairs. For
this reason, we produced two different maps with each scan and cross scan combination. We then
extracted PACS aperture photometry at the nominal coordinates of each object with the \emph{annular
  sky aperture photometry} task within HIPE, using aperture radii of 15'', 18'', and 22'' for 70,
100, and 160\,$\mu$m, respectively. These values were determined to be a good compromise based on
inspection of growth curves obtained in scan\,+\,cross scan maps. The background was estimated
within an annulus with radii of 25'' and 35'' centered around each object. We then applied the
corresponding aperture correction factors with the \emph{photApertureCorrectionPointSource} task,
corresponding to 0.83, 0.84, and 0.82 for 70, 100, and 160\,$\mu$m, respectively. Given the
different slopes of Class II SEDs in the PACS regime, we chose not to apply color corrections to
these fluxes---which, in any case, are significantly smaller than the assumed uncertainties (see
below).

SPIRE photometry was obtained at the three available wavelengths (250, 350, and 500\,$\mu$m) using
the recommended procedure of fitting sources in the timeline \citep{Pearson2014} within HIPE. The
level 1 data were previously corrected using the \emph{destriper} task. No extended emission gains
were applied because we do not expect any of these disks to be resolved in \emph{Herschel}/SPIRE
maps at their corresponding distances. The timeline fitting method does not require aperture
corrections, but we did apply color corrections in this case; at the longer SPIRE wavelengths, disks
are (at least partly) optically thin, and their emission at these wavelengths can therefore be
fitted with a power law. Based on mm spectral indexes by \citet{Ricci2010_Taurus}, we used an
intermediate power-law index of 2.3 and applied the corresponding color corrections. The uncertainty
from this parameter is, in any case, of only a few
percent\footnote{http://herschel.esac.esa.int/Docs/SPIRE/spire\_handbook.pdf}.

Reliable source detection in \emph{Herschel} maps of star-forming regions is a challenging task,
given the strong (and usually highly structured) background emission. We therefore performed visual
inspection of each source in all the available wavelengths to guarantee that we only include clean
point source detections. We discarded every source/band with extended objects, significant
contribution by nearby sources or the emission from the molecular clouds, or tentative/non
detections. For objects covered by more than one map, the median flux value was adopted. To account
for the effect of the aforementioned conspicuous background at \emph{Herschel} wavelengths, based on
previous \emph{Herschel} studies \citep[e.g.][]{Ribas2013,Rebollido2015}, we assigned a conservative
20\,\% uncertainty to each \emph{Herschel} photometric measurement. The resulting photometry,
together with objects that were discarded during the visual inspection process, are listed in
Appendix~B.

\subsubsection{Herschel/SPIRE Spectroscopy}

We also obtained SPIRE Fourier Transform Spectrometer (FTS) low-resolution
($\lambda/\triangle \lambda$ = 48 at 250\,$\mu$m) spectra for 113 objects (P.I.:
Catherine Espaillat, proposal ID: OT1\_cespaill\_1). These data cover wavelengths from 190
to 670\,$\mu$m, and were processed within HIPE 14 using the standard pipeline
\citep{Fulton2016}, which also reduces the long-wavelength artifacts produced when
operating the SPIRE FTS in low-resolution mode \citep{Marchili2017}. In addition to
standard processing, background subtraction is crucial at long wavelengths in star-forming
regions.  We inspected all SPIRE detectors for each object to discard undetected sources,
and removed those on top of isolated strong background emission that could yield an
overestimation of the true background flux. Once a reliable estimate of the background
flux was established, it was subtracted from the detector viewing the source. We applied
the pointing offset correction within HIPE \citep{Valtchanov2014} when possible, in order to
mitigate discontinuities between the two spectral windows. The extremities of the spectra
were then trimmed to avoid the lower S/N regions.  Finally, the resulting spectra were
compared with SPIRE data and archival photometry (see next section), and we discarded
those with obvious discrepancies with the photometry, mostly due to decreasing signal to
noise with increasing wavelengths. Thirty-four clean SPIRE spectra remained after this process,
and are available in the online version. The obsids of both clean and discarded SPIRE
  FTS spectra are listed in Appendix~A.

\floattable
\rotate
\begin{deluxetable*}{c c c c c}
\tablecolumns{5}
\tablewidth{0pt}
\tablecaption{Catalogs and Surveys Used in this Study \label{tab:catalogs}}
\tablehead{
\colhead{Catalog/Survey} &  \colhead{Telescope/Instrument(s)} & \colhead{Wavelength range} &
\colhead{Region}\tablenotemark{a} & \colhead{Reference}
}
\startdata
Sloan Digital Sky Survey (DR9) & SDSS telescope & 0.35 - 0.91\,$\mu$m & \ldots & \citet{Ahn2012} \\
AAVSO Photometric All Sky Survey (APASS) & Multiple telescopes & 0.44 - 0.76\,$\mu$m & \ldots & \citet{APASS}    \\
Carlsberg Meridian Catalog (DR15) & CMT & 0.62\,$\mu$m & \ldots & \ldots \\ 
Two Micron All Sky survey (2MASS) & 2MASS & 1.24 - 2.16\,$\mu$m & \ldots & \citet{2MASS}    \\
\emph{Wide Infrared Explorer (WISE)} & \emph{WISE} & 3.4 - 22\,$\mu$m & \ldots & \citet{WISE}  \\
Cores to disks (c2d) survey & \emph{Spitzer}/IRAC, MIPS & 3.6 - 24\,$\mu$m & Taurus, Ophiuchus & \citet{Evans2009}   \\
\emph{AKARI} mid-infrared survey & \emph{AKARI}/IRC & 9 - 18\,$\mu$m & \ldots & \citet{Ishihara2010}    \\
- & ALMA & 887\,$\mu$m  & Taurus & \citet{Ricci2014} \\
- & IRAM Plateau de Bure Interferometer & 3.2\,mm & Taurus & \citet{Pietu2014} \\
- & SCUBA/JCMT + literature & 350\,$\mu$m - 1.3\,mm & Ophiuchus & \citet{Andrews2007} \\
- & ALMA & 890\,$\mu$m  & Ophiuchus & \citet{Testi2016} \\
- & ATCA & 3.3\,mm & Ophiuchus & \citet{Ricci2010_Ophiuchus} \\
- & VLA & 4 - 7\,cm& Ophiuchus & \citet{Dzib2013} \\
- & \emph{Spitzer}/IRAC, MIPS & 3.4 - 24\,$\mu$m & Chamaeleon~I & \citet{Luhman2008a} \\
- & SEST & 1.3\,mm & Chamaeleon~I & \citet{Henning1993} \\
- & APEX/LABOCA & 870\,$\mu$m & Chamaeleon~I & \citet{Belloche2011} \\
- & ALMA & 887\,$\mu$m & Chamaeleon~I & \citet{Pascucci2016} \\
- & ATCA & 3\,mm - 6\,cm & Chamaeleon~I & \citet{Ubach2012} \\
\enddata
\tablenotetext{a}{Only for region-specific surveys.}  
\tablecomments{The majority of the data for Taurus
  objects were gathered from the compilation in \citet{Andrews2013}, and we refer the reader to this study for
  additional information.}
\end{deluxetable*}
\clearpage

\subsection{Archival data}

To complement \emph{Herschel} data, we queried a number of catalogs for photometry covering a broad wavelength
range. In the case of Taurus, data from the comprehensive compilation by \citet{Andrews2013} was used when
available. Ancillary photometry in the range of 60-160\,$\mu$m was found to be significantly noisy (possibly
due to the lower resolution and sensitivity of previous facilities/telescopes), and was excluded because
\emph{Herschel} data are now available. For the remaining Taurus objects, as well as for Ophiuchus and
  Chamaeleon, we cross-matched our sample with a number of surveys and catalogs, listed in
  Table~\ref{tab:catalogs}. The cross-match was performed by assigning the fluxes to the closest source within
  a radius of 3\arcsec, except for APEX/LABOCA, SCUBA, or VLA, where a 5\arcsec search radius was used due
  to their larger beam sizes.

We paid special attention to saturation magnitudes and the various flags (such as objects marked as extended)
in different observations.  It is likely that some of the compiled data suffer from undetected additional
problems (e.g., contamination by nearby sources) that may affect our analysis, particularly in the (sub)mm
domain; we therefore inspected each SED visually and discarded any photometric point clearly inconsistent with
the overall shape of the SED. This process also helps to identify and discard possible missmatches in
  the cross-matching process.

  \emph{Spitzer}/IRS \citep{IRS} spectra of these objects were also retrieved from the Cornell Atlas
  of \emph{Spitzer}/IRS Sources \citep[CASSIS,][]{CASSIS, CASSIS_HR}. CASSIS produces optimally
  extracted spectra (accounting for e.g. pointing shifts in the slit, local background), which are
  suitable for our purposes.  However, for some sources, we find issues in the automatic reduction
  by CASSIS (e.g. non-matching orders); in those cases we used the spectra from \citet{Furlan2011},
  \citet{McClure2010}, and \citet{Manoj2011}. If several spectra were available in CASSIS, we
  visually inspected them and chose those that better matched our compiled photometry, given that
  the mid-IR emission of disks can be variable \citep[e.g.][]{Espaillat2011, Morales2011}. Furthermore, we
  do not include the spectrum for T Tau, given the strong neighboring background emission and its
  inconsistency with the compiled SED.
 
  To our knowledge, this is the largest data compilation to date for Chamaeleon~I and Ophiuchus.  An
  example of one of the compiled, clean SEDs is presented in Table~\ref{tab:example_SED}. The whole
  data set (SEDs and available SPIRE and IRS spectra) is available for download in tar.gz
  packages. In addition, the entire data set is available in a Zenodo archive
  \dataset[10.5281/zenodo.889053]{https://doi.org/10.5281/zenodo.889053}.

\begin{deluxetable}{c c c}
\tablecolumns{4}
\tablewidth{0pt}
\tablecaption{Example of One of the Observed (Not De-reddened) SEDs\tablenotemark{a}: WX Cha \label{tab:example_SED}}
\tablehead{
\colhead{Wavelength} & \colhead{F$_\nu$} & \colhead{Reference}\\
\colhead{($\mu m$)} & \colhead{(mJy)}
}
\startdata
0.44 & 4.5$\pm$0.9 & \citet{APASS}\\
0.48 & 6$\pm$1 & \citet{APASS}\\
0.55 & 11$\pm$2 & \citet{APASS}\\
0.62 & 19$\pm$4 & \citet{APASS}\\
0.76 & 35$\pm$4 & \citet{APASS}\\
1.23 & 186$\pm$5 & \citet{2MASS}\\
1.66 & 320$\pm$10 & \citet{2MASS}\\
2.16 & 432$\pm$7 & \citet{2MASS}\\
3.6 & 450$\pm$20 & \citet{Luhman2008a}\\
4.5 & 450$\pm$20 & \citet{Luhman2008a}\\
4.6 & 430$\pm$10 & \citet{WISE}\\
5.8 & 400$\pm$20 & \citet{Luhman2008a}\\
8.0 & 400$\pm$20 & \citet{Luhman2008a}\\
9.0 & 440$\pm$20 & \citet{AKARI}\\
12 & 370$\pm$20 & \citet{WISE}\\
18 & 420$\pm$30 & \citet{AKARI}\\
22 & 410$\pm$20 & \citet{WISE}\\
24 & 390$\pm$20 & \citet{Luhman2008a}\\
70 & 330$\pm$70 & This work\\
100 & 220$\pm$40 & This work\\
160 & 180$\pm$40 & This work\\
160 & 120$\pm$20 & This work\\
250 & 100$\pm$20 & This work\\
350 & 110$\pm$20 & This work\\
500 & 120$\pm$20 & This work\\
887 & 21$\pm$2 & \citet{Pascucci2016}\\
\enddata
\tablenotetext{a}{The IRS spectrum is available in the online version of the manuscript.}
\tablecomments{Similar datasets, including \emph{Spitzer}/IRS and \emph{Herschel}/SPIRE spectra are available for each of the considered sources in the online version of the manuscript.}
\end{deluxetable}

\subsection{De-reddening and stellar parameters}

The data were de-reddened using $A_V$ values for Ophiuchus \citep{McClure2010} and $A_J$ values
for Taurus and Chamaeleon~I \citep[][]{Furlan2011,Manoj2011}. We followed the procedure adopted
in \citet{McClure2010} to select the extinction law to be used for each target:
\begin{enumerate}
\item For $A_V < 3$, we use the extinction law in \citet{Mathis1990} with $R_V$=3.1.
\item For cases $3 \leq A_V < 8$ and $A_V > 8$, we use the corresponding the extinction laws in
  \citet{McClure2009}.
\end{enumerate}
In the following, sources with $A_V \geq 15$ are excluded from the analysis: these objects
are either highly embedded in their parental cloud or located behind a significant amount
of dust. In both cases, their spectral types (SpTs) are more uncertain, and such large extinctions
may create important features in the SED shape that could alter the result of our
analysis. Moreover, the obscuring dust will emit at longer wavelengths (longward of
far-IR), and both \emph{Herschel} photometry and ancillary data may be contaminated by
this emission. After this stage of the analysis, the sample size has been reduced to 284
YSOs: 154 in Taurus, 70 in Ophiuchus, and 60 in Chamaeleon~I.

\begin{figure}
\includegraphics[width=\hsize]{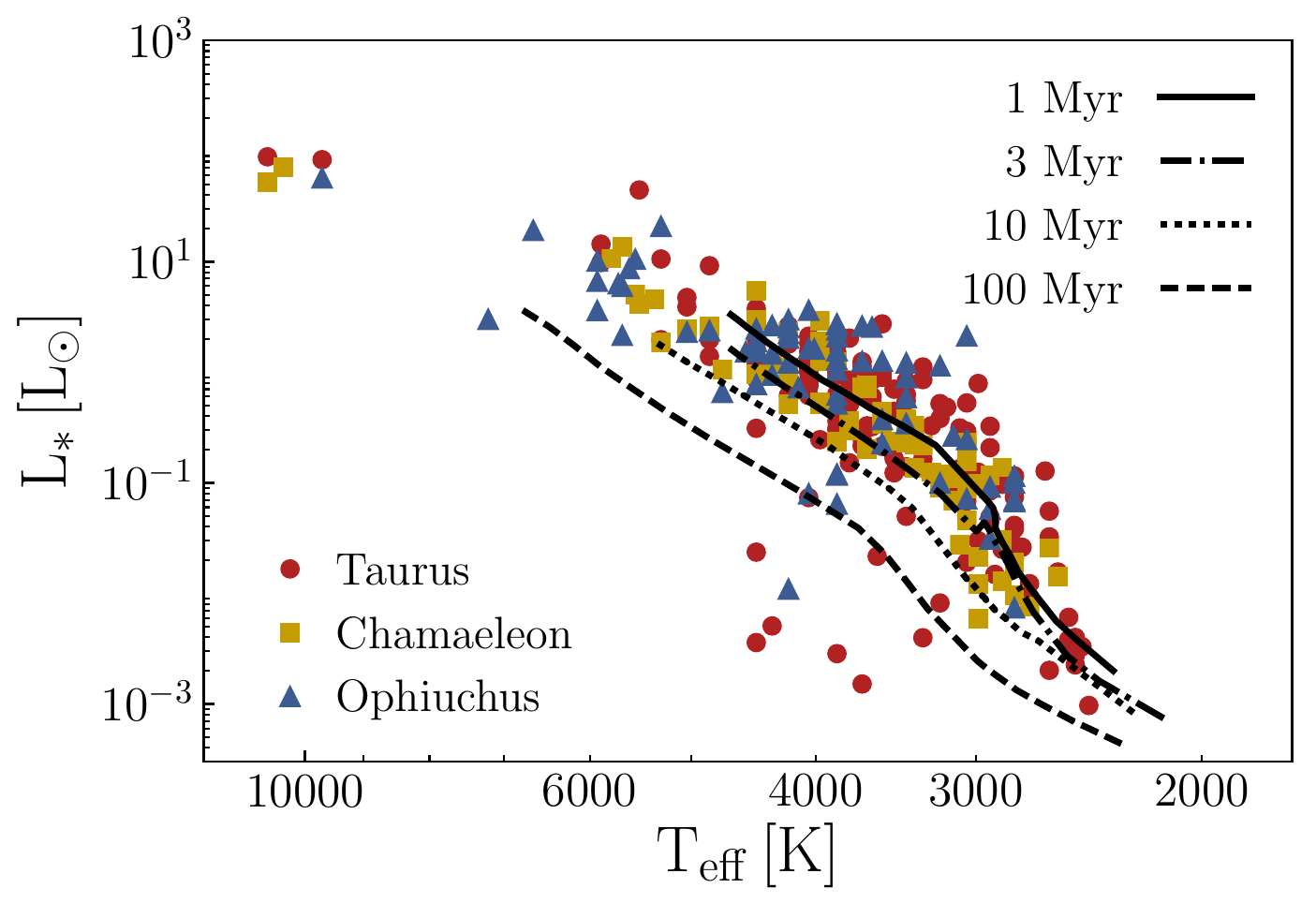}
\caption{HR diagram of the sample. Taurus objects are shown as red circles,
  Chamaeleon as yellow squares, and Ophiuchus members as blue
  triangles. Underluminous sources are likely YSOs with edge-on disks,
  self-extincting their stellar radiation. Isochrones from \citet{Baraffe2015} are
  also shown for comparison.}\label{fig:HR_diagram}
\end{figure}
 
To assign stellar parameters, we used SpTs listed in \citet{Furlan2011}, \citet{McClure2010}, and \citet{Manoj2011}. These were translated into stellar effective temperatures ($\rm T_{eff}$)
using the updated SpT--$\rm T_{eff}$ relation in \citet{Pecaut2013}. We then scaled the
corresponding BT-Settl photospheres \citep{Allard2012} to the de-reddened 2MASS $J$ fluxes
and computed the luminosities by integrating them in wavelength space at each region
distance. The resulting HR-diagram of the whole sample is shown in
Fig.~\ref{fig:HR_diagram}.  The adopted stellar parameters are available in the
  online version of the manuscript, and a reduced version can be found in
  Table~\ref{tab:sample_parameters}.

\floattable
\rotate
\begin{deluxetable}{c c c l c c c c}
\tablecolumns{8}
\tablewidth{0pt}
\tablecaption{Adopted Stellar Parameters for the Presented Sample \label{tab:sample_parameters}}
\tablehead{
\colhead{Name} & \colhead{R.A.} & \colhead{Decl.} & \colhead{Adopted. SpT} & \colhead{Adopted. Teff} & \colhead{Lum} & \colhead{Av} & \colhead{Region}\\
 & \colhead{(hh:mm:ss.ss)} & \colhead{(dd:mm:ss.s)} & & \colhead{(K)} & \colhead{(L$_\odot$)} & \colhead{(mag)}
}
\startdata
2MASS J04141188+2811535 & 04:14:11.88 & +28:11:53.5 & M6.25 & 2760 & 2.6e-02 & 2.5 & Taurus\\
2MASS J04153916+2818586 & 04:15:39.16 & +28:18:58.6 & M3.75 & 3250 & 3.3e-01 & 2.5 & Taurus\\
2MASS J04155799+2746175 & 04:15:57.99 & +27:46:17.5 & M5.5 & 2920 & 8.4e-02 & 1.9 & Taurus\\
2MASS J04163911+2858491 & 04:16:39.12 & +28:58:49.1 & M5.5 & 2920 & 4.6e-02 & 3.0 & Taurus\\
2MASS J04201611+2821325 & 04:20:16.11 & +28:21:32.6 & M6.5 & 2720 & 9.3e-03 & 0.8 & Taurus\\
2MASS J04202144+2813491 & 04:20:21.44 & +28:13:49.2 & M1 & 3680 & 1.5e-03 & 0.0 & Taurus\\
2MASS J04202606+2804089 & 04:20:26.07 & +28:04:09.0 & M3.5 & 3300 & 1.6e-01 & 0.0 & Taurus\\
2MASS J04210795+2702204 & 04:21:07.95 & +27:02:20.4 & M5.25 & 2990 & 3.0e-02 & 4.5 & Taurus\\
2MASS J04214631+2659296 & 04:21:46.31 & +26:59:29.6 & M5.75 & 2860 & 2.7e-02 & 4.2 & Taurus\\
2MASS J04230607+2801194 & 04:23:06.07 & +28:01:19.5 & M6 & 2800 & 3.8e-02 & 0.7 & Taurus\\
2MASS J04242090+2630511 & 04:24:20.90 & +26:30:51.2 & M6.5 & 2720 & 1.2e-02 & 0.8 & Taurus\\
2MASS J04242646+2649503 & 04:24:26.46 & +26:49:50.4 & M5.75 & 2860 & 2.5e-02 & 1.3 & Taurus\\
2MASS J04263055+2443558 & 04:26:30.55 & +24:43:55.9 & M8.75 & 2480 & 3.3e-03 & 0.2 & Taurus\\
2MASS J04284263+2714039 & 04:28:42.63 & +27:14:03.9 & M5.25 & 2990 & 1.2e-01 & 3.9 & Taurus\\
2MASS J04290068+2755033 & 04:29:00.68 & +27:55:03.4 & M8.25 & 2540 & 6.1e-03 & 0.2 & Taurus\\
2MASS J04295950+2433078 & 04:29:59.51 & +24:33:07.8 & M5 & 3050 & 2.4e-01 & 4.8 & Taurus\\
2MASS J04322415+2251083 & 04:32:24.15 & +22:51:08.3 & M4.5 & 3120 & 8.6e-02 & 1.3 & Taurus\\
2MASS J04330945+2246487 & 04:33:09.46 & +22:46:48.7 & M6 & 2800 & 4.1e-02 & 3.6 & Taurus\\
2MASS J04333905+2227207 & 04:33:39.05 & +22:27:20.7 & M1.75 & 3580 & 2.2e-02 & 0.0 & Taurus\\
2MASS J04334465+2615005 & 04:33:44.65 & +26:15:00.5 & M4.75 & 3090 & 3.1e-01 & 5.4 & Taurus\\
\enddata
\tablecomments{The complete version of Tables \ref{tab:sample_parameters},
  \ref{tab:mm_spectral_indices_individual}, \ref{tab:IR_and_silicates}, and
  \ref{tab:herschel_phot} are merged together in the Zenodo repository, also
  available in machine readable format in the online journal}
\end{deluxetable}

\section{Millimeter spectral indices and evidence for grain growth}\label{sec:obs_results}
The emission from protoplanetary disks at a given wavelength depends on several
factors, such as their morphology, dust composition, and stellar host properties. In
particular, the (sub)mm emission is informative of the mass and characteristics of
dust in disks. In this section, we investigate the observational evidence for grain
growth in the compiled data.

\subsection{Spectral indices versus wavelength}\label{sec:alpha_vs_lmb}
\begin{figure}
  \centering
  \includegraphics[width=\hsize]{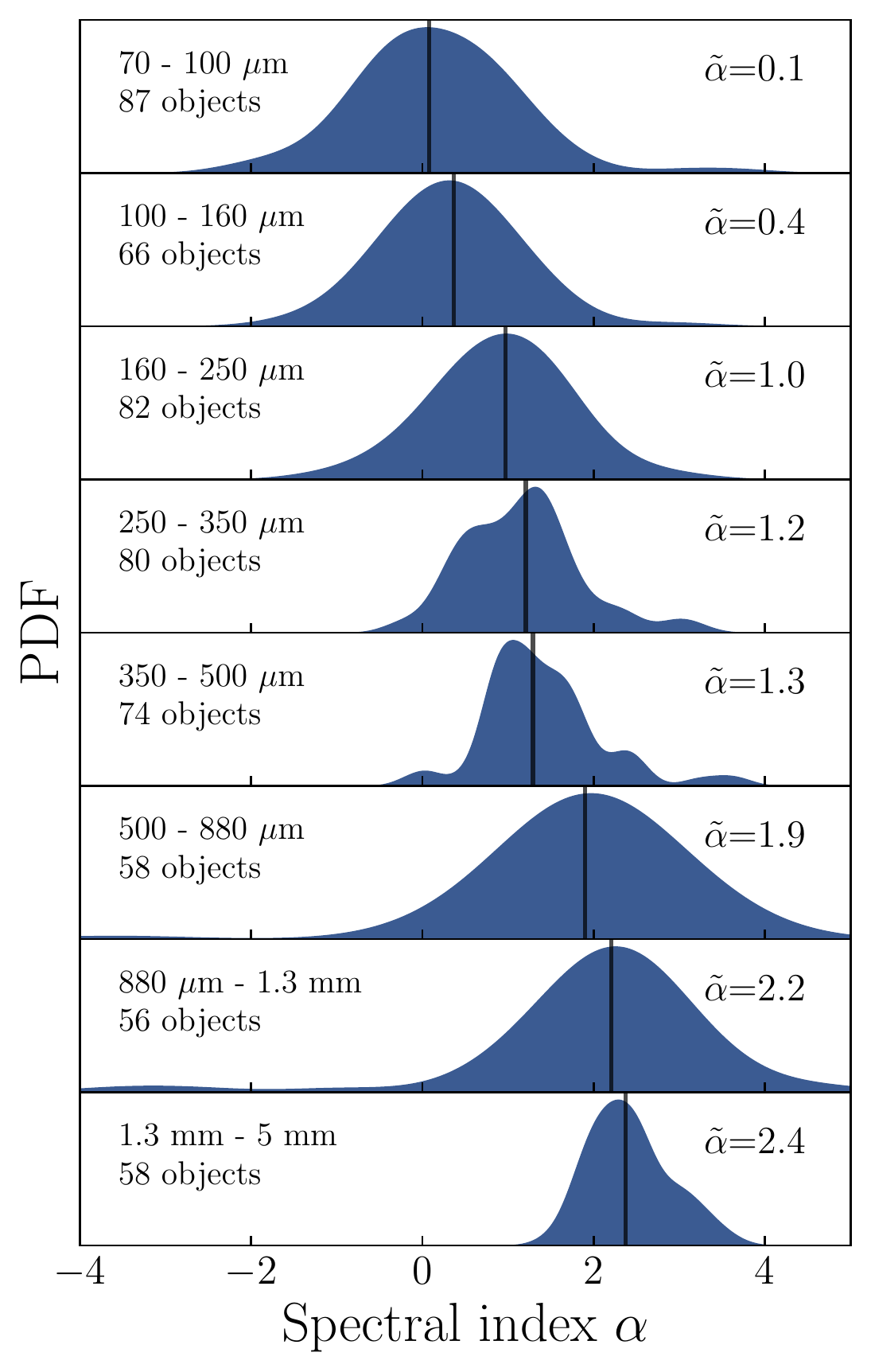}
  \caption{Probability density functions (PDFs) of the SED slopes ($\alpha$) of 
    considered sources measured at different wavelength ranges. The number
    of objects in each distribution and the median $\alpha$ values ($\tilde{\alpha}$,
    and black vertical line) are indicated in each case. Distributions shift to
    larger $\alpha$ values for increasing wavelengths, as emission approaches the
    Rayleigh Jeans regime and disks gradually become optically
    thin.}\label{fig:alpha_vs_lmb}
\end{figure}

Observations of protoplanetary disks in the (sub)mm range have two particularities: they
probe the Rayleigh Jeans (RJ) regime of the emission (unless the disk is abnormally cold),
and the opacity at these wavelengths is low enough for disks to be mostly optically
thin. When these two conditions are met (and assuming a power-law dependence of the
opacity with frequency), changing the wavelength does not affect the spectral index
($\alpha$) of the SED, and the emission from the disk follows $F_\nu \propto
\nu^\alpha$. We computed this spectral index
($\alpha = {\rm d} \log{F_\nu}/{\rm d} \log{\nu}$) at eight different wavelength ranges
for objects in the sample to investigate when it becomes independent of $\lambda$. The
wavelength ranges and corresponding $\alpha$ median values are listed in Table
\ref{tab:median_alpha_wavelengths}. These slopes were measured for each object with two or
more data points available in the corresponding range. Absolute $\alpha$ values larger
than five were discarded because they are unphysical (very likely they are the result of
individual problematic data). Figure~\ref{fig:alpha_vs_lmb} shows the obtained probability
distribution for each of these ranges \footnote{We prefer Gaussian Kernel Density Estimates
  (KDEs) over histograms, when possible, to present distributions, because the latter are
  sensitive to the choice of origin and widths of bins.}.  As expected, the median values
increase significantly from one range to the next for the shorter wavelength ranges, and
the distributions become very similar for $\alpha_{\rm 880-1.3}$ and $\alpha_{\rm 1.3-5}$
despite the significant change in wavelength. This suggests that the aforementioned
conditions (RJ regime and optically thin emission) are met for most disks in this range,
as typically assumed. The distribution of $\alpha_{\rm 500-880}$ is also close to those of
$\alpha_{\rm 880-1.3}$ and $\alpha_{\rm 1.3-5}$, implying that the deviations from these
conditions are small (at least for some disks) at these wavelengths.

\subsection{Measuring millimeter spectral indices}\label{sec:measure_alpha}

A significant number of protoplanetary disks lack enough (sub)mm data to estimate
$\alpha_{\rm mm}$. As suggested by Fig.~\ref{fig:alpha_vs_lmb}, it is possible that
the large amounts of SPIRE observations in the \emph{Herschel} Science Archive could
be used as an additional data set for this purpose, at the cost of introducing some
(systematic) uncertainty due to deviations from the RJ regime or optically thick
emission at these wavelengths.

\begin{figure*}
  \centering
  \includegraphics[width=\hsize]{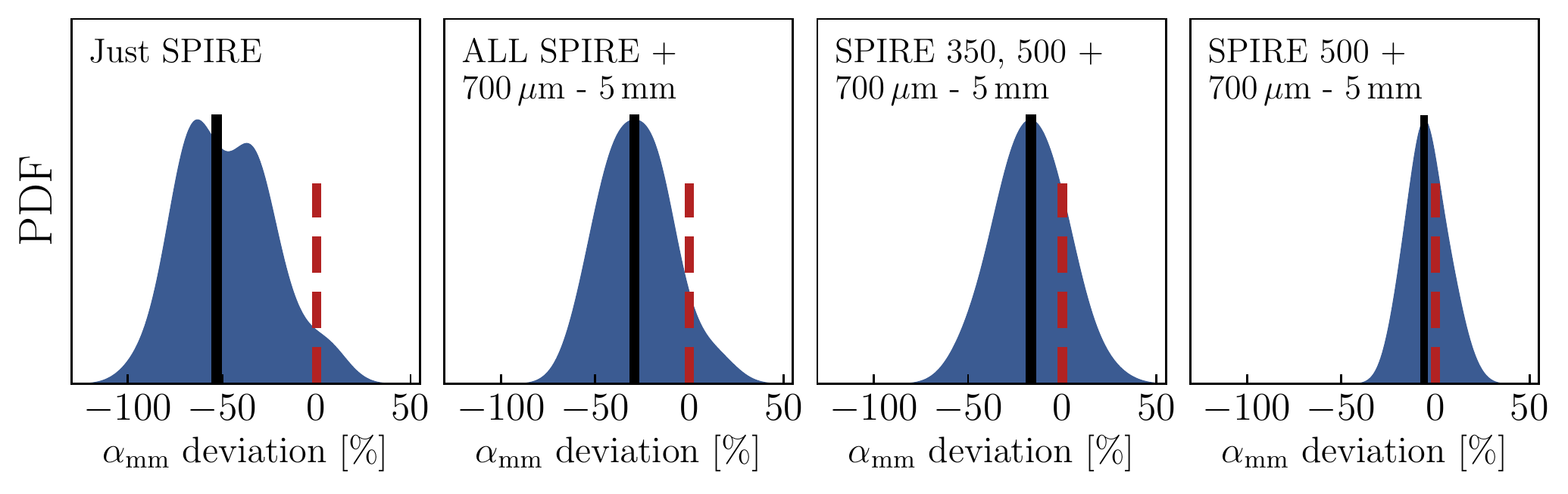}
  \caption{Probability density functions (PDF) of deviation from the ``true'' millimeter
    slope $\alpha_{\rm mm, true}$ (measured between 700\,$\mu$m and 5\,mm) for spectral
    indices computed in the four considered ranges: using SPIRE data only (left), using
    all SPIRE data + photometry between 700\,$\mu$m and 5\,mm (middle left), SPIRE 350 + SPIRE
    500 + photometry between 700\,$\mu$m and 5\,mm (middle right), and combining only SPIRE
    500\,$\mu$m with photometry between 700\,$\mu$m and 5\,mm (right). The median value of
    each distribution (-53\,\%, -29\,\%, -17\,\%, and -6\,\%, respectively) is shown as a
    black line, and the ideal case (no deviation) is shown as a red dashed line for
    reference. As expected, including shorter wavelengths in the analysis yields larger
    deviation in $\alpha_{\rm mm}$ estimates.}\label{fig:alpha_deviations}
\end{figure*}

The compiled data were used to quantify the deviation from the ``true''
$\alpha_{\rm mm}$ value produced by including SPIRE photometry in its
measurement. The ``true'' spectral index ($\alpha_{\rm mm, true}$) was defined as the
slope determined with all the available data between 700\,$\mu$m and 5\,mm; these
wavelengths are long enough to be mostly optically thin and in the RJ regime, yet
they include little contribution from other mechanisms such as free-free or
chromospheric emission \citep[][]{Pascucci2012}.  We computed $\alpha_{\rm mm, true}$
(when possible) for objects with at least three measurements in this
wavelength range, in order to make our estimates more robust against any problematic data point
that could have a significant effect in the results. We then computed spectral
indices at four different wavelength ranges to quantify their deviation from the
$\alpha_{\rm mm, true}$. The four wavelength ranges included:
\begin{enumerate}
\item SPIRE 250, 350, and 500 data,
\item SPIRE 250, 350, and 500 data + available photometry between 700\,$\mu$m and 5\,mm
\item SPIRE 350, and 500 data + available photometry between 700\,$\mu$m and 5\,mm, and
\item SPIRE 500 data + available photometry between 700\,$\mu$m and 5\,mm
\end{enumerate}
Values were obtained only for sources with at least three available data points in the
quoted regimes. The deviations of the different $\alpha_{\rm mm}$ values with respect to
$\alpha_{\rm mm, true}$ were computed as:
\begin{equation}
  Deviation = 100 \times \biggl(\frac{\alpha_{\rm mm, range}}{\alpha_{\rm mm, true}} - 1\biggr)
\end{equation}
where $\alpha_{\rm mm, range}$ is the slope measured for each of the four considered
cases. The results are shown in Fig.~\ref{fig:alpha_deviations}. Deviations are
largest when using SPIRE data only (ranging from $-93\,\%$ to $7\,\%$, and with a
median value of $-53\,\%$), as expected since this is the shortest wavelength range
considered. For cases combining SPIRE data with (sub)mm photometry from 700\,$\mu$m
to 5\,mm (as used to estimate $\alpha_{\rm true, mm}$), the most accurate values are
obtained excluding short SPIRE bands because the considered fluxes become closer to
the optically thin and RJ regimes. In particular, combining SPIRE 500\,$\mu$m
photometry only with (sub)mm data yields a median deviation of only 6\,\%, and in no
case more than 25\,\%.

\begin{deluxetable}{c c c}
\tablecolumns{4}
\tablewidth{0pt}
\tablecaption{Median Spectral Indices Computed at Different Wavelength Ranges \label{tab:median_alpha_wavelengths}}
\tablehead{
\colhead{Wavelength Range ($\mu$m)} & \colhead{Spectral Index} & \colhead{N. Objects}\\
\colhead{[$\mu$m]}
}
\startdata
65-105 & 0.1$_{-0.7}^{+0.9}$ & 87\\
95-165 & 0.4$_{-0.7}^{+0.7}$ & 66\\
155-255 & 1.0$_{-0.7}^{+0.7}$ & 82\\
245-355 & 1.2$_{-0.7}^{+0.5}$ & 80\\
345-505 & 1.3$_{-0.4}^{+0.6}$ & 74\\
495-890 & 1.9$_{-0.7}^{+0.9}$ & 58\\
860-1400 & 2.2$_{-0.9}^{+0.7}$ & 56\\
1200-5000 & 2.4$_{-0.4}^{+0.6}$ & 58\\
\enddata
\tablecomments{Uncertainties are derived from 16th and 84th percentiles.}
\end{deluxetable}

Based on these results, we chose to estimate millimeter slopes $\alpha_{mm}$ in the
following manner:
\begin{enumerate}
\item For objects with at least two data points between 800\,$\mu$m and 5\,mm, those
  data were used to estimate $\alpha_{\rm mm}$.
\item For objects without enough data in the previous case, but with at least two data
  points between 500 and 5\,mm, $\alpha_{\rm mm}$ was computed in that range, provided
  that at least one of the existing measurements was taken at wavelengths
  $\geq$\,1\,mm.
\item In all cases, the separation between the minimum and maximum available
  wavelengths was required to be at least 100\,$\mu$m to avoid issues when only two
  close measurements are available.
\end{enumerate}

With this criterion, millimeter spectral indices for 78 objects were obtained, as listed in
Table~\ref{tab:mm_spectral_indices_individual}. One problematic source, CU~Cha (one of the two
Herbig Ae/Be stars in Chamaeleon~I), was found to have a surprisingly high mm spectral index of
4.8. This value was computed using fluxes at 870\,$\mu$m and 1.3\,mm from \citet{Belloche2011} and
\citet{Henning1993}, and we were unable to identify any obvious problem in its SED. Such an $\alpha$
value is unlikely to be real \citep[e.g.][]{Dalessio2001, Ricci2010_Taurus}, and no other source in
our sample has a slope as steep as this object. More recent (sub)mm observations of this target by
\citet{Walsh2016}, using ALMA, obtained a totally different value of $\alpha_{\rm mm}$ value of 0.3
(unphysical for thermal emission in the millimeter), but they encountered calibration issues during
the observations and this value is highly uncertain. Later, \citet{vanderPlas2017} quoted an $\alpha$
value of 3.1\ measured from 1 to 10\,mm by including additional archival ATCA observations, which is
more similar to other values in Chamaeleon. For consistency, we chose to leave this source outside
the analysis.

\begin{figure}
  \centering
  \includegraphics[width=\hsize]{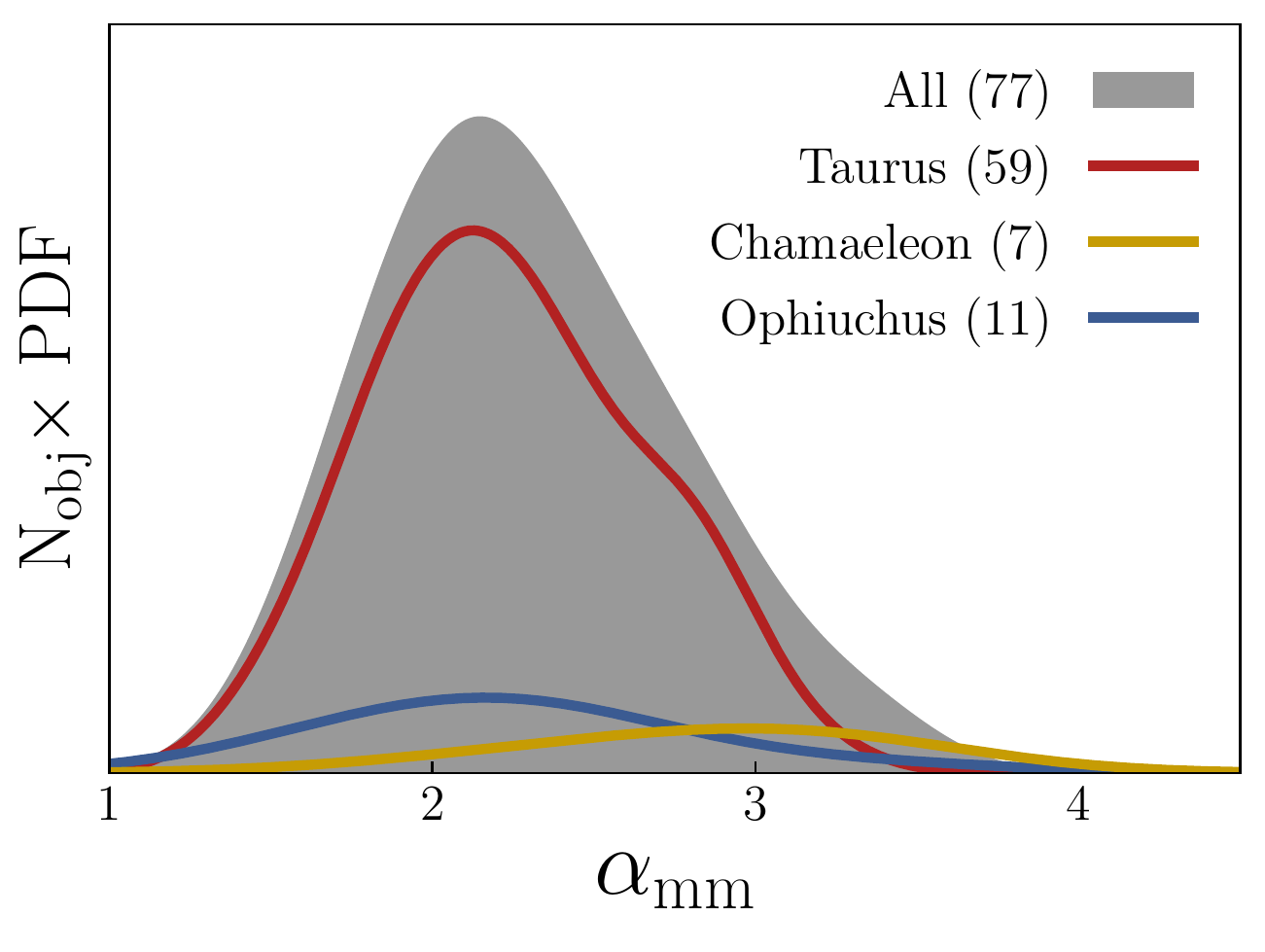}
  \caption{Distribution of millimeter SED slopes ($\alpha_{\rm mm}$, measured
    between 500/800\,$\mu$m and 5\,mm) for the sample. The overall distribution
    (gray), and the individual ones for Taurus (red), Chamaeleon~I (yellow), and
    Ophiuchus (blue) are shown, scaled to the corresponding number of objects as
    indicated in the legend.}\label{fig:alpha_distribution}
\end{figure}

The median of the distribution of millimeter spectral indices is 2.2, with values ranging
from 1.5 to 3.5. On a region-by-region basis, median values and ranges are 2.2 (1.5-3.2)
for Taurus (59 objects), 2.2 (1.7-3.3) for Ophiuchus (11 objects), and 3.0 (2.0-3.5) in
Chamaeleon~I (7 objects), in agreement with previous studies \citep[e.g.][]{Andrews2007,
  Ricci2010_Ophiuchus, Ricci2010_Taurus, Ubach2012, Andrews2013}. The distribution of
$\alpha_{\rm mm}$ is shown in Fig.~\ref{fig:alpha_distribution}. We note that the
  number of objects with measured spectral indices in Chamaeleon~I and Ophiuchus is
  significantly smaller than in Taurus due to the lack of enough (sub)mm data for many of
  their sources, and these results should be considered with caution for these two
  regions.

\begin{deluxetable*}{c c c c}
\tablecolumns{8}
\tablewidth{0pt}
\tablecaption{Adopted Millimeter Spectral Indices. \label{tab:mm_spectral_indices_individual}}
\tablehead{
\colhead{Name} & \colhead{$\alpha_{mm}$} & \colhead{Wavelength range ($\mu$m)} & \colhead{N. points}
}
\startdata
2MASS J04333905+2227207 & 2.2$_{-0.3}^{+0.3}$ & 500-1330 & 2\\
2MASS J04442713+2512164 & 2.0$_{-0.1}^{+0.1}$ & 869-3470 & 6\\
AA Tau & 2.0$_{-0.2}^{+0.2}$ & 863-2700 & 8\\
AB Aur & 2.9$_{-0.1}^{+0.1}$ & 850-2924 & 10\\
BP Tau & 2.7$_{-0.2}^{+0.2}$ & 869-3400 & 6\\
CFHT 4 & 2.1$_{-0.1}^{+0.1}$ & 869-3220 & 6\\
CIDA 1 & 2.0$_{-0.2}^{+0.2}$ & 887-3220 & 3\\
CI Tau & 2.1$_{-0.3}^{+0.3}$ & 869-2700 & 7\\
CW Tau & 2.9$_{-0.2}^{+0.2}$ & 1056-3560 & 4\\
CX Tau & 2.3$_{-0.2}^{+0.2}$ & 869-3477 & 3\\
\enddata
\tablecomments{The complete version of Tables \ref{tab:sample_parameters},
  \ref{tab:mm_spectral_indices_individual}, \ref{tab:IR_and_silicates}, and
  \ref{tab:herschel_phot} are merged together in the Zenodo repository, also
  available in machine readable format in the online journal. Uncertainties are
  derived from the 16th and 84th percentile levels from MCMC analysis.}
\end{deluxetable*}

\subsection{Dust growth in protoplanetary disks}\label{sec:dust_growth}

As previously mentioned, (sub)mm fluxes and spectral indices of protoplanetary disks
are related to their dust mass and properties. Under some assumptions (i.e. that the
wavelength is long enough for the disk to be optically thin, and that a single
temperature $T_c$ can be used to describe the emission at this wavelength), the
observed flux can be approximated as \citep{Hildebrand1983, Beckwith1990}:
\begin{equation}
F_{\nu} = \frac{B_\nu(T_c) M_{\rm d} \kappa_\nu}{d^2}
\end{equation}
where $B_\nu(T)$ is the Planck function at the characteristic temperature $T_c$,
$M_{\rm d}$ is the dust mass of the disk, $\kappa_\nu$ is the dust opacity at the
considered wavelength, and $d$ is the distance to the object. This has been routinely
used to estimate disk masses via (sub)mm surveys \citep[e.g.][]{Beckwith1990,
  Andre1994, Andrews2005, Andrews2007, Andrews2013}. In this equation, the opacity
value $\kappa_{\nu}$ is the main source of uncertainty, and the prescription of
\citet{Beckwith1990} is typically adopted:
\begin{equation}\label{eq:opacity}
\kappa_\nu = \kappa_0 \biggl(\frac{\nu}{\rm 10^{12} Hz}\biggl)^{\beta} {\rm cm^2 g^{-1}}
\end{equation}
where $\beta$ determines the change of the opacity with wavelength, and $\kappa_0$ is
the opacity value at $10^{12}$\,Hz
\citep[$\kappa_0$=0.01\,cm$^2$g$^{-1}$ in ][]{Beckwith1990}. If we also assume that
(sub)mm observations probe the RJ regime of $B_\nu(T)$, then $B_\nu(T)~\propto~\nu^2$
and the spectral index at these wavelengths is therefore:
\begin{equation}\label{eq:alpha}
\alpha = \frac{{\rm d} \log{F_\nu}}{{\rm d} \log{\nu}} = 2+\beta
\end{equation}
The interest of $\beta$ is that it depends on the properties of dust, i.e. the
particle size distribution (power-law index and maximum grain size $a_{\rm max}$) and
composition \citep{Dalessio2001, Natta2004a, Draine2006, Natta2007,
  Ricci2010_Taurus}. Therefore, $\beta$ can be used to probe grain growth in
protoplanetary disks \citep{Testi2014}, a crucial process in planet formation. For
ISM-like grains \citep[$a_{\rm min}$=0.005\,$\mu$m, $a_{\rm max}$=0.25\,$\mu$m,
$n(a) \propto a^{-p}$, $p=3.5$,][]{Mathis1977}, $\beta$ typically ranges between 1.6 and 1.8,
whereas increasing the maximum grain size to mm or cm sizes decreases $\beta$ to 0-1
(depending on the assumed power-law index). Low $\beta$ values have been found in
several protoplanetary disks \citep[e.g.][]{Calvet2002, Ricci2010_Ophiuchus,
  Ricci2010_Taurus, Ubach2012}, indicating dust growth from ISM sizes
\citep[][]{Beckwith1991, Miyake1993, Natta2004a}.

\begin{figure}
  \centering
  \includegraphics[width=\hsize]{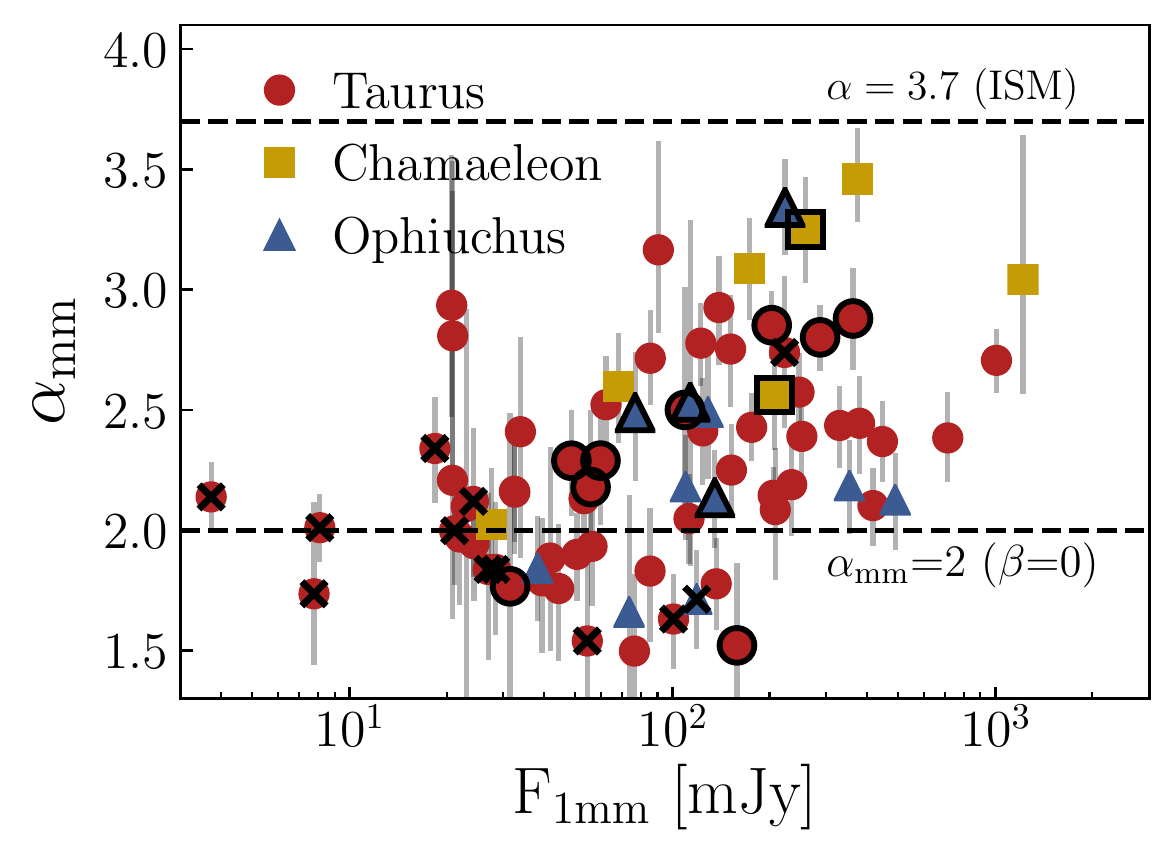}
  \caption{Predicted fluxes at 1\,mm (scaled to 140\,pc in the case of Chamaeleon~I) vs.
    spectral indices in the millimeter for objects in Taurus (red), Chamaeleon~I (yellow),
    and Ophiuchus (blue). M2 and later-type stars are marked with black crosses. Objects
    with a surrounding black border are classified as (pre)transitional disks by their
    13-31\,$\mu$m spectral index. The $\alpha$=2 line (corresponding to $\beta$=0 in the
    Rayleigh Jeans regime and for optically thin disks) and the $\alpha$ value of the ISM
    ($\beta=1.7$) are also shown.}\label{fig:Fmm_vs_alphmm}
\end{figure}

For sources with enough data to estimate $\alpha_{\rm mm}$ (see
  Sec.~\ref{sec:measure_alpha}), a line was fitted to their SEDs in
  $\log{\nu} - \log{F_{\nu}}$ space to predict the fluxes at 1\,mm for each
  source. Sources in Chamaeleon~I were scaled to 140\,pc to correct to its different
  distance (160\,pc) with respect to Taurus and
  Ophiuchus. Figure~\ref{fig:Fmm_vs_alphmm} shows these 1\,mm fluxes versus the
corresponding $\alpha_{\rm mm}$ values. Our results are very similar to those found in
previous studies \citep[e.g.][]{Ricci2010_Ophiuchus, Ricci2010_Taurus, Testi2014}. The
lack of sources with low $F_{\rm 1mm}$ and high $\alpha_{\rm mm}$ values is an
observational bias: for a given $F_{\rm 1mm}$, higher $\alpha_{\rm mm}$ values result in
more rapidly declining fluxes with increasing wavelength, and hence more challenging
detections \citep{Ricci2010_Taurus}.

An inspection of Figs.~\ref{fig:alpha_distribution} and~\ref{fig:Fmm_vs_alphmm} shows that
most objects have $\alpha_{\rm mm}$ values between 2 and 3. Following Eq.~\ref{eq:alpha},
this implies that most disks have $\beta \leq 1$, pointing to grain growth processes in
them. Given the young ages of sources in Ophiuchus and Taurus, this provides a robust
confirmation (with a larger sample and homogeneous treatment) of the fact that grain growth
from ISM-like to mm/cm sizes occurs quickly and early in the disk lifetimes, as already
found in previous studies \citep[e.g.][]{Rodmann2006,Ricci2010_Taurus}. The existence of
objects with $\alpha_{mm} < 2$ can not be explained with the relation
$\alpha = 2 + \beta$ because no physical dust model produces negative $\beta$ values
\citep[e.g.][]{Dalessio2001, Draine2006}. For those cases, it is likely that the
assumption of the RJ regime does not hold: emission from disks with a very cold mid-plane
\citep[e.g.][]{Guilloteau2016} may depart from $F_\nu \propto \nu^2$ significantly,
yielding flatter slopes.  Late-type stars (considered here arbitrarily as M2 or later type,
as a compromise; see crosses in Fig. ~\ref{fig:Fmm_vs_alphmm}) show $\alpha_{\rm mm}$
values below 2.5, with several of them even below 2. Although this may hint at important
dust growth around low-mass stars and brown dwarfs, disks around these objects can be both
colder and smaller than their counterparts in more massive stars, and therefore their
emission could be optically thick and/or outside the RJ regime. Resolved observations are
needed to unambiguously determine the origin of their low $\alpha_{\rm mm}$
values \citep[see discussion in][]{Ricci2014,Testi2016}. Additionally, Chamaeleon~I shows
an excess of high $\alpha_{\rm mm}$ - high $F_{\rm 1 mm}$ fluxes with respect to Taurus
and Ophiuchus, which is discussed later in the text (Sec.~\ref{sec:chamaeleon}).

\subsection{Millimeter indices and other tracers of disk evolution}\label{sec:other_tracers}

Using the compiled data, we also searched for correlations of the mm slopes measured in
Sec.~\ref{sec:measure_alpha} with other indicators of disk evolution; namely, the strength
and shape of the 10\,$\mu$m silicate feature probing dust growth in the upper layers of
disks, and tracers of cavities in them. The presence of gaps and cavities in the dust
spatial distribution in disks was first inferred in their SEDs due to a deficit of
near/mid-IR excess in some objects \citep{Strom1989}, and was later confirmed via direct
imaging \citep[e.g.][]{Andrews2011}. These particular disks with a hole are called
transitional disks (or pre-transitional disks, if they have ring-like gaps separating
their inner and outer regions), and they have gained significant attention in the last two
decades due to the exciting possibility of these gaps/holes being produced by forming
giant planets \citep[see][for a recent review]{Espaillat2014}. Because they lack material
in their inner regions, their near/mid-IR emission is reduced and they have steeper SEDs
at these wavelengths with respect to full disks, a fact that has been used in the past to
identify transitional disk candidates \citep[e.g.][]{Forrest2004,Brown2007, Merin2010,
  McClure2010}. The presence of giant planets directly implies that grains must have
suffered significant growth, and hence it is possible that dust in transitional disks may
have different properties. In the case of full disks, dust settling toward the disk
mid-plane also decreases the IR emission, producing a change in their slopes at these
wavelengths \citep[e.g.][]{Furlan2005}.

\begin{deluxetable*}{c c c c c}
\tablecolumns{5}
\tablewidth{0pt}
\tablecaption{IR spectral indices and 10\,$\mu$m silicate feature properties.\label{tab:IR_and_silicates}}
\tablehead{
\colhead{Name} & \colhead{$\alpha_{5.3-12.9}$} & \colhead{$\alpha_{12.9-31}$} & \colhead{Sil$_{\rm strength}$} & \colhead{Sil$_{\rm shape}$}
}
\startdata
2MASS J04141188+2811535 & -0.6$_{-0.1}^{+0.1}$ & -0.57$_{-0.08}^{+0.08}$ & 0.6$_{-0.3}^{+0.2}$ & 0.94$_{-0.07}^{+0.07}$\\
2MASS J04153916+2818586 & 0.47$_{-0.08}^{+0.09}$ & -1.43$_{-0.07}^{+0.07}$ & 0.11$_{-0.1}^{+0.08}$ & 0.96$_{-0.07}^{+0.08}$\\
2MASS J04155799+2746175 & 0.02$_{-0.08}^{+0.08}$ & 0.1$_{-0.1}^{+0.2}$ & 0.4$_{-0.11}^{+0.07}$ & 0.99$_{-0.05}^{+0.06}$\\
2MASS J04163911+2858491 & 0.5$_{-0.2}^{+0.2}$ & 0.0$_{-0.2}^{+0.2}$ & 0.1$_{-0.1}^{+0.2}$ & 1.0$_{-0.1}^{+0.1}$\\
2MASS J04201611+2821325 & -0.01$_{-0.08}^{+0.08}$ & 0.0$_{-0.3}^{+0.4}$ & 0.05$_{-0.05}^{+0.09}$ & 0.92$_{-0.05}^{+0.05}$\\
2MASS J04202144+2813491 & 0.0$_{-0.2}^{+0.3}$ & -1.2$_{-0.3}^{+0.3}$ & -0.3$_{-0.2}^{+0.2}$ & 2.1$_{-0.6}^{+1.7}$\\
2MASS J04202606+2804089 & -1.2$_{-0.1}^{+0.1}$ & -1.39$_{-0.06}^{+0.06}$ & 1.4$_{-0.2}^{+0.2}$ & 0.92$_{-0.05}^{+0.05}$\\
2MASS J04210795+2702204 & & & -0.1$_{-26.7}^{+0.7}$ \\
2MASS J04214631+2659296 & -0.0$_{-0.4}^{+0.4}$ & -0.2$_{-0.7}^{+3.2}$ & 0.2$_{-0.3}^{+0.7}$ & 0.9$_{-0.3}^{+0.3}$\\
2MASS J04230607+2801194 & 0.15$_{-0.07}^{+0.08}$ & -0.9$_{-0.1}^{+0.1}$ & 0.24$_{-0.07}^{+0.14}$ & 0.92$_{-0.05}^{+0.06}$\\
\enddata
\tablecomments{The complete version of Tables \ref{tab:sample_parameters},
  \ref{tab:mm_spectral_indices_individual}, \ref{tab:IR_and_silicates}, and \ref{tab:herschel_phot} are
  merged together in the Zenodo repository, also available in machine readable
  format in the online journal. Uncertainties derived from 16th and 84th
  percentile levels from 1000 bootstrapping iterations.}
\end{deluxetable*}

Dust growth is thought to occur mostly in the disk mid-plane, where the density is higher
and temperatures lower than in the upper layers \citep[see the review in][]{Testi2014},
where the 10\,$\mu$m silicate feature originates. Therefore, a relation between this
feature and $\alpha_{\rm mm}$ would imply a co-evolution of grains in the upper layers of
disks and their mid-plane. \citet{Lommen2007,Lommen2010} found a tentative correlation
between silicate strengths and shapes (a tracer of grain crystallinity) for YSOs in
different star-forming regions, but \citet{Ricci2010_Ophiuchus} did not find any in
Ophiuchus. A later study by \citet{Ubach2012} revealed only a weak, also tentative
correlation between the strength of this feature and the $\alpha_{\rm mm}$ for some
sources in Taurus, Ophiuchus, Chamaeleon, and Lupus.  We used the compiled IRS spectra
(when possible) to compute silicate strengths and shapes following \citet{Furlan2006} and
\citet{Kessler-Silacci2006} respectively. The resulting values are listed in
  Table~\ref{tab:IR_and_silicates}, and the process is described in more detail in
  Appendix~\ref{appendix:silicates}. Spearman rank tests revealed no significant
correlations between the strength/shape of the silicate feature and $\alpha_{\rm mm}$,
neither for any or these regions individually nor for the whole sample.

Near/mid-IR spectral indices $\alpha_{\rm IR}$ were also computed, following
\citet{McClure2010}, using the IRS spectra between 5.3 and 12.9\,$\mu$m (taken as the
median flux within a range of $\pm$0.2\,$\mu$m centered around each of these wavelengths),
and the slope between 12.9 and 31\,$\mu$m. Spearman rank tests between $\alpha_{\rm IR}$
and $\alpha_{\rm mm}$ showed these two quantities to be uncorrelated, both for the whole
sample and for each individual region. We also identified (pre)transitional disks by
selecting objects with spectral indices between 13 and 31\,$\mu$m $<$ -1.4 (in $F_\nu$
space, as computed from their IRS spectra), following the criterion in
\citet{McClure2010}. These sources are encircled in Fig.~\ref{fig:Fmm_vs_alphmm} for
comparison, but no obvious trend was found for them. These results suggest that the dust
population in the midplane/outer regions of transitional disks (at least those which gaps have a
detectable effect in the IR slope of their SEDs) is not substantially different than those
of their full counterparts.

\section{A simple disk model}\label{sec:models}

After the analysis of spectral indices in Sec.~\ref{sec:obs_results}, we applied the simple disk
model in \citet{Beckwith1990} to the compiled long-wavelength ($\geq 70\,\mu$m) data. This model
does not depict a physically self-consistent disk, but instead assumes that the emission arises from
a vertically isothermal one. The SED of such a disk can be written as:
\begin{equation}\label{eq:model_sed}
F_\nu = \frac{\cos{i}}{d^2} \int_{r_{in}}^{R_d}B_\nu(T(r))
(1-e^{-\tau_\nu(r) \sec{i}})2\pi r \,{\rm d}r
\end{equation}
where $i$ is the inclination, $d$ is the distance to the source, $r_{in}$ and $R_d$ are
the inner and outer radii of the disk, $B_\nu(T(r))$ is the Planck function at the
temperature $T(r)$, and $\tau_\nu(r)$ is the optical depth at the given frequency
$\nu$ and radius $r$. This optical depth is the product of the opacity at the
corresponding frequency ($\kappa_\nu$) and the radial surface density profile
($\Sigma(r)$). We assume the radial dependence of the temperature and surface density
to follow a power law:
\begin{equation}\label{eq:t_r}
T(r) = T_{0}\Biggl(\frac{r}{r_0}\Biggr)^{-q}
\end{equation} 

\begin{equation}\label{eq:sigma_r}
\Sigma(r) = \Sigma_0\Biggl(\frac{r}{r_0}\Biggr)^{-p}
\end{equation}
where $T_{0}$ and $\Sigma_0$ are the temperature and surface density at an arbitrary
radius $r_0$. The opacity law is also assumed to be a power law following
\citet{Beckwith1991}, as shown in Eq.~\ref{eq:opacity}. Therefore, the optical depth at a
given wavelength (frequency) and radius can be written as:
\begin{equation}
\tau_\nu(r) = \Sigma_0 \kappa_0 \biggl(\frac{r}{r_0}\biggr)^{-p} \biggl(\frac{\nu}{\rm 230\,GHz}\biggl)^{\beta}
\end{equation}
Here, $\kappa_0$ is the opacity value at 230\,GHz (we consider 230\,GHz instead of
  1000\,GHz as in Eq.~\ref{eq:opacity}, due to the common use of 1.3\,mm as the reference
  wavelength). However,
$\kappa_0$ also depends on the maximum grain size \citep[e.g.][]{Dalessio2001}, and it
should not be left constant when modeling while changing $\beta$ \citep[this could
introduce artificial trends in the modeling results, e.g.][]{Ricci2010_Taurus}. We
therefore combined $\Sigma_0$ and $\kappa_0$ into $\tau_{\nu, r_0}$, i.e. the optical
depth at the arbitrary radius $r_0$ (set to 10\,au in our study) and at 230\,GHz:
\begin{equation}
\tau_\nu(r) = \tau_{1.3\,mm, 10\,au} \biggl(\frac{r}{{\rm 10\,AU}}\biggr)^{-p} \biggl(\frac{\nu}{\rm 230\,GHz}\biggl)^{\beta}
\end{equation}
With this setup, there are a total number of eight free parameters in this model:
$\tau_{1.3\,mm, 10\,au}$, $r_{in}$, $R_d$, $T_{10 au}$, $p$, $q$, $i$, and
$\beta$. Because we will model far-IR and (sub)mm fluxes, the inner radius does not
have a crucial effect in our modeling and was fixed to 0.01\,au following
\citet{Andrews2005} - a rough estimate of where dust sublimation occurs
\citep{Dullemond2001, Muzerolle2003_sublimation}. Therefore, seven free parameters
remain.

Spectral indices $\alpha \sim 2$ can be produced both by
compact, optically thick disks (small $R_d$, typically $<$\,50\,au, high
$\tau_{1.3\,mm, 10\,au}$, and unconstrained $\beta$), or bigger, optically thin disks
with large dust grains (larger and unconstrained $R_d$, low $\tau_{1.3\,mm, 10\,au}$
and $\beta$ values). As a result, $\tau_{1.3\,mm, 10\,au}$, $R_d$, and $\beta$
estimates become degenerate in these cases from SED fitting alone. Observationally,
most resolved disks have been found to extend for about (or more than) 50-150\, au
\citep[][]{Andrews2010b, Ricci2010_Ophiuchus, Ricci2010_Taurus, Ricci2014}, but the
difficulty in resolving smaller and usually fainter disks introduces an important
bias. Recent high-resolution observations have identified a population of small disks
\citep[e.g.][]{Pietu2014, Osorio2016, Testi2016}, and consequently they cannot be
ruled out in the analysis. In an effort to break this degeneracy, we gathered disk
radii from \citet{Andrews2007} ,\citet{Ricci2010_Ophiuchus, Ricci2010_Taurus}, \citet{Pietu2014}, and
\citet{Pascucci2016}. In the last case, disk radii were estimated from FWHM
measurements converted to physical sizes using the distance to Chamaeleon~I, and
uncertainties of 25\,\% were assigned.

For each source, we aim at fitting data between 70\,$\mu$m and 5\,mm. We also included the
processed SPIRE spectra (when available) after binning them in five points to avoid
  giving them excessive weight, by simply dividing the corresponding wavelength range in five
  equal sub-ranges, and adopting the median flux value in each of them. For consistency
with our previous \emph{Herschel} data processing, we assigned 20\,\% uncertainties to
these data.  Inspection of uncertainties of the ancillary data revealed that many of them
were underestimated, probably due to a lack of the systematic contribution. We
circumvented the issue by assigning 20\,\% uncertainties to measurements with smaller
values. We note that the effect of this is to produce more conservative uncertainties in
our final estimates, and it should not affect our results.  We adopted a Bayesian methodology
and used the \emph{ensemble samplers with affine invariance} \citep{Goodman2010} variation
of the Markov Chain Monte Carlo (MCMC) method via the \emph{emcee} software
\citep{emcee}. Priors were chosen based on the interest of each parameter and typical
values in previous studies:

\begin{enumerate}

\item $R_d$: if the source had information about its radius from
  \citet{Andrews2007}, \citet{Ricci2010_Ophiuchus}, or \citet{Ricci2010_Taurus} where a range
  of values was quoted, a flat prior was assumed over the corresponding ranges. For
  resolved objects in \citet{Pietu2014} or \citet{Pascucci2016}, a Gaussian prior was
  used centered at the reported disk radii, with a standard deviation equal to the
  corresponding uncertainty. For objects with no resolved information, a flat prior
  from 10 to 300\,au was assumed.

\item $\tau_{1.3\,mm, 10\,AU}$: flat prior from $10^{-3}$ to $10^{3}$, considering
  extreme values of $\kappa_{1.3\,mm}$ and $\Sigma_{10\,au}$. Because the range extends
  for several orders of magnitude, this parameter was explored in logarithmic scale.

\item $T_{10 AU}$: flat prior from 5 to 500\,K.

\item $p$: flat prior from 0.5 to 1.5. This covers fiducial values used in modeling
  \citep[e.g.][]{Andrews2005}.

\item $q$: Gaussian prior centered at 0.5 with a standard deviation of
  0.1. This accounts for the typical spread obtained in models
  \citep[e.g.][]{Chiang1997, Dalessio1998}.

\item $i$: inclination values larger than 80 degrees were excluded to avoid issues at
  very high inclinations. For the remaining inclinations, a geometric prior
  sin($i$) was used.

\item $\beta$: flat prior from 0 to 2.5, based on $\beta$ measurements of disks
  \citep{Ricci2010_Ophiuchus, Ricci2010_Taurus, Ubach2012}.

\end{enumerate}

As already mentioned, the considered models have seven free parameters. In the
adopted approach, the use of restrictive priors for some of them (e.g. p, q,
inclination) provides additional information to the fitting process. We chose to
model objects with data available for at least seven different wavelengths, combining
photometry and the binned SPIRE spectra. We also required the minimum wavelength
available to be smaller than 200\,$\mu$m and the maximum one to be above 800\,$\mu$m
to guarantee a reasonable coverage of the far-IR/mm part of the SEDs. Sixty-three objects in
the sample meet this criterion: 40 in Taurus, 5 in Ophiuchus, and 14 in
Chamaeleon~I. From these, 28 had some information about their disk radii from
resolved high-resolution observations. In the \emph{emcee} setup for each source,
40,000 iterations with 50 walkers were run, and the last 10,000 steps were used to
generate our posterior distributions. The chains were visually inspected for
convergence, and we also checked that the adopted burn-in range (the discarded
initial 30,000 steps) was at least five times the corresponding autocorrelation time.

\begin{figure*}
  \centering
  \includegraphics[width=\hsize]{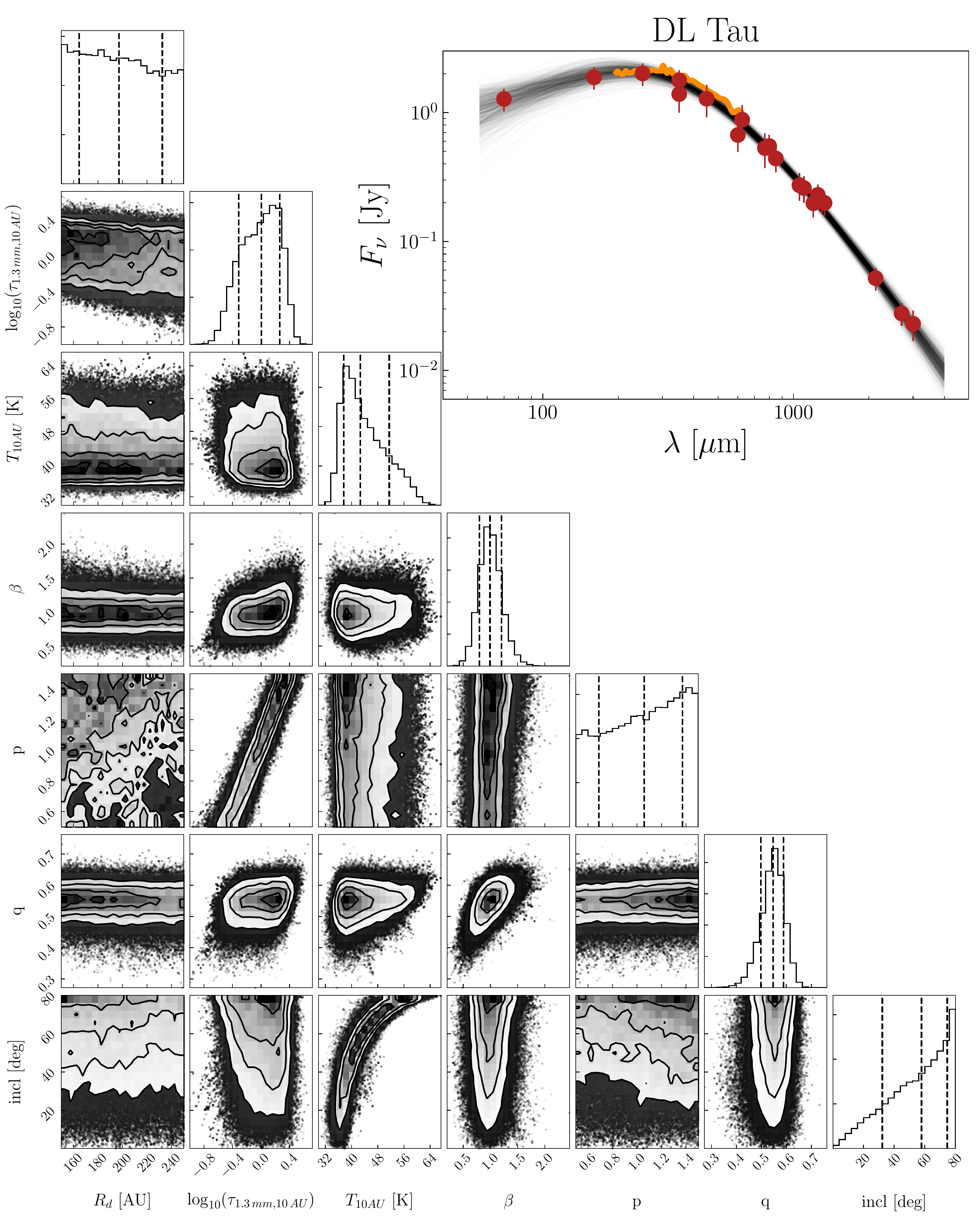}
  \caption{Fitting results for DL~Tau. The corner plot shows the posterior
    distributions for parameters and the corresponding 2D
    projections. Vertical dashed lines show the 16th, 50th, and 84th percentiles. The
    inset shows the fitted photometry (red circles) and SPIRE spectrum (yellow line),
    together with 1000 models randomly selected from the posterior distributions
    (dark area). This object is a well-behaved case for which $\tau_{1.3\,mm, 10\,AU}$,
     $T_{10\,AU}$, and $\beta$ do not become degenerate.}\label{fig:DL_Tau_models}
\end{figure*}

The adopted procedure yielded satisfactory fits in all cases, and the obtained
posterior functions revealed that $\tau_{1.3\,mm, 10\,au}$, $T_{10 au}$, and $\beta$
are generally constrained to some extent. As expected, the posteriors of $p$, $q$,
and $i$ follow the assumed priors because they are largely unconstrained with SEDs
alone. Despite our efforts to include resolved information, some objects displayed a
bi-modal behavior in their $R_d$, $\tau_{1.3\,mm, 10\,au}$, and $\beta$ posteriors,
as corresponds to the degenerate case formerly mentioned. Although the distributions
for $T_{10 au}$ are still informative (the Bayesian methodology naturally accounts
for the existence of degeneracies), the bi-modal posteriors of
$\tau_{1.3\,mm, 10\,au}$ makes them complex to analyze, and we excluded these objects
when focusing on these parameters in particular. The obtained results for
$\tau_{1.3\,mm, 10\,au}$, $T_{10 au}$, and $\beta$ are reported in
Table~\ref{tab:model_results}. An example of a well-behaved source (DL~Tau) is shown
in Fig.~\ref{fig:DL_Tau_models}.

\begin{deluxetable}{cccc}
\tablecolumns{4}
\tablewidth{0pt}
\tablecaption{Modeling Results (Median, 16th, and 84th percentiles). \label{tab:model_results}}
\tablehead{
\colhead{Name} & \colhead{$\tau_{1.3\,mm, 10\,AU}$} & \colhead{$\beta$} & \colhead{$T_{\rm 10\,AU}$ [K]}
}
\startdata
AA Tau$^\dagger$ & $-0.6_{-0.3} ^{+0.5} $ & $0.9_{-0.3} ^{+0.3} $ &$43_{-8} ^{+6} $\\
AB Aur & $-1.1_{-0.2} ^{+0.3} $ & $1.4_{-0.2} ^{+0.1} $ &$178_{-51} ^{+35} $\\
BP Tau & $-0.3_{-0.3} ^{+0.3} $ & $1.4_{-0.4} ^{+0.3} $ &$31_{-7} ^{+5} $\\
CIDA 7 & $-0.8_{-2.6} ^{+0.5} *$ & $0.7_{-0.9} ^{+0.5} *$ &$34_{-7} ^{+6} $\\
CIDA 9 & $-0.3_{-2.0} ^{+0.6} *$ & $0.9_{-0.9} ^{+0.6} *$ &$35_{-6} ^{+5} $\\
CI Tau$^\dagger$ & $-0.2_{-0.2} ^{+0.3} $ & $1.3_{-0.4} ^{+0.3} $ &$47_{-10} ^{+6} $\\
CW Tau$^\dagger$ & $-0.0_{-0.2} ^{+0.2} $ & $1.9_{-0.4} ^{+0.6} $ &$58_{-5} ^{+6} $\\
CY Tau$^\dagger$ & $0.1_{-0.3} ^{+0.4} $ & $0.7_{-0.4} ^{+0.4} $ &$25_{-5} ^{+3} $\\
DD Tau & $-1.4_{-1.9} ^{+0.6} $ & $0.3_{-0.6} ^{+0.2} $ &$66_{-49} ^{+16} $\\
DE Tau$^\dagger$ & $0.5_{-1.6} ^{+0.4} *$ & $1.5_{-0.7} ^{+0.9} *$ &$53_{-7} ^{+8} $\\
DG Tau & $0.1_{-0.4} ^{+0.4} $ & $0.7_{-0.4} ^{+0.2} $ &$94_{-20} ^{+13} $\\
DH Tau & $0.4_{-1.7} ^{+1.5} *$ & $0.5_{-1.1} ^{+0.4} *$ &$37_{-8} ^{+6} $\\
DK Tau & $0.6_{-1.7} ^{+1.9} *$ & $0.5_{-1.1} ^{+0.3} *$ &$50_{-20} ^{+9} $\\
DL Tau$^\dagger$ & $0.0_{-0.3} ^{+0.3} $ & $1.0_{-0.2} ^{+0.2} $ &$42_{-8} ^{+4} $\\
DN Tau$^\dagger$ & $-0.3_{-0.3} ^{+0.3} $ & $0.6_{-0.4} ^{+0.3} $ &$36_{-7} ^{+5} $\\
DO Tau$^\dagger$ & $-0.6_{-0.3} ^{+0.3} $ & $0.3_{-0.1} ^{+0.1} $ &$78_{-14} ^{+11} $\\
DQ Tau & $0.1_{-0.3} ^{+0.4} *$ & $1.8_{-0.4} ^{+0.7} *$ &$38_{-10} ^{+7} $\\
DS Tau & $-0.7_{-1.4} ^{+0.5} *$ & $0.6_{-0.7} ^{+0.4} *$ &$29_{-6} ^{+5} $\\
FM Tau$^\dagger$ & $1.5_{-1.0} ^{+1.1} *$ & $1.1_{-0.9} ^{+0.8} *$ &$34_{-6} ^{+6} $\\
FN Tau & $-1.5_{-2.0} ^{+0.6} $ & $0.2_{-0.5} ^{+0.1} $ &$85_{-66} ^{+25} $\\
FT Tau$^\dagger$ & $-0.4_{-1.9} ^{+0.8} $ & $0.5_{-0.6} ^{+0.3} $ &$43_{-11} ^{+6} $\\
FV Tau & $-1.6_{-0.3} ^{+0.5} *$ & $1.0_{-0.6} ^{+0.5} *$ &$57_{-42} ^{+13} $\\
GM Aur$^\dagger$ & $-0.3_{-0.3} ^{+0.4} $ & $1.5_{-0.2} ^{+0.2} $ &$54_{-10} ^{+6} $\\
GO Tau$^\dagger$ & $-0.3_{-0.2} ^{+0.3} $ & $1.5_{-0.3} ^{+0.2} $ &$30_{-5} ^{+3} $\\
Haro 6-13 & $-0.5_{-0.3} ^{+0.4} $ & $0.6_{-0.2} ^{+0.2} $ &$78_{-15} ^{+11} $\\
HK Tau & $-0.9_{-0.3} ^{+0.3} $ & $0.9_{-0.2} ^{+0.2} $ &$55_{-10} ^{+7} $\\
IRAS 04125+2902 & $-1.1_{-0.4} ^{+0.4} $ & $1.0_{-0.5} ^{+0.4} $ &$47_{-10} ^{+8} $\\
IRAS 04385+2550 & $-1.0_{-0.3} ^{+0.5} $ & $0.6_{-0.2} ^{+0.1} $ &$66_{-16} ^{+10} $\\
IP Tau & $-0.9_{-0.2} ^{+0.6} *$ & $1.7_{-0.5} ^{+1.2} *$ &$24_{-16} ^{+5} *$\\
IQ Tau$^\dagger$ & $-0.4_{-0.3} ^{+0.3} $ & $0.8_{-0.3} ^{+0.3} $ &$37_{-7} ^{+5} $\\
LkCa 15 & $-0.0_{-0.3} ^{+0.3} $ & $1.4_{-0.3} ^{+0.2} $ &$42_{-8} ^{+4} $\\
RW Aur & $-1.2_{-3.1} ^{+0.7} *$ & $0.1_{-1.4} ^{+0.1} *$ &$91_{-80} ^{+23} $\\
RY Tau & $-0.3_{-0.5} ^{+0.5} $ & $0.6_{-0.6} ^{+0.2} $ &$92_{-19} ^{+13} $\\
UX Tau A+C & $-0.7_{-0.3} ^{+0.4} $ & $0.8_{-0.4} ^{+0.3} $ &$59_{-12} ^{+8} $\\
UY Aur & $-1.3_{-0.3} ^{+0.4} $ & $0.9_{-0.2} ^{+0.2} $ &$78_{-18} ^{+13} $\\
UZ Tau A$^\dagger$ & $-0.3_{-0.3} ^{+0.4} $ & $0.7_{-0.3} ^{+0.3} $ &$48_{-10} ^{+6} $\\
V710 Tau & $-0.4_{-0.2} ^{+0.3} *$ & $1.7_{-0.6} ^{+0.5} *$ &$30_{-5} ^{+3} $\\
V807 Tau & $-1.6_{-0.5} ^{+0.5} $ & $0.6_{-0.5} ^{+0.4} $ &$51_{-32} ^{+13} $\\
V836 Tau$^\dagger$ & $-0.2_{-0.7} ^{+0.3} $ & $0.5_{-1.1} ^{+0.3} $ &$34_{-5} ^{+6} $\\
V892 Tau & $-0.6_{-0.3} ^{+0.4} $ & $0.6_{-0.2} ^{+0.1} $ &$153_{-44} ^{+28} $\\
ZZ Tau IRS & $-0.3_{-0.2} ^{+0.3} $ & $2.2_{-0.2} ^{+0.4} $ &$54_{-12} ^{+7} $\\
DOAR16AB & $-1.1_{-1.7} ^{+0.5} *$ & $0.6_{-0.8} ^{+0.4} *$ &$53_{-23} ^{+10} $\\
DOAR25$^\dagger$ & $-0.1_{-0.3} ^{+0.3} $ & $0.6_{-0.2} ^{+0.2} $ &$49_{-7} ^{+5} $\\
GSS39 & $0.4_{-0.4} ^{+0.4} $ & $0.9_{-0.4} ^{+0.3} $ &$39_{-8} ^{+5} $\\
SR21AB$^\dagger$ & $-1.3_{-0.3} ^{+0.4} $ & $1.4_{-0.2} ^{+0.1} $ &$106_{-22} ^{+14} $\\
IRS48 & $-1.4_{-0.3} ^{+0.4} $ & $0.7_{-0.3} ^{+0.2} $ &$236_{-133} ^{+73} $\\
IRS49 & $-0.9_{-2.8} ^{+0.8} *$ & $0.3_{-1.1} ^{+0.2} *$ &$65_{-41} ^{+13} $\\
WSB60$^\dagger$ & $-0.6_{-0.4} ^{+0.6} $ & $0.6_{-0.3} ^{+0.3} $ &$44_{-8} ^{+7} $\\
ROX-44$^\dagger$ & $-1.1_{-0.4} ^{+0.4} $ & $0.1_{-0.2} ^{+0.1} $ &$97_{-41} ^{+25} $\\
SX Cha & $-1.2_{-0.8} ^{+0.5} $ & $0.4_{-0.5} ^{+0.3} $ &$50_{-29} ^{+12} $\\
SZ Cha & $-0.2_{-0.2} ^{+0.3} $ & $1.8_{-0.4} ^{+0.3} $ &$62_{-13} ^{+8} $\\
TW Cha$^\dagger$ & $-0.2_{-1.9} ^{+0.4} *$ & $0.5_{-1.1} ^{+0.4} *$ &$35_{-5} ^{+6} $\\
CR Cha$^\dagger$ & $0.1_{-0.2} ^{+0.2} $ & $2.2_{-0.2} ^{+0.3} $ &$52_{-7} ^{+6} $\\
CS Cha$^\dagger$ & $0.1_{-0.3} ^{+0.4} *$ & $1.6_{-0.7} ^{+0.9} *$ &$67_{-9} ^{+11} $\\
CT Cha & $-0.0_{-2.1} ^{+0.6} *$ & $0.7_{-1.1} ^{+0.5} *$ &$41_{-7} ^{+6} $\\
CU Cha & $-0.2_{-0.3} ^{+0.3} $ & $2.3_{-0.2} ^{+0.3} $ &$156_{-39} ^{+29} $\\
T33A$^\dagger$ & $-0.5_{-0.4} ^{+0.3} *$ & $1.1_{-1.1} ^{+0.3} *$ &$83_{-19} ^{+15} $\\
VZ Cha & $0.2_{-1.7} ^{+0.6} *$ & $1.1_{-0.9} ^{+0.8} *$ &$31_{-7} ^{+4} $\\
B43 & $0.9_{-1.4} ^{+1.0} *$ & $1.0_{-1.0} ^{+0.7} *$ &$26_{-4} ^{+3} $\\
T42$^\dagger$ & $-1.3_{-0.2} ^{+0.2} $ & $2.1_{-0.3} ^{+0.3} $ &$88_{-18} ^{+11} $\\
WW Cha$^\dagger$ & $0.2_{-0.2} ^{+0.3} $ & $1.8_{-0.5} ^{+0.5} $ &$99_{-15} ^{+12} $\\
T47$^\dagger$ & $-0.6_{-1.5} ^{+0.3} $ & $0.3_{-0.8} ^{+0.2} $ &$41_{-7} ^{+7} $\\
CV Cha$^\dagger$ & $0.3_{-1.9} ^{+1.0} *$ & $1.0_{-1.0} ^{+0.7} *$ &$80_{-15} ^{+15} $\\
T56$^\dagger$ & $-0.7_{-1.4} ^{+0.3} *$ & $0.6_{-1.1} ^{+0.4} *$ &$46_{-7} ^{+8} $\\
\enddata
\ \\
{\bf Notes.}
\tablenotetext{\dagger}{ Object with R$_d$ constraints from resolved observations.}
\tablenotetext{*}{ Unconstrained/bimodal distribution.}\end{deluxetable}

Before analyzing these results, it is important to mention that the disk model used here
is a very simplistic approximation. It assumes a fixed inner radius and an axisymmetric
geometry.  More importantly, it does not include a vertical temperature gradient or dust
mixing/settling, which produce flared disks required to properly explain far-IR fluxes of
disks \citep[e.g.][]{Kenyon1987, Calvet1992}. We have also assumed a power-law opacity law
longward of $\sim$\,70\,$\mu$m, which is not realistic in the presence of different dust
species \citep[e.g.][]{Dalessio2001, Draine2006}. These two last issues combined are
especially relevant for $\beta$ estimates, which may therefore be higher than the expected
$\alpha = 2 + \beta$ relation. Thus, although they can provide interesting insights and
comparisons, the results from the modeling should therefore be considered with caution.

\vspace{0.2cm}
\subsection{Optical depth and $\beta$ values}

Despite having included disk size estimates from the literature, some objects lacked that
information, or the measured size ranges were not restrictive enough to avoid the
degeneracy in the fitting process. Here, we limit our analysis to non-degenerate
  $\tau_{1.3\,mm, 10\,au}$ and $\beta$ distributions, as revealed by their well-behaved
  distributions (i.e. constrained and not bi-modal). Therefore, 40 objects were used to
study these parameters, 28 in Taurus, 6 in Ophiuchus, and 6 in Chamaeleon~I.

\begin{figure}
  \centering
  \includegraphics[width=\hsize]{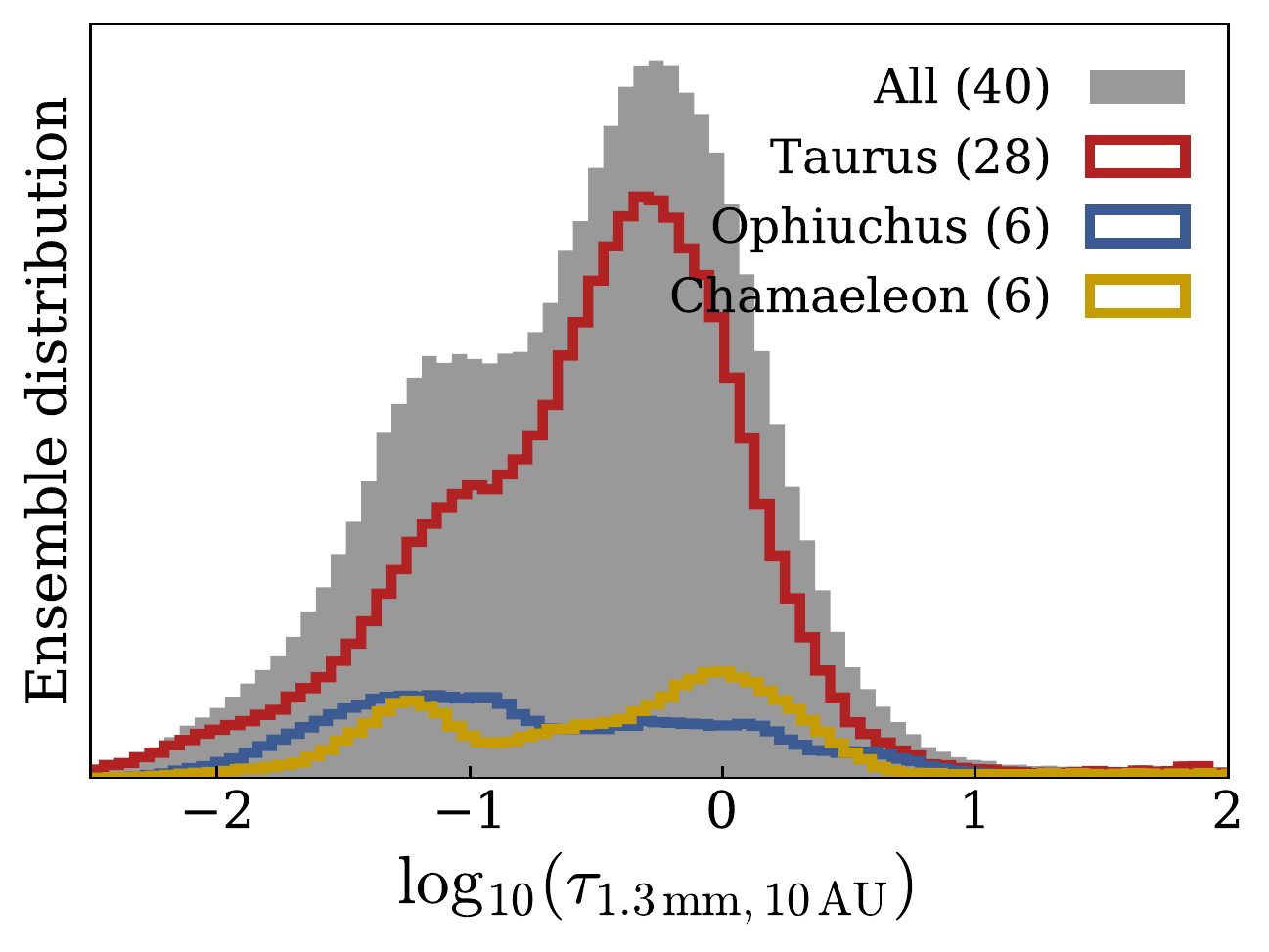}\\
  \includegraphics[width=\hsize]{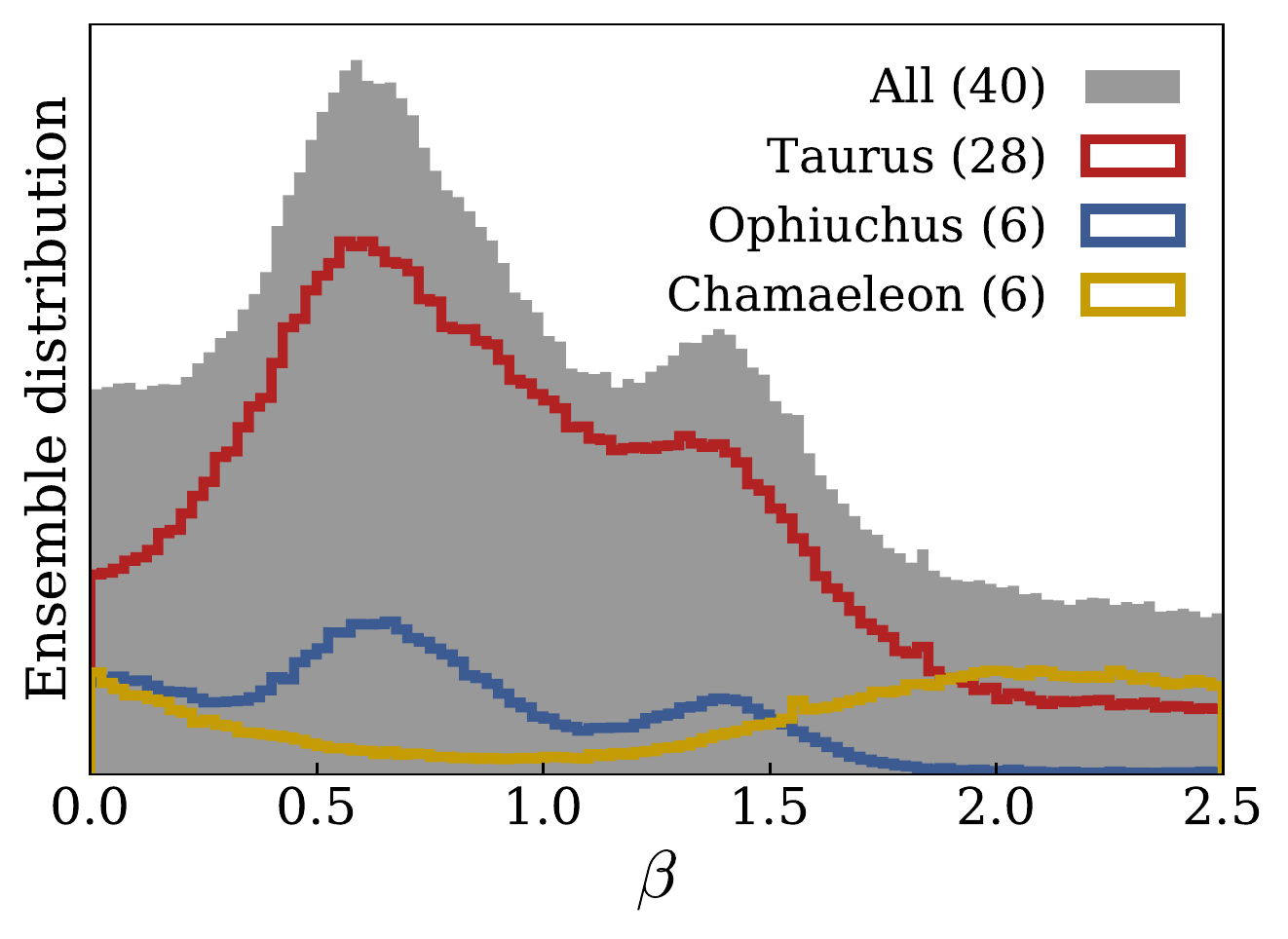}\\
  \includegraphics[width=\hsize]{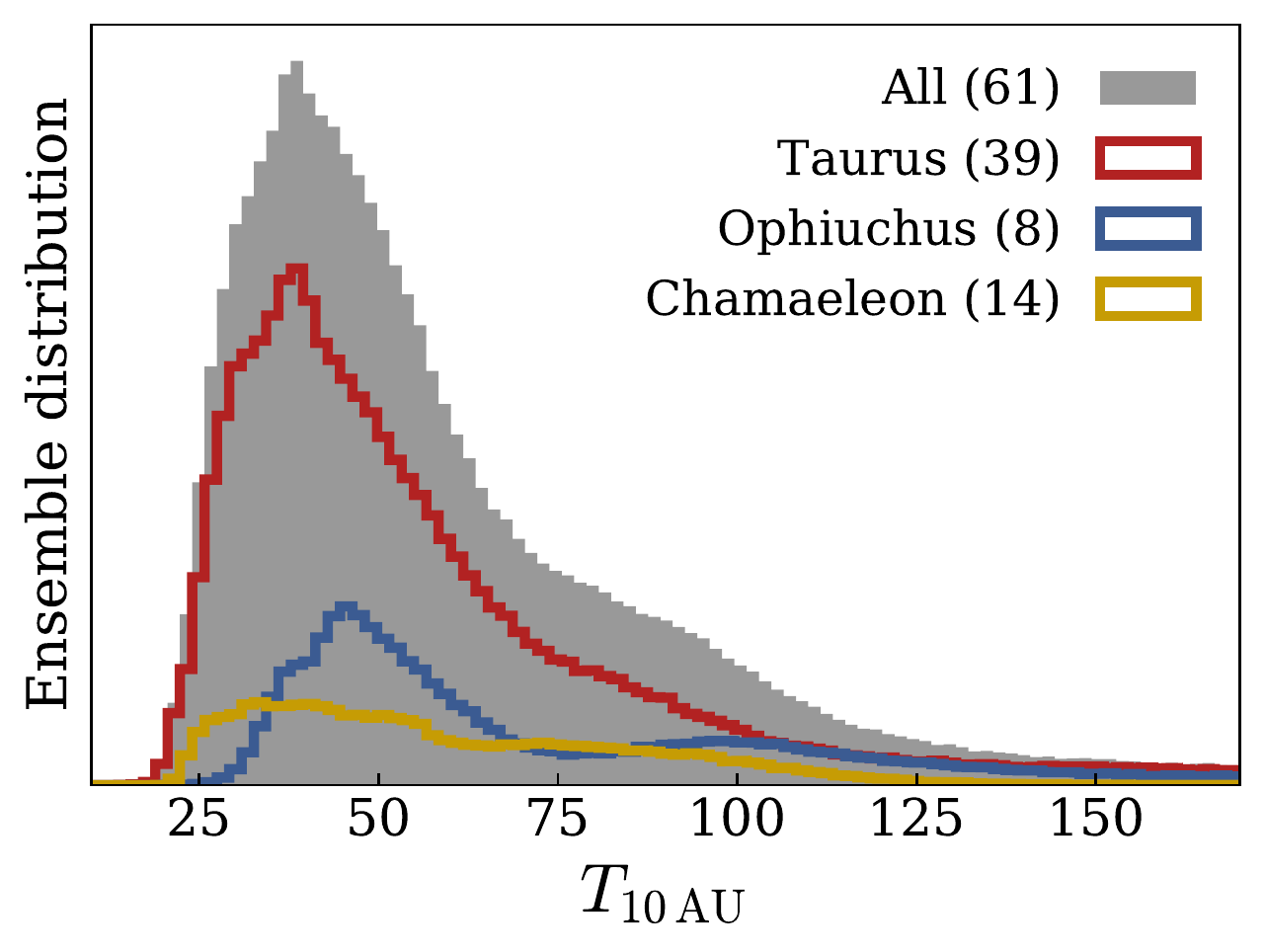}
  \caption{Ensemble distributions for $\tau_{1.3\,mm, 10\,au}$ (top), $\beta$
    (middle), and $T_{\rm 10\,AU}$ (bottom) for each region, normalized to the number
    of objects in each association. Sources with bi-modal (degenerate) and flat
    (uninformative) distributions have been excluded. The number of objects in each
    case is indicated in the legend.}\label{fig:model_distributions}
\end{figure}

The ensemble distributions of $\tau_{1.3\,mm, 10\,au}$, and $\beta$ for each region
(produced by randomly selecting 1 million positions from the individual posteriors) are
shown in the top and middle panels in Fig.~\ref{fig:model_distributions}\footnote{Given
  the large number of MCMC steps and the computational requirement to compute KDEs in
  these particular cases, we displayed these distributions using histograms with a large
  (100) number of bins.}. We first note that the low number of sources remaining after the
adopted curation processes both in Ophiuchus and Chamaeleon~I is an obvious caveat to our
interpretation, and therefore it cannot be extrapolated to the whole
sample. However, they can still be used to investigate possible differences among regions,
under the assumption that these distributions are not very different from the underlying
ones (or at least that they are different in similar ways). In the following, we assume
this to be the case, bearing in mind that additional observations may improve and modify
some of these results.

The distribution of optical depth values at 10\,au and 1.3mm
(Fig.~\ref{fig:model_distributions}, top) has its maximum at
  $\log{\tau_{1.3\,mm, 10\,au}}$ = -0.25 (corresponding to $\tau_{1.3\,mm, 10\,au} \sim$
  0.5), and a secondary peak at $\log{\tau_{1.3\,mm, 10\,au}}=-1$. We note that the shape
  of this distribution is determined mostly by Taurus, given the lack of sufficient
  long-wavelength data for most objects in Chamaeleon~I and Ophiuchus, and the
  distributions in these regions appear to be broader than the one in Taurus (again, this
  interpretation is limited by the small number statistics in these regions). For
comparison, reasonable assumptions about the dust opacity and surface density based on
observations of the solar system bodies yield $\tau_{\rm 1\,mm}$=1 at $\sim$\,10\,au for
the Minimum Mass Solar Nebula \citep{Davis2005}, suggesting that several of the modeled
protoplanetary disks may have optical depth profiles (and hence possibly surface
densities) similar to that of the parental disk of the solar system. In the case of
$\beta$, values smaller than the one measured for the ISM \citep[$\sim$1.6-2; see
e.g.][and references therein]{Draine2006} imply some degree of grain growth. Almost the
entirety of the Taurus and Ophiuchus distributions (and part of Chamaeleon~I) are constrained within
that value, in correspondence with the observational result discussed in
Sec.~\ref{sec:dust_growth}. As with $\alpha_{\rm mm}$, Chamaeleon~I shows a different
behavior (an excess of high $\beta$ values) that will be discussed further on. We note
that the distributions of $\beta$ should be considered with caution, not only due to the
aforementioned caveats, but also because degenerate cases have been removed from the
analysis. Because these occur when $\alpha=2$, this procedure inevitably discards objects
with $\beta \sim 0$. There is a tentative bimodality in both distributions, and especially
in the case of $\beta$, with a tentative secondary peak occurring at
$\sim$\,1.4-1.5. Given the limited size of the modeled sample and the simplicity of the
models used, we do not investigate this issue in detail here. However, we speculate that, if real,
it may hint at a quick transition from micron-sized grains (large $\beta$ values) to
mm/cm-sized dust (smaller $\beta$).

\begin{figure}
  \centering
  \includegraphics[width=\hsize]{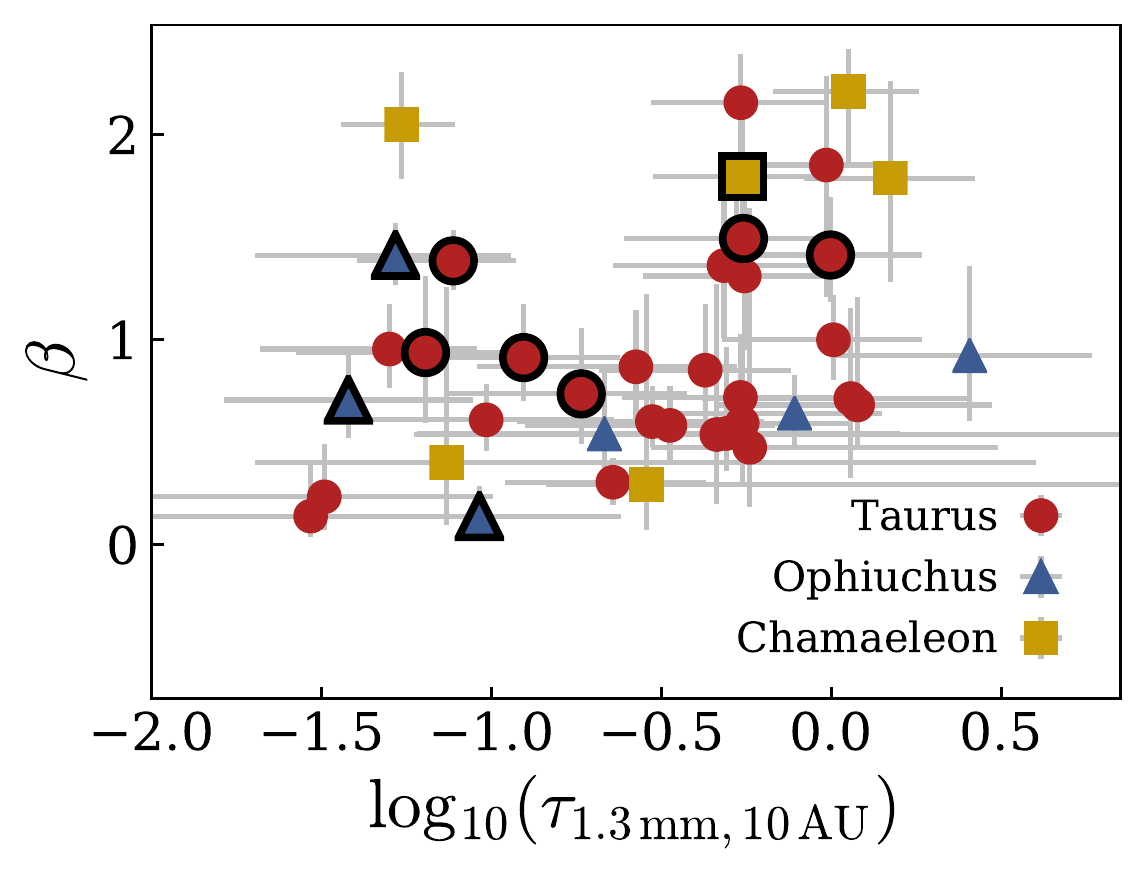}
  \caption{$\beta$ vs. optical depths at 1.3\,mm and 10\,au for Taurus (red), Ophiuchus
    (blue), and Chamaeleon (yellow) objects. Symbols with black borders are transitional
    disks.}\label{fig:tau_vs_beta}
\end{figure}

We also inspected our results in the $\tau_{1.3\,mm, 10\,au}$ versus $\beta$ space. Individually,
these two parameters affect the optical depth---and are therefore correlated---but a more general
correlation may also exist as a result of disk evolution. Fig.~\ref{fig:tau_vs_beta} shows that
objects with low $\beta$ values ($\lesssim 1$) spread through optical depth values from
$\log_{10}(\tau_{1.3\,mm, 10\,au})=-1.5$ to $0.5$. However, a lack of low optical depths is found
for $\beta$s above that threshold, an expected effect from an observational bias toward bright
sources: mm fluxes decrease faster with increasing wavelengths for steeper $\beta$ values, and only
massive disks (likely to be optically thicker) are detectable. Although such an effect could also be
partially produced by disk evolution (disk masses decrease with time, and dust growth leads to
smaller $\beta$s), more complex models are required to quantify how much (if any) of this paucity of
low $\beta$ - low optical depth values is due to disk evolution itself. Like
Fig.~\ref{fig:Fmm_vs_alphmm}, Fig.~\ref{fig:tau_vs_beta} also shows the position of
(pre)transitional disks (as classified in Sec.~\ref{sec:other_tracers}), with no obvious difference
between these sources and full disks.

\subsection{Disk temperatures at 10\,au}

The modeling process also yielded estimates of the disk temperature at 10\,au. For
this parameter---even when the disk radii, optical depth, and $\beta$ are degenerate---
the posterior $T_{\rm 10\,au}$ is constrained to some extent, in most cases. Only two
of the modeled sources displayed an uninformative ($\sim$ flat) posterior distribution
and were excluded from the analysis, leaving 39 objects in Taurus, 8 in Ophiuchus,
and 14 in Chamaeleon~I. Fig.~\ref{fig:model_distributions} (bottom) shows the results for the
three regions, all of them showing a distribution that peaks at $\sim$40-50\,K with a
lower probability tail extending to $\sim$\,100-150\,K due to the effect of high
inclinations (see corner plot in Fig.~\ref{fig:DL_Tau_models}).

\begin{figure}
  \centering
  \includegraphics[width=\hsize]{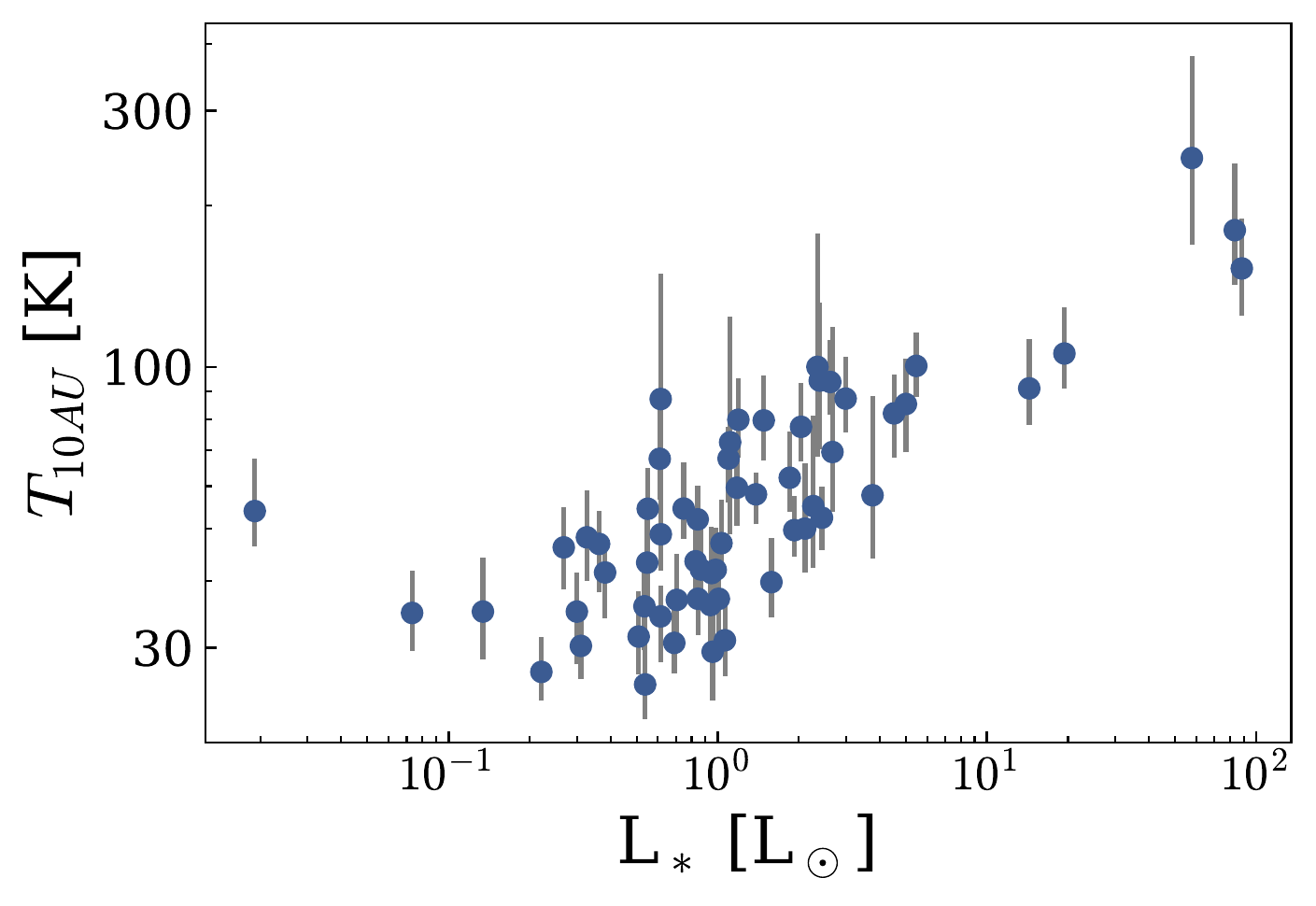}
  \includegraphics[width=\hsize]{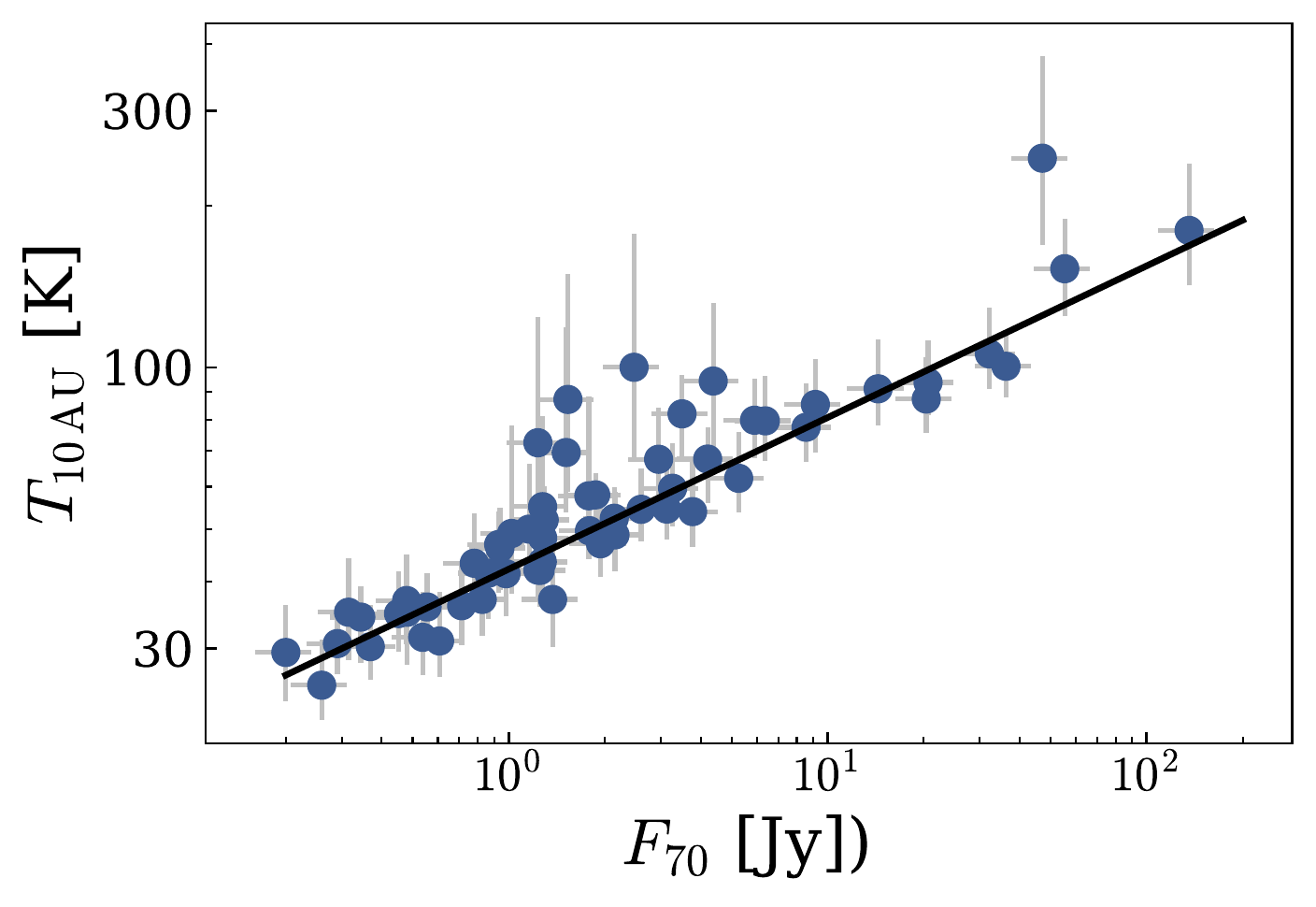}
  \caption{Top: stellar luminosities vs disk temperatures at 10\,au as determined by the
    models. Bottom: Observed 70\,$\mu$m fluxes (scaled to 140\,pc in the case of Chamaeleon~I)
    vs. disk temperatures at 10\,au as derived from the models. The relation derived in
    Eq.~\ref{eq:f70_T10} is also shown as a black solid line.}\label{fig:T0s_vs_Lstar_and_F70}
\end{figure}

The obtained $T_{\rm 10\,au}$ values can be used to test the general performance of
these models by comparing them with the luminosity of their host star: despite the
former parameter not being part of the adopted model, a clear correlation among these
two is found (Fig.~\ref{fig:T0s_vs_Lstar_and_F70}, top panel), showing that the obtained
disk temperatures are higher for more luminous stars, as expected.

The temperature of the disk at 10\,au is also related to far-IR fluxes; they trace
both areas from which most of this emission originates, and wavelengths at which the
deviation from the RJ regime is significant and provides information on the
temperature of the region. Figure~\ref{fig:T0s_vs_Lstar_and_F70} shows the observed
fluxes at 70\,$\mu$m with respect to the corresponding $T_{\rm 10\,AU}$, with an
obvious correlation between these two parameters. The best fit for this trend yields:
\begin{multline}\label{eq:f70_T10}
\log_{10}(T_{\rm 10\,au}[K]) = (0.28\pm 0.01) \log_{10}(F_{70} [Jy]) \\+ (1.625\pm 0.005)
\end{multline}
We note that this tight relation is likely produced by the simplicity of model used: far-IR (mostly
PACS) fluxes determine the value of $T_{\rm 10\,AU}$ to a great extent. Such a close correlation is
unlikely when considering more complex models with vertical temperature gradients and dust mixing
\citep[e.g.][]{Dalessio1998,Dalessio2006}, which would introduce significant scatter in the
70\,$\mu$m fluxes. Nevertheless, this relation provides a rough estimate of the disk temperature at
a few astronomical units from the star using far-IR observations, and it could have applications for
comparative studies of disk samples.

\subsection{The peculiar case of Chamaeleon~I}\label{sec:chamaeleon}

As mentioned in previous sections, Chamaeleon~I shows a different behavior than Taurus and
Ophiuchus; the spectral indices in its sources appear to be systematically larger with
respect to the other regions. We applied both the Kolmogorov-Smirnov (KS) and 
  Anderson-Darling (AD) tests to the estimated millimeter indices and found no significant
  differences between Taurus and Ophiuchus, while Chamaeleon~I showed clear indications of
  a different distribution ($p$-values $\sim 0.005$ and $<0.01$ for the KS and AD tests,
  respectively). This was already mentioned in \citet{Testi2014} based on the results of
\citet{Ubach2012}, who measured $\alpha_{\rm mm}$=2.9-3.8 for eight sources in this
association. Assuming that the relation $\beta = \alpha - 2$ can be used for these
objects, the corresponding $\beta$ values (0.9-1.8) hint at different dust properties of
sources in these regions. Although this may be simply a result of the small number of
sources with mm spectral index estimates in the region, here we discuss some of the
plausible explanations for this phenomenon provided it is real.

Our results are in agreement with the aforementioned findings: after applying the various data
curation processes, $\alpha_{\rm mm}$ values were measured for seven disks in
Chamaeleon~I. Out of these seven, four (CR~Cha, SZ~Cha, WW~Cha, and T33~A) have values
larger than $\alpha > 3$ \citep[and the number would go up to five if we were to
include the $\alpha$ value for CU~Cha in][]{vanderPlas2017}. This represents 55-60\,\%
of the sample with available spectral index estimates in this region. For comparison, the
fractions of objects with $\alpha > 3$ in Taurus and Ophiuchus are 2\,\% (1/58) and
9\,\% (1/11), respectively. A similar (but more uncertain) result is found when
analyzing the $\beta$ distributions from the models, where Chamaeleon~I shows an
excess of high values ($> 1.5$). Although an in-depth analysis of each data set would
be required to completely rule out systematic problems in any of the sources of
ancillary (sub)mm data used for Chamaeleon~I, our inspection of the corresponding SEDs
revealed no apparent issue---and this result is still present when using individual
studies to estimate spectral indices. We therefore discard the possibility of this
being a data-processing problem.

Chamaeleon~I is estimated to be the oldest region of the considered sample, and hence
its steeper (sub)mm slopes could be the result of dust evolution in its disks. These
mm spectral indices imply high $\beta$ values, suggestive of a decrease in the
maximum grain size in the outer regions of these disks. Such a trend is expected from
the inward drift of mm/cm-sized grains in the presence of gas: at these sizes, dust
particles start experiencing a head wind from the gas, losing angular
momentum and spiraling inward \citep{Adachi1976, Weidenschilling1977,
  Takeuchi2002}. This mechanism clears out the outer regions of the disk of
mm/cm-sized grains (if they are not replenished by other means) and effectively
decreases the value of $a_{max}$, thus increasing $\beta$ as a function of radius
\citep[e.g.][]{Birnstiel2010, Pinilla2012}. Disk lifetimes are on the order of
5-10\,Myr \citep{Haisch2001, Hernandez2007_25Ori, Ribas2015}, but the inward
migration of mm/cm grains is estimated to be much faster \citep[10$^4$-10$^5$\,years,
e.g.][]{Takeuchi2005, Brauer2007}; braking mechanisms, such as dust accumulation at
pressure bumps, are required to slow down this inward migration
\citep[e.g.][]{Zhu2011, Pinilla2012} and explain observed mm spectral indices in disks
\citep{Testi2014}. A possible explanation for the higher $\alpha_{\rm mm}$ values in
Chamaeleon~I is that dust migration has significantly altered disks in this region, but
is still not as important in Taurus and Ophiuchus. However, Chamaeleon~I sources with
steep $\beta$ values are also among the brightest objects at mm wavelengths in the
sample (Fig.~\ref{fig:Fmm_vs_alphmm}), in contradiction with the expected decline of
disk masses with time (produced by viscous evolution and/or the decrease in mm
opacities caused by inward migration of dust). If the fact that disks in Chamaeleon~I
are older was the only reason for differences in spectral indices, then these disks
would have needed to be initially more massive than their counterparts in Taurus and
Ophiuchus for them to have their current brightness. 

\begin{figure}
  \centering
  \includegraphics[width=\hsize]{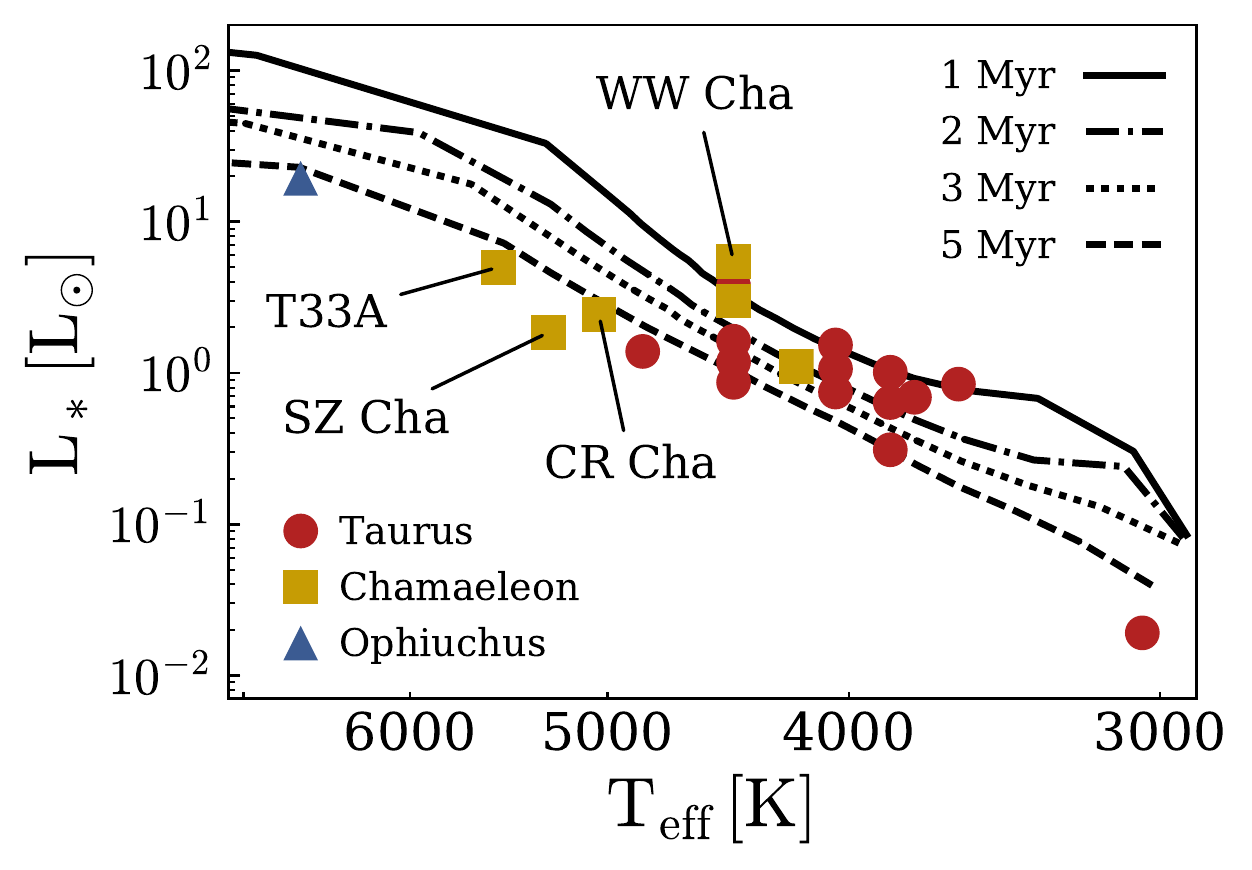}
  \caption{H-R diagram for sources with $\alpha_{\rm mm} > 2.5$ and $T_*$ between 3000 and
    7000\,K. Objects in Taurus (red circles), Chamaeleon~I (yellow squares), and Ophiuchus
    (blue triangles) are shown. \citet{Siess2000} isochrones are also shown for 1,\,2,\,3,
    and 5\,Myr. The four labeled Chamaeleon~I sources have $\alpha_{\rm mm} > 3$; three of
    these appear to be older than most of their Taurus counterparts.}\label{fig:HRdiagram_mmslopes}
\end{figure}

A different explanation for the abnormal millimeter spectral indices in Chamaeleon~I could
be a strong selection bias toward its youngest members. Chamaeleon~I is both further and
older than the other regions, meaning that its disks are likely dimmer, on average, than
those of Taurus and Ophiuchus. As a result, it is possible that only the brightest (and
therefore probably youngest) population of its disks is detected, where grain growth may
not have reached advanced stages. One example of this is WW~Cha, which is surrounded by a
substantial amount of extended emission in the \emph{Herschel} maps and possibly a young
source still embedded in its parental cloud. To test this idea,
  Fig.~\ref{fig:HRdiagram_mmslopes} shows the H-R diagram for objects with
  $\alpha_{\rm mm} > 2.5$. Several of these sources have $T_* > 4000$\,K, and therefore we
  used isochrones from \citet{Siess2000}, which cover this temperature range. However, the
  fact that three out of the four sources with $\alpha_{\rm mm} > 3$ in Chamaeleon~I
  appear to be older than most of their Taurus counterparts implies that a hypothetical bias
  toward the youngest sources is probably not enough to explain the systematically larger
  $\alpha_{\rm mm}$ values observed in Chamaeleon.

  Nevertheless, we remark here that the number of objects with available $\alpha_{\rm mm}$
  measurements in Chamaeleon~I is still small; although the recent ALMA survey by
  \citet{Pascucci2016} has yielded 887\,$\mu$m flux measurements for several sources in
  this region, measurements at longer wavelengths are still scarce. Therefore, these
  results should be considered with caution. Follow-up surveys in the mm will provide
  spectral slopes for several additional sources and determine whether the difference in spectral
  indices in Chamaeleon~I is real or an observational/small-sample
  effect.

\section{Median SEDs}\label{sec:median_seds}

Using the compiled data set, we produced and compared the median SEDs of Class~II objects
in each region. (Pre)transitional disks and Class~III sources were first discarded using
the procedure in \citet{McClure2010}; Class~II objects have $-0.8 < \alpha_{5-12} < 1.25$
(in $F_\nu$ versus $\nu$ space). We required at least one detection at
70\,$\mu$m or longer wavelengths for any source to be included, in order to mitigate different
completeness levels at different wavelengths. A total of 114 objects met these criteria:
70 in Taurus, 26 in Chamaeleon, and 18 in Ophiuchus. The median SEDs, together with the
25th and 75th percentiles, were then computed using the dereddened SEDs after scaling
each object to its 2MASS $J$ flux. Only photometric bands with at least five measurements are
available were included in the median SED calculation. The results are provided in
Tables~\ref{tab:median_Taurus}--\ref{tab:median_Chamaeleon}.

\begin{deluxetable*}{cccccc}
\tablecaption{Median SED, Upper, and Lower Quartiles of Taurus. SEDs are Normalized to the $J$ Band \label{tab:median_Taurus}}
\tablewidth{0pt}
\tablehead{
\colhead{$\lambda$ [$\mu$m]} & \colhead{F$_\nu$,  Percentile 25\,\%} & \colhead{F$_\nu$,  Median} & \colhead{F$_\nu$, Percentile
  75\,\%} & \colhead{N points} &  \colhead{Reference}
}
 \decimalcolnumbers
\startdata
0.35 & 1.05e-02 & 2.59e-02 & 6.16e-02 & 19 & a\\
0.36 & 3.42e-02 & 7.56e-02 & 1.74e-01 & 42 & a\\
0.44 & 7.35e-02 & 1.83e-01 & 2.47e-01 & 42 & a\\
0.48 & 2.95e-02 & 6.47e-02 & 1.63e-01 & 23 & a\\
0.55 & 1.18e-01 & 2.29e-01 & 3.62e-01 & 51 & a\\
0.62 & 9.22e-02 & 2.10e-01 & 3.29e-01 & 25 & a\\
0.64 & 3.06e-01 & 4.88e-01 & 6.68e-01 & 43 & a\\
0.76 & 2.53e-01 & 3.25e-01 & 5.35e-01 & 16 & a\\
0.79 & 5.18e-01 & 6.48e-01 & 7.46e-01 & 37 & a\\
0.91 & 5.28e-01 & 5.94e-01 & 7.66e-01 & 19 & a\\
1.24 & 1.00e+00 & 1.00e+00 & 1.00e+00 & 67 & a\\
1.66 & 1.12e+00 & 1.22e+00 & 1.40e+00 & 67 & a\\
2.16 & 1.01e+00 & 1.15e+00 & 1.59e+00 & 67 & a\\
3.35 & 6.39e-01 & 1.08e+00 & 1.70e+00 & 39 & a\\
3.36 & 7.55e-01 & 1.05e+00 & 1.42e+00 & 24 & a\\
3.55 & 6.80e-01 & 1.08e+00 & 1.59e+00 & 28 & a\\
3.55 & 7.11e-01 & 9.78e-01 & 1.51e+00 & 63 & a\\
4.49 & 6.43e-01 & 9.21e-01 & 1.65e+00 & 65 & a\\
4.60 & 6.05e-01 & 9.61e-01 & 1.60e+00 & 63 & a\\
5.7 & 5.18e-01 & 7.97e-01 & 1.58e+00 & 66 & a\\
7.9 & 5.61e-01 & 8.60e-01 & 1.91e+00 & 65 & a\\
9.0 & 7.76e-01 & 1.32e+00 & 2.73e+00 & 49 & a\\
10.5 & 8.55e-01 & 1.11e+00 & 2.24e+00 & 13 & a\\
11.6 & 5.73e-01 & 1.08e+00 & 2.00e+00 & 63 & a\\
12.0 & 9.57e-01 & 1.94e+00 & 3.56e+00 & 31 & a\\
18 & 1.19e+00 & 2.37e+00 & 4.07e+00 & 41 & a\\
22 & 9.12e-01 & 1.75e+00 & 3.60e+00 & 63 & a\\
24 & 7.60e-01 & 1.37e+00 & 3.17e+00 & 56 & a\\
25 & 1.34e+00 & 3.29e+00 & 5.47e+00 & 30 & a\\
70 & 1.20e+00 & 2.24e+00 & 3.87e+00 & 56 & PACS 70 (this work)\\
100 & 8.73e-01 & 1.75e+00 & 3.36e+00 & 33 & PACS 100 (this work)\\
160 & 1.00e+00 & 1.93e+00 & 3.86e+00 & 54 & PACS 160 (this work)\\
250 & 9.33e-01 & 2.05e+00 & 4.25e+00 & 40 & SPIRE 250 (this work)\\
350 & 1.49e+00 & 2.38e+00 & 3.70e+00 & 14 & a\\
350 & 8.36e-01 & 1.63e+00 & 3.90e+00 & 35 & SPIRE 350 (this work)\\
443 & 8.30e-01 & 1.29e+00 & 1.81e+00 & 13 & a\\
450 & 1.29e+00 & 1.87e+00 & 2.71e+00 & 7 & a\\
500 & 5.65e-01 & 1.21e+00 & 2.45e+00 & 30 & SPIRE 500 (this work)\\
600 & 7.15e-01 & 8.97e-01 & 1.30e+00 & 6 & a\\
624 & 7.60e-01 & 1.23e+00 & 2.16e+00 & 7 & a\\
769 & 5.78e-01 & 8.46e-01 & 1.13e+00 & 13 & a\\
800 & 3.42e-01 & 4.28e-01 & 6.26e-01 & 11 & a\\
850 & 3.99e-01 & 6.40e-01 & 9.12e-01 & 6 & a\\
869 & 6.80e-02 & 2.15e-01 & 4.53e-01 & 38 & a\\
880 & 5.65e-02 & 1.36e-01 & 3.80e-01 & 17 & a\\
1056 & 2.86e-01 & 3.93e-01 & 5.96e-01 & 15 & a\\
1100 & 2.57e-01 & 2.70e-01 & 4.27e-01 & 6 & a\\
1200 & 4.63e-02 & 1.67e-01 & 2.49e-01 & 10 & a\\
1250 & 5.59e-02 & 1.82e-01 & 2.39e-01 & 27 & a\\
1300 & 1.06e-01 & 1.87e-01 & 3.09e-01 & 18 & a\\
1330 & 1.09e-01 & 2.56e-01 & 4.10e-01 & 14 & a\\
1360 & 4.62e-02 & 6.91e-02 & 9.65e-02 & 7 & PdBI1.36mm \citep{Pietu2014}\\
2126 & 6.51e-02 & 7.64e-02 & 1.17e-01 & 5 & a\\
2700 & 2.08e-02 & 3.60e-02 & 6.36e-02 & 17 & a\\
2974 & 4.53e-03 & 7.50e-03 & 2.08e-02 & 5 & a\\
\enddata
\tablecomments{a: data from \citet{Andrews2013}.}
\end{deluxetable*}

\begin{deluxetable*}{cccccc}
\tablecaption{Median SED, Upper, and Lower Quartiles of Chamaeleon. SEDs are Normalized to the $J$
  Band \label{tab:median_Chamaeleon}}
\tablewidth{0pt}
\tablehead{
\colhead{$\lambda$ [$\mu$m]} & \colhead{F$_\nu$,  Percentile 25\,\%} & \colhead{F$_\nu$,  Median} & \colhead{F$_\nu$, Percentile
  75\,\%} & \colhead{N points} &  \colhead{Reference}
}
 \decimalcolnumbers
\startdata
0.44 & 1.04e-01 & 1.73e-01 & 2.29e-01 & 17 & B Johnson (APASS)\\
0.48 & 1.37e-01 & 1.86e-01 & 2.80e-01 & 18 & g SDSS (APASS) \\
0.55 & 1.75e-01 & 2.65e-01 & 4.25e-01 & 17 & V Johnson (APASS)\\
0.62 & 3.38e-01 & 3.93e-01 & 6.58e-01 & 17 & r SDSS (APASS) \\
0.76 & 3.40e-01 & 4.67e-01 & 7.05e-01 & 17 & i SDSS (APASS) \\
1.24 & 1.00e+00 & 1.00e+00 & 1.00e+00 & 26 & 2MASS $J$\\
1.66 & 1.25e+00 & 1.28e+00 & 1.39e+00 & 25 & 2MASS $H$\\
2.16 & 1.19e+00 & 1.26e+00 & 1.54e+00 & 25 & 2MASS $K$\\
3.35 & 7.87e-01 & 1.02e+00 & 1.69e+00 & 17 & \emph{WISE 1}\\
3.55 & 7.35e-01 & 9.43e-01 & 1.52e+00 & 18 & IRAC 1\\
4.49 & 7.12e-01 & 8.97e-01 & 2.02e+00 & 19 & IRAC 2\\
4.60 & 7.79e-01 & 1.06e+00 & 1.67e+00 & 18 & \emph{WISE 2}\\
5.7 & 5.81e-01 & 7.60e-01 & 2.14e+00 & 22 & IRAC 3\\
7.9 & 6.14e-01 & 8.91e-01 & 2.42e+00 & 24 & IRAC 4\\
9.0 & 8.62e-01 & 1.31e+00 & 2.45e+00 & 22 & AKARI 9\\
11.6 & 8.52e-01 & 1.20e+00 & 2.15e+00 & 21 & \emph{WISE 3}\\
18 & 1.17e+00 & 1.75e+00 & 2.96e+00 & 20 & \emph{AKARI 18}\\
22 & 1.25e+00 & 1.68e+00 & 2.33e+00 & 21 & \emph{WISE 4}\\
24 & 1.03e+00 & 1.42e+00 & 2.05e+00 & 21 & MIPS 1\\
70 & 1.37e+00 & 2.29e+00 & 5.94e+00 & 19 & PACS 70 (this work) \\
100 & 1.35e+00 & 2.10e+00 & 4.70e+00 & 25 & PACS 100 (this work) \\
160 & 1.12e+00 & 2.12e+00 & 2.78e+00 & 12 & PACS 160 (this work) \\
250 & 4.83e-01 & 1.65e+00 & 2.04e+00 & 8 & SPIRE 250 (this work) \\
350 & 1.12e+00 & 1.20e+00 & 1.94e+00 & 6 & SPIRE 350 (this work) \\
500 & 5.08e-01 & 7.55e-01 & 1.41e+00 & 6 & SPIRE 500 (this work) \\
870 & 9.35e-02 & 1.78e-01 & 3.02e-01 & 6 & LABOCA 870 \citep{Belloche2011}\\
887 & 4.42e-02 & 1.05e-01 & 2.53e-01 & 25 & ALMA 887 \citep{Pascucci2016}\\
\enddata
\end{deluxetable*}

\begin{deluxetable*}{cccccc}
\tablecaption{Median SED, Upper, and Lower Quartiles of Ophiuchus. SEDs are Normalized to the $J$
  Band \label{tab:median_Ophiuchus}}
\tablewidth{0pt}
\tablehead{
\colhead{$\lambda$ [$\mu$m]} & \colhead{F$_\nu$,  Percentile 25\,\%} & \colhead{F$_\nu$,  Median} & \colhead{F$_\nu$, Percentile
  75\,\%} & \colhead{N points} &  \colhead{Reference}
}
 \decimalcolnumbers
\startdata
0.44 & 7.59e-02 & 1.96e-01 & 2.21e-01 & 7 & B Johnson (APASS) \\
0.48 & 1.12e-01 & 2.82e-01 & 3.12e-01 & 9 & g SDSS (APASS)\\
0.55 & 1.50e-01 & 3.81e-01 & 4.18e-01 & 7 & V Johnson (APASS)\\
0.62 & 2.58e-01 & 3.95e-01 & 4.83e-01 & 7 & R CMC15\\
0.62 & 2.94e-01 & 5.12e-01 & 6.65e-01 & 9 & r SDSS (APASS)\\
0.76 & 4.83e-01 & 5.68e-01 & 7.05e-01 & 8 & i SDSS (APASS)\\
1.24 & 1.00e+00 & 1.00e+00 & 1.00e+00 & 18 & 2MASS $J$\\
1.66 & 1.11e+00 & 1.18e+00 & 1.22e+00 & 18 & 2MASS $H$\\
2.16 & 9.97e-01 & 1.07e+00 & 1.23e+00 & 18 & 2MASS $K$\\
3.35 & 6.80e-01 & 8.22e-01 & 1.01e+00 & 17 & \emph{WISE 1}\\
3.55 & 6.17e-01 & 8.07e-01 & 1.10e+00 & 14 & IRAC 1\\
4.49 & 4.43e-01 & 6.68e-01 & 9.10e-01 & 16 & IRAC 2\\
4.60 & 4.42e-01 & 6.99e-01 & 8.99e-01 & 17 & \emph{WISE 2}\\
5.7 & 3.70e-01 & 6.25e-01 & 7.60e-01 & 17 & IRAC 3\\
7.9 & 4.62e-01 & 7.57e-01 & 9.44e-01 & 16 & IRAC 4\\
9.0 & 7.86e-01 & 1.04e+00 & 1.44e+00 & 12 & \emph{AKARI 9}\\
11.6 & 5.04e-01 & 7.22e-01 & 1.15e+00 & 17 & \emph{WISE 3}\\
18 & 8.26e-01 & 1.15e+00 & 1.23e+00 & 7 & \emph{AKARI 18}\\
22 & 6.92e-01 & 1.22e+00 & 1.85e+00 & 16 & \emph{WISE 4}\\
24 & 7.98e-01 & 1.33e+00 & 1.73e+00 & 17 & MIPS 1\\
70 & 4.39e-01 & 9.50e-01 & 2.83e+00 & 14 & PACS 70 (this work)\\
100 & 6.68e-01 & 1.16e+00 & 2.93e+00 & 11 & PACS 100 (this work)\\
160 & 7.65e-01 & 2.34e+00 & 2.58e+00 & 5 & PACS 160 (this work)\\
250 & 2.86e-01 & 1.54e+00 & 3.11e+00 & 7 & SPIRE 250 (this work)\\
850 & 6.09e-02 & 2.07e-01 & 9.00e-01 & 5 & SCUBA 850 \citep{Andrews2007} \\
1300 & 2.46e-02 & 4.31e-02 & 2.36e-01 & 7 & SCUBA 1300 \citep{Andrews2007} \\
3300 & 4.50e-03 & 1.06e-02 & 4.22e-02 & 6 & ATCA 3.3 \citep{Ricci2010_Ophiuchus}\\
\enddata
\end{deluxetable*}

\begin{figure*}
\includegraphics[width=.8\hsize]{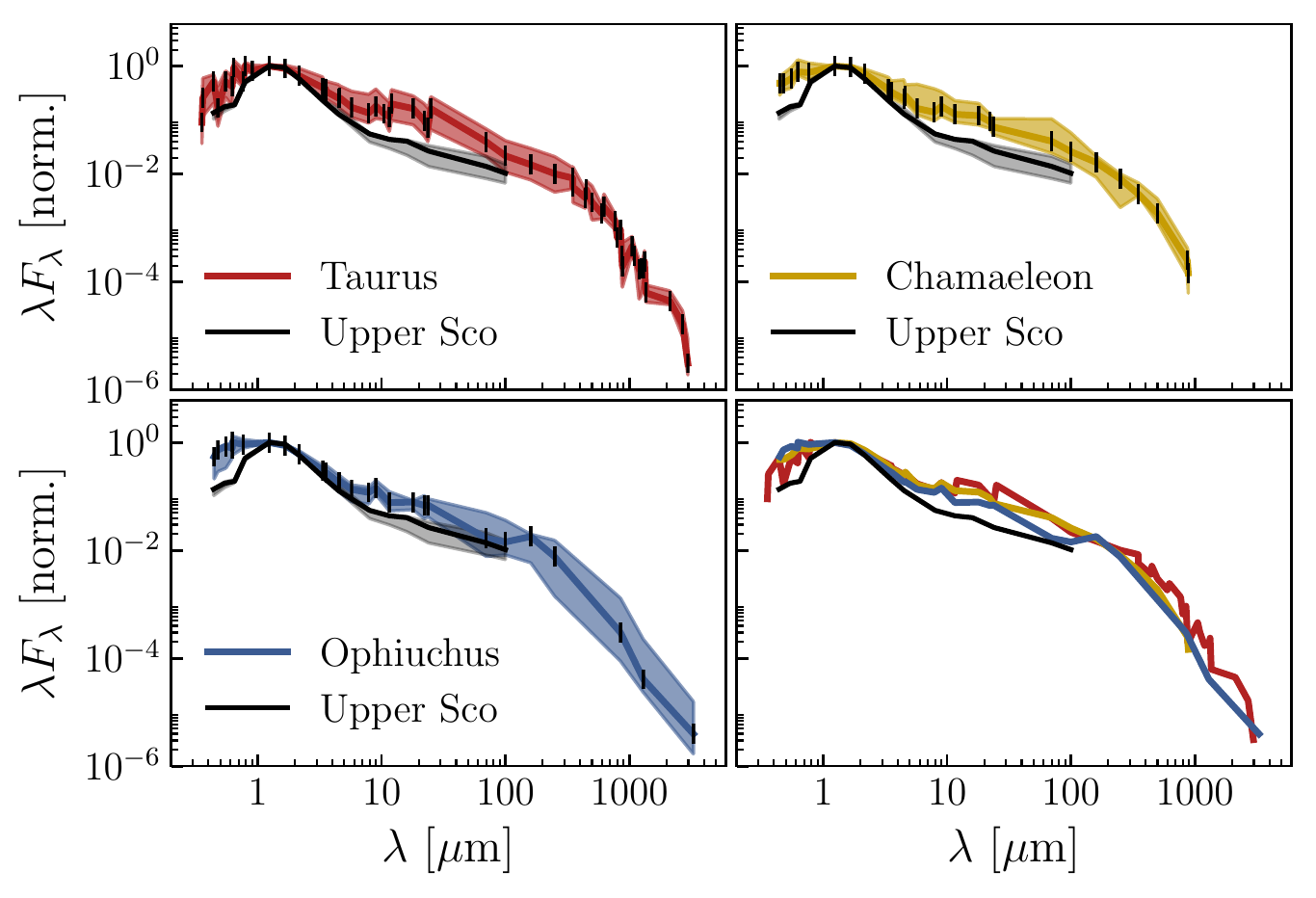}
  \centering
  \caption{Obtained median SEDs for Taurus (top left), Chamaeleon (top right), and
    Ophiuchus (bottom left). These data sets are available as the Data behind the Figure. Shaded areas represent the first and third
    quartiles. The median SEDs are normalized to $J$ band. Small black vertical lines
    mark the available wavelengths in each case. The bottom right panel compares the
    median SEDs of the three regions. The black line shows the median SED and quartiles of
  the older Upper Sco association from \citet{Mathews2013}.}\label{fig:median_seds}
\end{figure*}

The obtained SEDs are shown in Fig.~\ref{fig:median_seds}. We found them to be identical
within the quartiles down to the (sub)mm regime. This comparison expands the result in
\citet{Furlan2009}, where the median IRS spectra of Taurus, Chamaeleon, and Ophiuchus were
already found to be very similar. This is somewhat surprising, given that the three regions
have different ages and more evolved disks are expected in Chamaeleon~I than in
Taurus and Ophiuchus. Additionally, the extent of the upper and lower quartiles also
  appear to be similar for the three regions, suggesting a similar spread in disk
  morphologies in the three regions. We speculate that the intrinsic scatter in disk
morphologies, together with the selection of Class~II disks formerly applied (which
effectively includes disks with certain spectral indices only), may erase existing trends
with age. Including objects classified as (pre)transitional disks does not change
  this result, but the median SEDs compiled here do not include Class~III or diskless
  sources; thus, they represent the typical SED of disks in the regions more so than the median
  SED of all objects. We also note that our sample is not complete, and these median SEDs
are biased toward bright objects.  Despite this, the lack of differences between the
estimated medians, combined with the similar median IRS spectra of these associations,
suggest that either Class~II disks have very similar structures/properties in different
star-forming regions (at least to the extent traceable with SEDs), or the intrinsic
variations in their morphologies are broad enough to create a ``typical protoplanetary
disk'' SED. High-resolution observations have often revealed complex structures in disks,
such as multiple rings, spiral arms, or dust traps \citep[e.g.][]{Muto2012,
  vanderMarel2013, HLTau}, favoring the second explanation, but the number of objects per
region is still significantly small, particularly in Ophiuchus and Chamaeleon. 
Future surveys of large samples of disks, especially in the (sub)mm regime, will produce
more complete and sensitive median SEDs, which may reveal unseen region-to-region
differences and shed some light on the ``typical'' median SED of protoplanetary
disks. For comparison, in Fig.~\ref{fig:median_seds} we also include the median SED
  of K/M-type stars in the Upper Scorpius (Upper Sco) region from \citet{Mathews2013}. The
  absolute age of this region is still under debate, but there is strong evidence for
  Upper Sco being older than Taurus, Chamaeleon~I, and Ophiuchus \citep[4-13\,Myr,
  e.g.][]{Preibisch2002, Pecaut2012, Herczeg2015}. Although the optical part of the
  SEDs cannot be directly compared due to the different treatment in \citet{Mathews2013},
  the median SED of Upper Sco shows a deficit of near/mid-IR excess with respect to the
  younger ones, indicating that disks in Upper Sco are more evolved (e.g. more
  settled). In the future, extending the median SED of disks in Upper Sco to longer
  wavelengths will enable one to search for evidence of dust growth signatures in such
  older regions.

\vspace{0.1cm}
\section{Conclusions}\label{sec:conclusions}

We have compiled multiwavelength data (including \emph{Herschel} photometry and
spectroscopy when available) for 284 Class~II disks in the Taurus, Ophiuchus, and
Chamaeleon~I star-forming regions. These SEDs have been used to study different
aspects of dust growth and properties of protoplanetary disks:

\begin{enumerate}

\item We investigated the spectral index of SEDs as a function of wavelength from the
    far-IR (70\,$\mu$m) to the millimeter, and determined that a small (5-20\,\%)
  systematic shift is introduced in the calculation of the millimeter spectral index when
  combining SPIRE 500\,$\mu$m and (sub)mm observations.

\item We estimated millimeter spectral indices of disks in the three considered
  regions and found their values to be indicative of dust growth in disks in Taurus
  and Ophiuchus, in agreement with previous studies \citep[e.g.][]{Lommen2007,
    Ricci2010_Ophiuchus, Ricci2010_Taurus, Ubach2012}. In contrast, part of the disk
  population in Chamaeleon~I seems to have smaller dust grains.

\item No correlations were found between the mm slopes and other tracers of disk
  evolution (near/mid-IR spectral indices or properties of 10\,$\mu$m silicate
  feature). In particular, the dust properties of transitional disks show no
  appreciable difference with respect to full disks.

\item We used a Bayesian approach to fit the long wavelengths (longward of
  70\,$\mu$m) of sources with sufficient data, using a simple disk model. This allows
  us to estimate posterior distributions for the disk opacity at 10\,au and 1.3\,mm
  ($\tau_{1.3\,mm, 10\,au}$), their temperature at the same distance
  ($T_{\rm 10\,au}$), and the opacity-law exponent ($\beta$). Disk radii from
  high-resolution observations were considered when possible to mitigate the
  degeneracy between compact, optically thick disks, and less dense disks with
  large ($\gtrsim$\,mm) grains.

\item The optical depth values obtained are comparable to estimates for the Minimum Mass
  Solar Nebula. Given that large dust grains produce small $\beta$ values, the obtained
    $\beta$ values ($<1$) in most disks imply some dust growth with respect to the
    ISM. The individual analysis of the regions shows that, as already indicated by
  spectral indices, Chamaeleon contains disks with large $\beta$ values.

\item The distribution of temperatures at 10\,AU ($T_{\rm 10\,AU}$) peaks around
  40-50\,K and shows no significant difference among regions. A strong correlation of
  this parameter with observed 70\,$\mu$m fluxes is also found.

\item There is evidence for a different distribution of spectral indices and $\beta$s in
  Taurus and Ophiuchus with respect to Chamaeleon~I, the latter having steeper values. If
  this difference in spectral indices is caused by its older age, then the high millimeter
  fluxes of its disks require that they were initially more massive than their
  Taurus/Ophiuchus counterparts. After inspecting their location in the HR diagram,
    we find that a bias toward the youngest sources cannot explain all the steep
    slopes. Given the small number of sources with millimeter spectral index measurements
    in Chamaeleon~I, this result should be confirmed with a larger sample.

\item We built the median SED of protoplanetary disks in each region and found them
  to be indistinguishable down to mm wavelengths within their quartiles, suggesting
  that either disks are quite similar in these associations or they display such a large
  variety of morphologies that produces a ``typical'' median SED.

\end{enumerate}

Some of these results are tentative due to the small sample sizes, despite the
large data compilation presented here. Future (sub)mm surveys of disks, especially in
Ophiuchus and Chamaeleon~I, will be critical to obtain better estimates of spectral
indices at these wavelengths and, in particular, determine the origin of the
apparently different spectral indices in Chamaeleon~I.  Additionally, analyzing
the compiled data with detailed disk models such as the ones presented in
\citet{Dalessio1998, Dalessio1999, Dalessio2001, Dalessio2006}, especially when
combined with a Bayesian approach, will inform us of important processes such as the
dust settling.
\vspace{1cm}

We thank the referee for their thorough review and comments that helped improve the quality of this
manuscript. We also thank Melissa McClure and Manoj Puravankara for providing IRS spectra of objects
in Ophiuchus and Chamaeleon, and Greg Schwarz for helping with the data formatting.  This material
is based upon work supported by the National Science Foundation under Grant No. AST-1455042. It is
also funded by \emph{Herschel} Grant No. JPL RSA 1500040. D.A.N. and M.H.D.vdW. acknowledge support
from the CSA and NSERC.  PACS has been developed by a consortium of institutes led by MPE (Germany)
and including UVIE (Austria); KU Leuven, CSL, IMEC (Belgium); CEA, LAM (France); MPIA (Germany);
INAF-IFSI/OAA/OAP/OAT, LENS, SISSA (Italy); IAC (Spain). This development has been supported by the
funding agencies BMVIT (Austria), ESA-PRODEX (Belgium), CEA/CNES (France), DLR (Germany), ASI/INAF
(Italy), and CICYT/MCYT (Spain).  SPIRE has been developed by a consortium of institutes led by
Cardiff University (UK) and including Univ. Lethbridge (Canada); NAOC (China); CEA, LAM (France);
IFSI, Univ. Padua (Italy); IAC (Spain); Stockholm Observatory (Sweden); Imperial College London,
RAL, UCL-MSSL, UKATC, Univ. Sussex (UK); and Caltech, JPL, NHSC, Univ. Colorado (USA). This
development has been supported by national funding agencies: CSA (Canada); NAOC (China); CEA, CNES,
CNRS (France); ASI (Italy); MCINN (Spain); SNSB (Sweden); STFC, UKSA (UK); and NASA (USA).  The
CMC15 Data Access Service is the result of a collaboration agreement between the Centro de
Astrobiolog\'ia (CAB, INTA-CSIC) and the Real Instituto y Observatorio de la Armada en San Fernando
(ROA). It has been developed in the framework of the Spanish Virtual Observatory project, supported
by the Spanish MINECO through grant AYA 2011-14052 and the CoSADIE FP7 project (Call INFRA-2012-3.3
Research Infrastructures, project 312559). The system is maintained by the Data Archive Unit of the
CAB (CSIC -INTA). Based on data from CMC15 Data Access Service at CAB (INTA-CSIC). Funding for
SDSS-III has been provided by the Alfred P. Sloan Foundation, the Participating Institutions, the
National Science Foundation, and the U.S. Department of Energy Office of Science. The SDSS-III web
site is http://www.sdss3.org/. SDSS-III is managed by the Astrophysical Research Consortium for the
Participating Institutions of the SDSS-III Collaboration including the University of Arizona, the
Brazilian Participation Group, Brookhaven National Laboratory, Carnegie Mellon University,
University of Florida, the French Participation Group, the German Participation Group, Harvard
University, the Instituto de Astrofisica de Canarias, the Michigan State/Notre Dame/JINA
Participation Group, Johns Hopkins University, Lawrence Berkeley National Laboratory, Max Planck
Institute for Astrophysics, Max Planck Institute for Extraterrestrial Physics, New Mexico State
University, New York University, Ohio State University, Pennsylvania State University, University of
Portsmouth, Princeton University, the Spanish Participation Group, University of Tokyo, University
of Utah, Vanderbilt University, University of Virginia, University of Washington, and Yale
University.

 \software{HIPE v14 \citep{Ott2010}, emcee \citep{emcee}, Matplotlib \citep{Matplotlib},
  SciPy \citep{Scipy}, Numpy \citep{Numpy}, pandas \citep{Scipy}}

\bibliography{biblio}

\appendix
\newpage
\section{Herschel observations used in this study}\label{appendix:Herschel_data}

\begin{table*}[!hp]
\begin{center}
\caption{Summary of \emph{Herschel} OBSIDs of Large Maps} \label{tab:obsids_large}
\begin{tabular}{cccc}
\tablewidth{0pt}
\hline
\hline
OBSID pair & Wavelengths & \multicolumn{2}{c}{Central Coordinates}\\
& ($\mu$m) & (hh:mm:ss) & (dd:mm:ss)\\
\hline
\multicolumn{4}{c}{Taurus}\\\hline 
1342202088, 1342202089 &  PACS: 70, 160            & 04 03 59 & +26 17 52\\
                                           &  SPIRE: 250, 350, 500  & 04 04 44 & +26 20 01\\
1342190617, 1342190618 &  PACS: 70, 160            & 04 13 07 & +25 01 43\\
                                           &  SPIRE: 250, 350, 500  & 04 12 20 & +25 00 05\\
1342202254, 1342202090* &PACS: 70, 160            & 04 15 05 & +28 27 06\\
                                           &  SPIRE: 250, 350, 500  & 04 15 51 & +28 28 24\\
1342216549, 1342216550 &  PACS: 100, 160          & 04 16 29 & +28 22 56\\
1342204860, 1342204861 &  PACS: 70, 160            & 04 20 38 & +25 13 03\\
                                           &  SPIRE: 250, 350, 500  & 04 21 23 & +25 14 33\\
1342227304, 1342227305 &  PACS: 100, 160          & 04 21 21 & +25 14 14\\
1342202254, 1342190616* & PACS: 70, 160            & 04 25 40 & +27 13 29\\
                                           &  SPIRE: 250, 350, 500  & 04 24 54 & +27 11 58\\
1342202250, 1342202251 &  PACS: 70, 160            & 04 28 45 & +18 31 00\\
                                           &  SPIRE: 250, 350, 500  & 04 29 29 & +18 33 11\\
1342190654, 1342190655 &  PACS: 70, 160            & 04 32 01 & +24 26 31\\
                                           &  SPIRE: 250, 350, 500  & 04 31 15 & +24 25 05\\
1342228005, 1342228006 &  PACS: 100, 160          & 04 31 16 & +24 25 06\\
1342239280, 1342239281 &  PACS: 70, 160            & 04 34 22 & +25 13 23\\
                                           &  SPIRE: 250, 350, 500  & 04 33 38 & +25 12 29\\
1342190652, 1342190653 &  PACS: 70, 160            & 04 35 49 & +23 01 41\\
                                           &  SPIRE: 250, 350, 500  & 04 35 00 & +23 00 04\\
1342202252, 1342202253 &  PACS: 70, 160            & 04 36 49 & +26 00 28\\
                                           &  SPIRE: 250, 350, 500  & 04 37 30 & +26 01 15\\
1342228174, 1342228175 &  PACS: 100, 160          & 04 38 25 & +25 50 14\\
1342204843, 1342204844 &  PACS: 70, 160            & 04 52 35 & +30 45 29\\
                                           &  SPIRE: 250, 350, 500  & 04 53 26 & +30 46 25\\
1342204841, 1342204842 &  PACS: 70, 160            & 05 01 00 & +25 53 28\\
                                           &  SPIRE: 250, 350, 500  & 05 01 46 & +25 54 56\\
\hline\multicolumn{4}{c}{Ophiuchus}\\\hline 
1342205093, 1342205094 &  PACS: 70, 160            & 16 27 28 & -24 12 03\\
                                           &  SPIRE: 250, 350, 500  & 16 26 45 & -24 11 15\\
1342227148, 1342227149 &  PACS: 100, 160          & 16 26 23 & -24 12 09\\
 1342238816, 1342238817 &  PACS: 70, 160          & 16 27 06 & -24 28 42\\
\hline
\multicolumn{4}{c}{Chamaeleon}\\\hline
1342213178, 1342213179 &  PACS: 70, 160             & 10 58 56 & -77 10 45 \\
                                           &  SPIRE: 250, 350, 500  & 11 01 03 & -77 18 50\\
1342224782, 1342224783 &  PACS: 100, 160           & 11 07 48 & -77 25 01\\
\hline
\end{tabular}\\
\end{center}
$*$: obsids 1342202254 (scan), 1342202090 (cross scan 1), 1342190616
(cross scan 2) covered a similar region, but were
processed separately as two sub-maps due to processing limitations, each of them being one scan + cross scan combination.
\end{table*}

\begin{table*}[h!]
\begin{center}
  \caption{Additional \emph{Herschel} Observations of Sources Outside Large Maps} \label{tab:obsids_small}
\begin{tabular}{lcclcc}
\tablewidth{0pt}
\hline
\hline
Object & \multicolumn{1}{c}{OBSIDs} &  Wavelengths & \multicolumn{2}{c}{Object Coordinates} \\
& & ($\mu$m) & (hh:mm:ss) & (dd:mm:ss) \\
\hline
2MASS J04390525+2337450 & 1342243086, 1342243087 & 70, 160 & 04 39 05 & +23 37 45 \\
2MASS J04393364+2359212 & 1342243438, 1342243439 & 70, 160 & 04 39 34 & +23 59 21 \\
2MASS J04400067+2358211 & 1342243436, 1342243437 & 70, 160 & 04 40 01 & +23 58 21 \\
CoKu Tau/4                            & 1342193136                      & 70, 160* & 04 41 17 & +28 40 00 \\
                                              & 1342217520, 1342217521 & 100, 160* &               &               \\
CX Tau                                   & 1342216545, 1342216546 & 70, 160 & 04 14 48 & +26 48 11 \\
                                              & 1342216547, 1342216548 & 100, 160 &               &               \\
DQ Tau                                   & 1342217462, 1342217463 & 70, 160 & 04 46 53 & +17 00 00 \\
                                              & 1342217464, 1342217465 & 100, 160 &               &               \\
DS Tau                                   & 1342193135                       & 70, 160 & 04 47 49 & +29 25 11 \\
                                             & 1342217518, 1342217519 & 100, 160 &               &               \\
FP Tau                                   & 1342216545, 1342216546 & 70, 160 & 04 14 47 & +26 46 26 \\
                                             & 1342216547, 1342216548 & 100, 160 &               &                \\
Haro 6-37                             & 1342193141                     & 70, 160 & 04 46 59 & +17 02 38 \\
                                             & 1342217466, 1342217467 & 100, 160 &               &                \\
LkCa 15                                 & 1342217470, 1342217471 & 70, 160 & 04 39 18 & +22 21 04 \\
                                             & 1342217472, 1342217473 & 100, 160 &               &               \\
RW Aur                                  & 1342193130                     & 70, 160 & 05 07 50 & +30 24 05 \\
                                             & 1342217508, 1342217509 & 100, 160 &               &               \\
\hline
\end{tabular}\\
\end{center}
$*$: 160 $\mu$m flux measurements were discarded in the case of CoKu/Tau 4 due to extended
emission.
\tablecomments{For single observations, only one OBSID is listed.}
\end{table*}

\begin{deluxetable}{c l}
\tablecolumns{2}
\tablewidth{0pt}
\tablecaption{SPIRE FTS spectra Used (after Quality Check) \label{tab:spire_obsids}}
\tablehead{
\colhead{Name} & \colhead{OBSIDs}
}
\startdata
16156-2358AB & 1342242600 \\
16201-2410 & 1342262850 \\
AA Tau & 1342265818 \\
CI Tau & 1342265815 \\
CR Cha & 1342257313 \\
CS Cha & 1342257315 \\
CU Cha & 1342224750 \\
CW Tau & 1342249051 \\
DG Tau & 1342265852 \\
DL Tau & 1342265849 \\
DOAR25 & 1342262851 \\
DQ Tau & 1342228736 \\
DR Tau & 1342243598 \\
GG Tau & 1342265813 \\
GM Aur & 1342228740 \\
Haro 6-37 & 1342228737 \\
IRAS 04385+2550 & 1342243595 \\
IRS48 & 1342262840 \\
LkCa 15 & 1342265848 \\
ROX-44 & 1342262828 \\
RW Aur & 1342243599 \\
SR21AB & 1342262844 \\
SU Aur & 1342228741 \\
SZ Cha & 1342257314 \\
T33A & 1342224749 \\
T42 & 1342248248 \\
T56 & 1342248247 \\
UX Tau A+C & 1342249050 \\
UY Aur & 1342228742 \\
UZ Tau A & 1342265857 \\
V892 Tau & 1342265825 \\
WSB60 & 1342262834 \\
WW Cha & 1342257327 \\
ZZ Tau IRS & 1342265850 \\
\enddata
\end{deluxetable}

\begin{deluxetable}{l l}
\tablecolumns{2}
\tablewidth{0pt}
\tablecaption{SPIRE FTS Spectra (Non-detections and Those Discarded after Quality Check). \label{tab:spire_obsids_discarded}}
\tablehead{
\colhead{Name} & \colhead{OBSIDs}
}
\startdata
BP Tau & 1342250506 \\
CIDA 9 & 1342227782 \\
CY Tau & 1342250504 \\
DD Tau & 1342250505 \\
DE Tau & 1342250507 \\
DH Tau & 1342265854 \\
DK Tau & 1342265856 \\
DM Tau & 1342265814 \\
DN Tau & 1342265817 \\
DO Tau & 1342265859 \\
FN Tau & 1342250503 \\
FQ Tau & 1342239355 \\
FS Tau & 1342250502 \\
FT Tau & 1342265823 \\
FV Tau & 1342265851 \\
FX Tau & 1342265822 \\
GH Tau & 1342250496 \\
GI Tau & 1342250498 \\
GK Tau & 1342265819 \\
GO Tau & 1342243597 \\
Haro 6-13 & 1342265820 \\
GY314 & 1342262838 \\
HK Tau & 1342265821 \\
HP Tau & 1342227449 \\
IRAS 04154+2823 & 1342265826 \\
IRAS 04216+2603 & 1342250501 \\
IRAS 04260+2642 & 1342265853 \\
IP Tau & 1342265824 \\
IQ Tau & 1342265855 \\
MHO 3 & 1342249052 \\
V710 Tau & 1342250495 \\
V807 Tau & 1342250497 \\
V836 Tau & 1342227783 \\
16193-2314 & 1342251283 \\
DOAR16AB & 1342262825 \\
16225-2607 & 1342262853 \\
DOAR21 & 1342262849 \\
DOAR24 & 1342262848 \\
GSS39 & 1342262846 \\
DOAR28 & 1342262824 \\
IRS49 & 1342262839 \\
GY314 & 1342262838 \\
16289-2457 & 1342262833 \\
ROX-43A1 & 1342262829 \\
SX Cha & 1342251309 \\
TW Cha & 1342248245 \\
CT Cha & 1342231974 \\
ISO 52 & 1342257316 \\
T21 & 1342257318 \\
CHXR 20 & 1342231082 \\
UY Cha & 1342257319 \\
UZ Cha & 1342257332 \\
CHXR 22E & 1342251310 \\
T25 & 1342231973 \\
CHXR 30B & 1342257320 \\
VW Cha & 1342224752 \\
T35 & 1342257321 \\
VZ Cha & 1342257331 \\
C7-1 & 1342231081 \\
B43 & 1342231080 \\
T43 & 1342257330 \\
WX Cha & 1342257317 \\
WY Cha & 1342257329 \\
Hn 11 & 1342257328 \\
T47 & 1342257322 \\
WZ Cha & 1342257326 \\
XX Cha & 1342251291 \\
T51 & 1342257325 \\
CV Cha & 1342257324 \\
T54 & 1342257323 \\
2M J11241186-7630425 & 1342243638 \\
\enddata
\end{deluxetable}

\newpage
\section{Herschel photometry}\label{appendix:Herschel_phot}
\begin{deluxetable}{lcccccccc}
\tabletypesize{\scriptsize}
\tablecolumns{9}
\tablewidth{0pt}
\tablecaption{\emph{Herschel} Photometry Obtained in this Study \label{tab:herschel_phot}}
\tablehead{
\colhead{Name} & \colhead{R.A.} & \colhead{Decl.} & \colhead{$F$70\,$\mu$m} & \colhead{$F$100\,$\mu$m} & \colhead{$F$160\,$\mu$m} & \colhead{$F$250\,$\mu$m} & \colhead{$F$350\,$\mu$m} & \colhead{$F$500\,$\mu$m} \\
\colhead{}& \colhead{(hh:mm:ss)} & \colhead{(dd:mm:s)} & \colhead{(mJy)} & \colhead{(mJy)} & \colhead{(mJy)} & \colhead{(mJy)} & \colhead{(mJy)} & \colhead{(mJy)}
}
\startdata
IRAS 04108+2910 & 04:13:57.38 & +29:18:19.3 & 500$\pm$100 & \ldots & 290$\pm$60 & 170$\pm$30 & 110$\pm$20 & 70$\pm$10\\
V773 Tau & 04:14:12.92 & +28:12:12.4 & 800$\pm$200 & 500$\pm$100 & 310$\pm$60 & \ldots & \ldots & \ldots\\
FM Tau & 04:14:13.58 & +28:12:49.2 & 500$\pm$100 & 330$\pm$70 & 280$\pm$60 & 130$\pm$30 & \ldots & \ldots\\
FN Tau & 04:14:14.59 & +28:27:58.1 & 1500$\pm$300 & 1100$\pm$200 & 700$\pm$100 & 390$\pm$80 & 200$\pm$40 & 110$\pm$20\\
CW Tau & 04:14:17.00 & +28:10:57.8 & 1900$\pm$400 & 2000$\pm$400 & 1800$\pm$400 & 1300$\pm$300 & 800$\pm$200 & \ldots\\
CIDA 1 & 04:14:17.61 & +28:06:09.7 & 300$\pm$60 & 410$\pm$80 & 250$\pm$50 & \ldots & \ldots & \ldots\\
MHO 3 & 04:14:30.55 & +28:05:14.7 & 3700$\pm$700 & 3100$\pm$600 & 3500$\pm$700 & \ldots & \ldots & \ldots\\
FP Tau & 04:14:47.31 & +26:46:26.4 & 330$\pm$70 & 380$\pm$80 & 420$\pm$80 & \ldots & \ldots & \ldots\\
CX Tau & 04:14:47.86 & +26:48:11.0 & 310$\pm$60 & 260$\pm$50 & 240$\pm$50 & \ldots & \ldots & \ldots\\
FO Tau & 04:14:49.29 & +28:12:30.6 & 500$\pm$100 & 470$\pm$90 & 120$\pm$20 & \ldots & \ldots & \ldots\\
2MASS J04153916+2818586 & 04:15:39.16 & +28:18:58.6 & 500$\pm$100 & 450$\pm$90 & 500$\pm$100 & 280$\pm$60 & 210$\pm$40 & 180$\pm$40\\
IRAS 04125+2902 & 04:15:42.78 & +29:09:59.7 & 1300$\pm$300 & 1300$\pm$300 & 1000$\pm$200 & 700$\pm$100 & 380$\pm$80 & 200$\pm$40\\
CY Tau & 04:17:33.73 & +28:20:46.9 & 260$\pm$50 & 210$\pm$40 & 280$\pm$60 & 400$\pm$80 & 340$\pm$70 & 250$\pm$50\\
KPNO 10 & 04:17:49.55 & +28:13:31.9 & 120$\pm$20 & 150$\pm$30 & \ldots & \ldots & \ldots & \ldots\\
DD Tau & 04:18:31.13 & +28:16:29.0 & 1200$\pm$200 & 1100$\pm$200 & 700$\pm$100 & 370$\pm$70 & 230$\pm$50 & 130$\pm$30\\
CZ Tau & 04:18:31.59 & +28:16:58.5 & 380$\pm$80 & 100$\pm$20 & \ldots & \ldots & \ldots & \ldots\\
IRAS 04154+2823 & 04:18:32.03 & +28:31:15.4 & 1400$\pm$300 & 1600$\pm$300 & 1200$\pm$200 & 800$\pm$200 & 600$\pm$100 & \ldots\\
V410 X-ray 2 & 04:18:34.44 & +28:30:30.2 & 800$\pm$200 & 600$\pm$100 & 470$\pm$90 & 370$\pm$70 & \ldots & \ldots\\
V892 Tau & 04:18:40.62 & +28:19:15.5 & 50000$\pm$10000 & 40000$\pm$8000 & 23000$\pm$5000 & 9000$\pm$2000 & 5000$\pm$1000 & 2100$\pm$400\\
LR 1 & 04:18:41.33 & +28:27:25.0 & 800$\pm$200 & 1000$\pm$200 & 1200$\pm$200 & 900$\pm$200 & 600$\pm$100 & 500$\pm$100\\
V410 X-ray 6 & 04:19:01.11 & +28:19:42.0 & 500$\pm$100 & 430$\pm$90 & 500$\pm$100 & \ldots & \ldots & \ldots\\
FQ Tau & 04:19:12.81 & +28:29:33.1 & \ldots & 190$\pm$40 & 150$\pm$30 & \ldots & \ldots & \ldots\\
BP Tau & 04:19:15.84 & +29:06:26.9 & 600$\pm$100 & \ldots & 500$\pm$100 & 500$\pm$100 & 400$\pm$80 & 300$\pm$60\\
IRAS 04173+2812 & 04:20:25.84 & +28:19:23.6 & 250$\pm$50 & 90$\pm$20 & \ldots & \ldots & \ldots & \ldots\\
2MASS J04202606+2804089 & 04:20:26.07 & +28:04:09.0 & 320$\pm$60 & \ldots & 190$\pm$40 & 47$\pm$9 & \ldots & \ldots\\
2MASS J04210795+2702204 & 04:21:07.95 & +27:02:20.4 & 3200$\pm$600 & \ldots & 3700$\pm$700 & 1600$\pm$300 & 1500$\pm$300 & 1500$\pm$300\\
DE Tau & 04:21:55.64 & +27:55:06.1 & 1300$\pm$300 & \ldots & 800$\pm$200 & 400$\pm$80 & 270$\pm$50 & 140$\pm$30\\
RY Tau & 04:21:57.40 & +28:26:35.5 & 14000$\pm$3000 & \ldots & 9000$\pm$2000 & 5000$\pm$1000 & 3200$\pm$600 & 1800$\pm$400\\
IRAS F04192+2647 & 04:22:16.76 & +26:54:57.1 & 350$\pm$70 & \ldots & \ldots & \ldots & \ldots & \ldots\\
IRAS 04196+2638 & 04:22:47.87 & +26:45:53.0 & 500$\pm$100 & \ldots & 430$\pm$90 & 340$\pm$70 & 300$\pm$60 & 220$\pm$40\\
IRAS 04200+2759 & 04:23:07.77 & +28:05:57.3 & 500$\pm$100 & \ldots & 270$\pm$50 & 230$\pm$50 & 220$\pm$40 & 160$\pm$30\\
FT Tau & 04:23:39.19 & +24:56:14.1 & 800$\pm$200 & 1000$\pm$200 & 1200$\pm$200 & 1000$\pm$200 & 800$\pm$200 & 600$\pm$100\\
IRAS 04216+2603 & 04:24:44.58 & +26:10:14.1 & 800$\pm$200 & \ldots & 1000$\pm$200 & 700$\pm$100 & 470$\pm$90 & 310$\pm$60\\
IP Tau & 04:24:57.08 & +27:11:56.5 & 500$\pm$100 & \ldots & 160$\pm$30 & 200$\pm$40 & 130$\pm$30 & 110$\pm$20\\
FV Tau & 04:26:53.53 & +26:06:54.4 & 1700$\pm$300 & \ldots & 1000$\pm$200 & \ldots & \ldots & \ldots\\
DF Tau & 04:27:02.80 & +25:42:22.3 & 700$\pm$100 & \ldots & 120$\pm$20 & 70$\pm$10 & \ldots & \ldots\\
DG Tau & 04:27:04.70 & +26:06:16.3 & 21000$\pm$4000 & \ldots & 16000$\pm$3000 & 8000$\pm$2000 & 5000$\pm$1000 & 2900$\pm$600\\
IRAS 04260+2642 & 04:29:04.98 & +26:49:07.3 & 1100$\pm$200 & \ldots & 1500$\pm$300 & 1000$\pm$200 & 800$\pm$200 & 470$\pm$90\\
IRAS 04263+2654 & 04:29:21.65 & +27:01:25.9 & 350$\pm$70 & \ldots & \ldots & \ldots & \ldots & \ldots\\
XEST 13-010 & 04:29:36.07 & +24:35:55.7 & 250$\pm$50 & 200$\pm$40 & 260$\pm$50 & \ldots & \ldots & \ldots\\
DH Tau & 04:29:41.56 & +26:32:58.3 & 500$\pm$100 & \ldots & 500$\pm$100 & 330$\pm$70 & 240$\pm$50 & 140$\pm$30\\
IQ Tau & 04:29:51.56 & +26:06:44.9 & 800$\pm$200 & \ldots & 900$\pm$200 & 800$\pm$200 & 700$\pm$100 & 500$\pm$100\\
2MASS J04295950+2433078 & 04:29:59.51 & +24:33:07.8 & \ldots & 47$\pm$9 & 13$\pm$3 & \ldots & \ldots & \ldots\\
UX Tau A+C & 04:30:04.00 & +18:13:49.4 & 3300$\pm$700 & \ldots & 2600$\pm$500 & 1700$\pm$300 & 1000$\pm$200 & 500$\pm$100\\
FX Tau & 04:30:29.61 & +24:26:45.0 & 350$\pm$70 & 240$\pm$50 & 170$\pm$30 & 100$\pm$20 & \ldots & \ldots\\
DK Tau & 04:30:44.25 & +26:01:24.5 & 1200$\pm$200 & \ldots & 900$\pm$200 & 600$\pm$100 & 360$\pm$70 & 240$\pm$50\\
IRAS 04278+2253 & 04:30:50.28 & +23:00:08.8 & \ldots & \ldots & \ldots & 1100$\pm$200 & 600$\pm$100 & 470$\pm$90\\
ZZ Tau IRS & 04:30:51.71 & +24:41:47.5 & 3800$\pm$800 & 3100$\pm$600 & 3300$\pm$700 & 2600$\pm$500 & 2100$\pm$400 & 1500$\pm$300\\
XZ Tau & 04:31:40.07 & +18:13:57.2 & 8000$\pm$2000 & \ldots & \ldots & \ldots & \ldots & \ldots\\
HK Tau & 04:31:50.57 & +24:24:18.1 & 2600$\pm$500 & 2600$\pm$500 & 2100$\pm$400 & 1400$\pm$300 & 900$\pm$200 & 500$\pm$100\\
V710 Tau & 04:31:57.80 & +18:21:35.1 & 290$\pm$60 & \ldots & 700$\pm$100 & 700$\pm$100 & 500$\pm$100 & 370$\pm$70\\
Haro 6-13 & 04:32:15.41 & +24:28:59.7 & 8000$\pm$2000 & 8000$\pm$2000 & 7000$\pm$1000 & 4100$\pm$800 & 2700$\pm$500 & 2100$\pm$400\\
MHO 6 & 04:32:22.11 & +18:27:42.6 & \ldots & \ldots & 180$\pm$40 & 190$\pm$40 & 170$\pm$30 & 110$\pm$20\\
UZ Tau A & 04:32:43.04 & +25:52:31.1 & 2200$\pm$400 & \ldots & 2100$\pm$400 & 1600$\pm$300 & 1300$\pm$300 & 900$\pm$200\\
JH 112 A & 04:32:49.11 & +22:53:02.8 & 600$\pm$100 & \ldots & 440$\pm$90 & \ldots & \ldots & \ldots\\
V807 Tau & 04:33:06.64 & +24:09:55.0 & 700$\pm$100 & 700$\pm$100 & 600$\pm$100 & 190$\pm$40 & 150$\pm$30 & 100$\pm$20\\
IRAS 04303+2240 & 04:33:19.07 & +22:46:34.2 & 1600$\pm$300 & \ldots & 700$\pm$100 & 320$\pm$60 & \ldots & \ldots\\
2MASS J04333905+2227207 & 04:33:39.05 & +22:27:20.7 & 320$\pm$60 & \ldots & 500$\pm$100 & 430$\pm$90 & 350$\pm$70 & 260$\pm$50\\
DL Tau & 04:33:39.06 & +25:20:38.2 & 1300$\pm$300 & \ldots & 1900$\pm$400 & 2000$\pm$400 & 1800$\pm$400 & \ldots\\
HN Tau & 04:33:39.35 & +17:51:52.4 & \ldots & \ldots & \ldots & 280$\pm$60 & 190$\pm$40 & 100$\pm$20\\
2MASS J04334465+2615005 & 04:33:44.65 & +26:15:00.5 & \ldots & 170$\pm$30 & 280$\pm$60 & 230$\pm$50 & \ldots & \ldots\\
DM Tau & 04:33:48.72 & +18:10:10.0 & \ldots & \ldots & \ldots & 800$\pm$200 & 700$\pm$100 & 600$\pm$100\\
CI Tau & 04:33:52.00 & +22:50:30.2 & 1900$\pm$400 & \ldots & 2000$\pm$400 & 2000$\pm$400 & 1800$\pm$400 & 1300$\pm$300\\
IT Tau & 04:33:54.70 & +26:13:27.5 & 300$\pm$60 & 350$\pm$70 & 430$\pm$90 & 180$\pm$40 & 110$\pm$20 & \ldots\\
AA Tau & 04:34:55.42 & +24:28:53.2 & 1300$\pm$300 & 1000$\pm$200 & 1200$\pm$200 & 1100$\pm$200 & 900$\pm$200 & 600$\pm$100\\
HO Tau & 04:35:20.20 & +22:32:14.6 & \ldots & \ldots & \ldots & 140$\pm$30 & 120$\pm$20 & 90$\pm$20\\
DN Tau & 04:35:27.37 & +24:14:58.9 & 700$\pm$100 & 800$\pm$200 & 800$\pm$200 & 600$\pm$100 & 500$\pm$100 & 400$\pm$80\\
HQ Tau & 04:35:47.34 & +22:50:21.7 & 1400$\pm$300 & \ldots & 600$\pm$100 & 200$\pm$40 & 70$\pm$10 & \ldots\\
2MASS J04381486+2611399 & 04:38:14.87 & +26:11:39.7 & \ldots & 50$\pm$10 & \ldots & \ldots & \ldots & \ldots\\
DO Tau & 04:38:28.58 & +26:10:49.4 & 6000$\pm$1000 & 6000$\pm$1000 & 5000$\pm$1000 & 2500$\pm$500 & 1800$\pm$400 & 1300$\pm$300\\
2MASS J04390525+2337450 & 04:39:05.25 & +23:37:45.0 & 420$\pm$80 & \ldots & 340$\pm$70 & \ldots & \ldots & \ldots\\
VY Tau & 04:39:17.41 & +22:47:53.4 & 240$\pm$50 & \ldots & 220$\pm$40 & \ldots & \ldots & \ldots\\
LkCa 15 & 04:39:17.80 & +22:21:03.5 & 1200$\pm$200 & 1500$\pm$300 & 1800$\pm$400 & \ldots & \ldots & \ldots\\
GN Tau & 04:39:20.91 & +25:45:02.1 & 100$\pm$20 & 170$\pm$30 & \ldots & \ldots & \ldots & \ldots\\
2MASS J04393364+2359212 & 04:39:33.64 & +23:59:21.2 & 80$\pm$20 & \ldots & 60$\pm$10 & \ldots & \ldots & \ldots\\
ITG 15 & 04:39:44.88 & +26:01:52.8 & 350$\pm$70 & 240$\pm$50 & \ldots & \ldots & \ldots & \ldots\\
2MASS J04400067+2358211 & 04:40:00.68 & +23:58:21.2 & 70$\pm$10 & \ldots & 60$\pm$10 & \ldots & \ldots & \ldots\\
IRAS 04370+2559 & 04:40:08.00 & +26:05:25.4 & 600$\pm$100 & 420$\pm$80 & 410$\pm$80 & \ldots & \ldots & \ldots\\
JH 223 & 04:40:49.51 & +25:51:19.2 & 120$\pm$20 & 80$\pm$20 & \ldots & \ldots & \ldots & \ldots\\
ITG 33A & 04:41:08.26 & +25:56:07.5 & \ldots & 140$\pm$30 & \ldots & \ldots & \ldots & \ldots\\
CoKu Tau/4 & 04:41:16.81 & +28:40:00.1 & 1000$\pm$200 & 1100$\pm$200 & \ldots & \ldots & \ldots & \ldots\\
IRAS 04385+2550 & 04:41:38.82 & +25:56:26.8 & 2900$\pm$600 & 3000$\pm$600 & 2600$\pm$500 & 1300$\pm$300 & 700$\pm$100 & \ldots\\
CIDA 7 & 04:42:21.02 & +25:20:34.4 & 310$\pm$60 & 400$\pm$80 & 410$\pm$80 & 230$\pm$50 & 130$\pm$30 & 90$\pm$20\\
DP Tau & 04:42:37.70 & +25:15:37.5 & 600$\pm$100 & 340$\pm$70 & 150$\pm$30 & \ldots & \ldots & \ldots\\
GO Tau & 04:43:03.09 & +25:20:18.8 & 370$\pm$70 & 380$\pm$80 & 600$\pm$100 & 600$\pm$100 & 600$\pm$100 & 500$\pm$100\\
DQ Tau & 04:46:53.05 & +17:00:00.2 & 1400$\pm$300 & 1500$\pm$300 & 900$\pm$200 & \ldots & \ldots & \ldots\\
Haro 6-37 & 04:46:58.98 & +17:02:38.2 & 1100$\pm$200 & 1100$\pm$200 & 1300$\pm$300 & \ldots & \ldots & \ldots\\
DS Tau & 04:47:48.59 & +29:25:11.2 & 200$\pm$40 & 280$\pm$60 & 240$\pm$50 & \ldots & \ldots & \ldots\\
UY Aur & 04:51:47.38 & +30:47:13.5 & 6000$\pm$1000 & \ldots & 3600$\pm$700 & 1500$\pm$300 & 700$\pm$100 & 370$\pm$70\\
GM Aur & 04:55:10.98 & +30:21:59.5 & 3100$\pm$600 & \ldots & 4300$\pm$900 & 4300$\pm$900 & 3300$\pm$700 & 2000$\pm$400\\
AB Aur & 04:55:45.83 & +30:33:04.4 & 140000$\pm$30000 & \ldots & 70000$\pm$10000 & 22000$\pm$4000 & 8000$\pm$2000 & 2500$\pm$500\\
V836 Tau & 05:03:06.60 & +25:23:19.7 & 340$\pm$70 & \ldots & 430$\pm$90 & 320$\pm$60 & 240$\pm$50 & 140$\pm$30\\
CIDA 9 & 05:05:22.86 & +25:31:31.2 & 450$\pm$90 & \ldots & 600$\pm$100 & 450$\pm$90 & 330$\pm$70 & 210$\pm$40\\
RW Aur & 05:07:49.54 & +30:24:05.1 & 2500$\pm$500 & 2900$\pm$600 & 1500$\pm$300 & \ldots & \ldots & \ldots\\
\hline
16156-2358AB & 16:18:37.25 & -24:05:18.19 & \ldots & \ldots & \ldots & 2500$\pm$500 & 1200$\pm$200 & 500$\pm$100\\
16193-2314 & 16:22:18.55 & -23:21:45.36 & 1400$\pm$300 & \ldots & 1400$\pm$300 & 700$\pm$100 & 450$\pm$90 & 250$\pm$50\\
16201-2410 & 16:23:09.23 & -24:17:04.69 & 2400$\pm$500 & 2600$\pm$500 & 2100$\pm$400 & 1000$\pm$200 & 600$\pm$100 & 410$\pm$80\\
16220-2452AB & 16:25:02.13 & -24:59:31.85 & 900$\pm$200 & 500$\pm$100 & 440$\pm$90 & 290$\pm$60 & 160$\pm$30 & 100$\pm$20\\
DOAR16AB & 16:25:10.45 & -23:19:11.96 & 1300$\pm$300 & 1000$\pm$200 & 900$\pm$200 & 420$\pm$80 & 280$\pm$60 & 170$\pm$30\\
IRS2AB & 16:25:36.75 & -24:15:42.12 & 2100$\pm$400 & \ldots & \ldots & \ldots & \ldots & \ldots\\
16225-2607 & 16:25:38.48 & -26:13:53.99 & \ldots & \ldots & \ldots & 90$\pm$20 & 60$\pm$10 & \ldots\\
SR4 & 16:25:56.18 & -24:20:48.22 & 8000$\pm$2000 & 10000$\pm$2000 & 8000$\pm$2000 & \ldots & \ldots & \ldots\\
DOAR24 & 16:26:17.09 & -24:20:21.41 & 1200$\pm$200 & \ldots & \ldots & \ldots & \ldots & \ldots\\
GSS31AB & 16:26:23.38 & -24:20:59.69 & 5000$\pm$1000 & 7000$\pm$1000 & \ldots & \ldots & \ldots & \ldots\\
DOAR25 & 16:26:23.678 & -24:43:13.86 & 1800$\pm$400 & 3200$\pm$600 & 5000$\pm$1000 & 2800$\pm$600 & 2300$\pm$500 & 1500$\pm$300\\
GSS39 & 16:26:45.05 & -24:23:07.72 & \ldots & \ldots & 1900$\pm$400 & 3400$\pm$700 & 2600$\pm$500 & \ldots\\
VSS27AB & 16:26:46.44 & -24:11:59.99 & 700$\pm$100 & 1300$\pm$300 & \ldots & \ldots & \ldots & \ldots\\
DOAR28 & 16:26:47.49 & -23:14:54.79 & 1000$\pm$200 & 1000$\pm$200 & 1000$\pm$200 & 800$\pm$200 & 600$\pm$100 & 400$\pm$80\\
16237-2349 & 16:26:48.66 & -23:56:33.98 & 360$\pm$70 & 380$\pm$80 & \ldots & \ldots & \ldots & \ldots\\
WL18AB & 16:26:48.99 & -24:38:25.1 & 600$\pm$100 & 1400$\pm$300 & \ldots & \ldots & \ldots & \ldots\\
VSSG5AB & 16:26:54.45 & -24:26:20.56 & 70$\pm$10 & \ldots & \ldots & \ldots & \ldots & \ldots\\
GY204 & 16:27:06.61 & -24:41:48.8 & 110$\pm$20 & \ldots & \ldots & \ldots & \ldots & \ldots\\
WL10 & 16:27:09.12 & -24:34:08.29 & 900$\pm$200 & 600$\pm$100 & 340$\pm$70 & \ldots & \ldots & \ldots\\
SR21AB & 16:27:10.28 & -24:19:12.61 & 31000$\pm$6000 & 26000$\pm$5000 & 16000$\pm$3000 & 8000$\pm$2000 & 3900$\pm$800 & 1500$\pm$300\\
IRS36 & 16:27:15.9 & -24:25:14.03 & 200$\pm$40 & 500$\pm$100 & \ldots & \ldots & \ldots & \ldots\\
VSSG25AB & 16:27:27.4 & -24:31:16.57 & 600$\pm$100 & 700$\pm$100 & \ldots & \ldots & \ldots & \ldots\\
GY289 & 16:27:32.67 & -24:33:24.15 & 70$\pm$10 & \ldots & \ldots & \ldots & \ldots & \ldots\\
GY292 & 16:27:33.11 & -24:41:15.14 & 2400$\pm$500 & \ldots & \ldots & \ldots & \ldots & \ldots\\
IRS48 & 16:27:37.18 & -24:30:35.2 & 42000$\pm$8000 & 31000$\pm$6000 & 19000$\pm$4000 & 3700$\pm$700 & 1400$\pm$300 & \ldots\\
IRS49 & 16:27:38.31 & -24:36:58.73 & 1400$\pm$300 & 1600$\pm$300 & 900$\pm$200 & 500$\pm$100 & \ldots & \ldots\\
GY314 & 16:27:39.43 & -24:39:15.51 & 2300$\pm$500 & 2700$\pm$500 & 3300$\pm$700 & 1100$\pm$200 & 500$\pm$100 & \ldots\\
SR9AB & 16:27:40.28 & -24:22:04.31 & 900$\pm$200 & 600$\pm$100 & \ldots & \ldots & \ldots & \ldots\\
GY352 & 16:27:47.08 & -24:45:34.79 & 60$\pm$10 & \ldots & \ldots & \ldots & \ldots & \ldots\\
GY397 & 16:27:55.24 & -24:28:39.72 & 150$\pm$30 & \ldots & \ldots & \ldots & \ldots & \ldots\\
GY463 & 16:28:04.65 & -24:34:56.15 & 39$\pm$8 & \ldots & \ldots & \ldots & \ldots & \ldots\\
WSB60 & 16:28:16.51 & -24:36:57.95 & 900$\pm$200 & 1200$\pm$200 & 1000$\pm$200 & 1300$\pm$300 & 1000$\pm$200 & \ldots\\
ROX-42Cab & 16:31:15.75 & -24:34:02.21 & 220$\pm$40 & 320$\pm$60 & \ldots & \ldots & \ldots & \ldots\\
ROX-43A1 & 16:31:20.12 & -24:30:05.03 & 1100$\pm$200 & 900$\pm$200 & 280$\pm$60 & 180$\pm$40 & \ldots & \ldots\\
IRS-60 & 16:31:30.88 & -24:24:39.88 & 1200$\pm$200 & \ldots & 800$\pm$200 & 390$\pm$80 & 230$\pm$50 & 70$\pm$10\\
ROX-44 & 16:31:33.45 & -24:27:37.11 & 4300$\pm$900 & \ldots & 2600$\pm$500 & 1500$\pm$300 & 900$\pm$200 & 500$\pm$100\\
16289-2457 & 16:31:54.74 & -25:03:23.82 & 2000$\pm$400 & \ldots & \ldots & \ldots & \ldots & \ldots\\
\hline
SX Cha & 10:55:59.73 & -77:24:39.9 & 800$\pm$200 & 600$\pm$100 & 420$\pm$80 & 280$\pm$60 & 190$\pm$40 & 130$\pm$30\\
T5 & 10:57:42.20 & -76:59:35.7 & 220$\pm$40 & 210$\pm$40 & 200$\pm$40 & \ldots & \ldots & \ldots\\
SZ Cha & 10:58:16.77 & -77:17:17.1 & 4000$\pm$800 & 3800$\pm$800 & 3600$\pm$700 & 2800$\pm$600 & 1900$\pm$400 & 1100$\pm$200\\
TW Cha & 10:59:01.09 & -77:22:40.7 & 420$\pm$80 & 340$\pm$70 & 400$\pm$80 & 310$\pm$60 & 210$\pm$40 & 110$\pm$20\\
CR Cha & 10:59:06.99 & -77:01:40.4 & 1600$\pm$300 & 2300$\pm$500 & 2700$\pm$500 & 2400$\pm$500 & 1800$\pm$400 & 1200$\pm$200\\
CS Cha & 11:02:24.91 & -77:33:35.7 & 3200$\pm$600 & 2900$\pm$600 & 2200$\pm$400 & 1400$\pm$300 & 900$\pm$200 & 500$\pm$100\\
CT Cha & 11:04:09.09 & -76:27:19.4 & 700$\pm$100 & 700$\pm$100 & 800$\pm$200 & 500$\pm$100 & 380$\pm$80 & 270$\pm$50\\
ISO 52 & 11:04:42.58 & -77:41:57.1 & \ldots & 200$\pm$40 & \ldots & \ldots & \ldots & \ldots\\
UY Cha & 11:06:59.07 & -77:18:53.6 & \ldots & 170$\pm$30 & \ldots & \ldots & \ldots & \ldots\\
UZ Cha & 11:07:12.07 & -76:32:23.2 & 270$\pm$50 & 340$\pm$70 & 260$\pm$50 & \ldots & \ldots & \ldots\\
T25 & 11:07:19.15 & -76:03:04.8 & 500$\pm$100 & \ldots & 390$\pm$80 & 300$\pm$60 & 180$\pm$40 & 70$\pm$10\\
T28 & 11:07:43.66 & -77:39:41.1 & 460$\pm$90 & 600$\pm$100 & 470$\pm$90 & \ldots & \ldots & \ldots\\
T29 & 11:07:57.93 & -77:38:44.9 & 13000$\pm$3000 & 14000$\pm$3000 & \ldots & \ldots & \ldots & \ldots\\
VW Cha & 11:08:01.49 & -77:42:28.8 & 1400$\pm$300 & 900$\pm$200 & 600$\pm$100 & 220$\pm$40 & \ldots & \ldots\\
CU Cha & 11:08:03.30 & -77:39:17.4 & 120000$\pm$20000 & 70000$\pm$10000 & 60000$\pm$10000 & 39000$\pm$8000 & 24000$\pm$5000 & 12000$\pm$2000\\
T33A & 11:08:15.10 & -77:33:53.2 & 7000$\pm$1000 & 5000$\pm$1000 & 3900$\pm$800 & 2200$\pm$400 & 1300$\pm$300 & 600$\pm$100\\
T35 & 11:08:39.05 & -77:16:04.2 & 400$\pm$80 & 320$\pm$60 & 160$\pm$30 & 120$\pm$20 & \ldots & \ldots\\
VY Cha & 11:08:54.64 & -77:02:13.0 & 340$\pm$70 & 150$\pm$30 & \ldots & \ldots & \ldots & \ldots\\
C1-6 & 11:09:22.67 & -76:34:32.0 & 1500$\pm$300 & 1600$\pm$300 & 1600$\pm$300 & \ldots & \ldots & \ldots\\
VZ Cha & 11:09:23.79 & -76:23:20.8 & 410$\pm$80 & 350$\pm$70 & 370$\pm$70 & 430$\pm$90 & 360$\pm$70 & 270$\pm$50\\
B43 & 11:09:47.42 & -77:26:29.1 & 200$\pm$40 & 160$\pm$30 & 340$\pm$70 & 360$\pm$70 & 350$\pm$70 & 250$\pm$50\\
T42 & 11:09:53.41 & -76:34:25.5 & 15000$\pm$3000 & 19000$\pm$4000 & 17000$\pm$3000 & \ldots & \ldots & \ldots\\
T43 & 11:09:54.08 & -76:29:25.3 & 400$\pm$80 & 500$\pm$100 & \ldots & \ldots & \ldots & \ldots\\
WX Cha & 11:09:58.74 & -77:37:08.9 & 330$\pm$70 & 220$\pm$40 & 120$\pm$20 & 100$\pm$20 & \ldots & \ldots\\
WW Cha & 11:10:00.11 & -76:34:57.9 & 27000$\pm$5000 & 37000$\pm$7000 & 30000$\pm$6000 & 15000$\pm$3000 & 9000$\pm$2000 & 5000$\pm$1000\\
T45a & 11:10:07.04 & -76:29:37.7 & 260$\pm$50 & 300$\pm$60 & \ldots & \ldots & \ldots & \ldots\\
ISO 237 & 11:10:11.42 & -76:35:29.3 & 2900$\pm$600 & 6000$\pm$1000 & \ldots & \ldots & \ldots & \ldots\\
CHXR 47 & 11:10:38.02 & -77:32:39.9 & 220$\pm$40 & 120$\pm$20 & \ldots & \ldots & \ldots & \ldots\\
T47 & 11:10:49.60 & -77:17:51.7 & 700$\pm$100 & 500$\pm$100 & 430$\pm$90 & 320$\pm$60 & 190$\pm$40 & 120$\pm$20\\
WZ Cha & 11:10:53.34 & -76:34:32.0 & \ldots & 160$\pm$30 & \ldots & \ldots & \ldots & \ldots\\
ISO 256 & 11:10:53.59 & -77:25:00.5 & \ldots & 40$\pm$8 & \ldots & \ldots & \ldots & \ldots\\
XX Cha & 11:11:39.66 & -76:20:15.2 & 190$\pm$40 & 180$\pm$40 & \ldots & 100$\pm$20 & 90$\pm$20 & 100$\pm$20\\
T50 & 11:12:09.85 & -76:34:36.5 & \ldots & 120$\pm$20 & \ldots & \ldots & \ldots & \ldots\\
CV Cha & 11:12:27.72 & -76:44:22.3 & 2700$\pm$500 & 2200$\pm$400 & 1300$\pm$300 & 600$\pm$100 & 340$\pm$70 & 150$\pm$30\\
T56 & 11:17:37.01 & -77:04:38.1 & 700$\pm$100 & 600$\pm$100 & 450$\pm$90 & 290$\pm$60 & 180$\pm$40 & 100$\pm$20\\
\enddata
\tablecomments{The complete version of Tables \ref{tab:sample_parameters},
  \ref{tab:mm_spectral_indices_individual}, \ref{tab:IR_and_silicates}, and
  \ref{tab:herschel_phot} are merged
  together in the Zenodo repository, also available in machine readable format in the\\
  online journal.}
\end{deluxetable}

\clearpage
\begin{deluxetable}{l l}
\tablecolumns{2}
\tablewidth{0pt}
\tablecaption{Objects with Nearby Sources/Extended Emission in \emph{Herschel}
  Maps \label{tab:Herschel_problematic}}
\tablehead{
\colhead{Name} & \colhead{Notes}
}
\startdata
CoKu Tau/4 & No flux estimate for $\lambda>$70\,$\mu$m due to extended emission\\
FS Tau & Nearby source \\
FY Tau &  Nearby source (FZ Tau) \\
FZ Tau &  Nearby source (FY Tau) \\
GH Tau &  Nearby source\\
GI Tau & Nearby source (GK Tau)\\
GK Tau &  Nearby source (GI Tau)\\
Haro 6-5B &  Nearby source \\
HP Tau &  Extended emission/nearby source?\\
IC 2087 IR &  Extended emission\\
ITG 40 &  Nearby source\\
JH 112 B &  Emission attributed to the A component, based on SED\\
KPNO 10 &  No flux estimate for $\lambda>$100\,$\mu$m, due to extended emission\\
LkHa 358 &  Nearby source\\
SU Aur &  Extended emission / Nearby source?\\
V807 Tau &  Nearby source, blended longward to 250\,$\mu$m \\
V955 Tau &  Extended emission\\
IRS2AB & No flux estimate for $\lambda>$70\,$\mu$m, due to extended emission\\
SR4 &  Background emission \\
DOAR21 &  Extended emission\\
VSSG1 & Background emission \\
GY12 &  Extended emission\\
VSSG27AB &  Background emission (no detection) \\
GSS37AB & Background emission \\
CRBR51 &  Background emission (no detection) \\
GY262 &  Nearby source \\
GY292 &  Extended emission\\
SR9AB &  No flux estimate for $\lambda>$100\,$\mu$m, due to extended emission\\
ROX-43A2 &  ROX-43 system unresolved in \emph{Herschel} maps. \emph{Herschel} fluxes assigned to A, based on its SED\\
T21 &  Extended emission\\
DI Cha &  Extended emission\\
CHXR 30B &  Nearby source (CHXR 30A)\\
T29 &  Nearby sources (measurements only at 70 and 100\,$\mu$m)\\
CHXR 30A &  Nearby source (CHXR 30B)\\
C1-6 &   No flux estimate for $\lambda>$100\,$\mu$m, due to extended emission\\
Hn 10E &  Background emission (no detection) \\
HD 97300 &  Extended emission\\
XX Cha &  No flux estimate for $\lambda>$100\,$\mu$m due, to nearby source\\
T54 &  Extended emission\\
\enddata
\end{deluxetable}

\clearpage
\section{Silicate feature characterization}\label{appendix:silicates}

As mentioned in Sec.~\ref{sec:other_tracers}, the compiled IRS spectra were used to
characterize the 10\,$\mu$m silicate feature of these disks when possible. Here, we
describe the procedure in more detail.

We adopted the feature strength and shape definitions in \citet{Furlan2006} and
\citet{Kessler-Silacci2006}, respectively. The strength is defined as:
\begin{equation}
{\rm Sil_{strength}} = \frac{\int{(F_{obs} - F_{cont}){\rm d}\lambda}}{\int{F_{cont} {\rm d}\lambda}}
\end{equation}
where $F_{\rm obs}$ and $F_{\rm cont}$ are the observed and continuum fluxes, and the
    integral goes from 8 to 12.4\,$\mu$m, covering the extent of the 10\,$\mu$m feature.
    In the case of the silicate shape, a normalized spectrum ($S_{\rm norm}$) is first estimated:
\begin{equation}
{\rm S_{norm}} = 1 + \frac{F_{obs} - F_{cont}}{<F_{cont}>}
\end{equation}
where $<F_{cont}>$ is the frequency-averaged continuum estimated from 5 to 16\,$\mu$m
\citep{Kessler-Silacci2006}. The shape ($\rm Sil_{shape}$) is then the ratio of fluxes around 9.8
and 11.3 and 9.8\,$\mu$m (${\rm S_{11.3}}/{\rm S_{9.8}}$). The ${\rm S_{11.3}}$ and ${\rm S_{9.8}}$
fluxes were computed as the median flux for wavelengths $\pm$0.2\,$\mu$m around the central
wavelengths, and only if at least three points were available to ensure a robust estimate. Following
\citet{Furlan2011}, this procedure was performed using the observed (i.e. not extinction-corrected)
spectra. The value and uncertainties listed in Table~\ref{tab:IR_and_silicates} are the median, 16th
and 84th percentiles of 1000 bootstrapping iterations, randomly changing flux values in the IRS
spectra within their uncertainties. Given the uncertainty in estimating the continuum level
(especially in the presence of strong silicate features), we followed a similar procedure as in
\citet[][third-order polynomial fit]{Furlan2006} but allowed the polynomial degree to change from 3
to 5 during the bootstrapping, in order to account for this in the final uncertainty estimates.

%
%
%
%



\listofchanges

\end{document}